\documentclass[%
prx, notitlepage,
 amsmath,amssymb,
reprint,%
onecolumn
]{revtex4-1}

\usepackage{CJK}
\usepackage[colorlinks=true, linkcolor=cyan, citecolor=blue, urlcolor =black]{hyperref}
\usepackage{hyperref,color}
\usepackage{amsmath,mathtools}
\usepackage{amsthm}
\usepackage{bbold}
\usepackage[utf8]{inputenc} 
\usepackage{graphicx}
\usepackage{braket}
\usepackage{ragged2e}
\usepackage{wrapfig}
\usepackage{color}
\usepackage{pifont}
\usepackage{siunitx}
\usepackage[english]{babel}

\usepackage[colorinlistoftodos]{todonotes}

\newcommand{\circledone}[1]{\large \ding{192} \normalsize}
\newcommand{\circledtwo}[1]{\large \ding{193} \normalsize}

\begin{document}
\begin{CJK*}{GB}{} 
\title{\sffamily Line-Graph Lattices: Euclidean and Non-Euclidean Flat Bands, and Implementations in Circuit Quantum Electrodynamics} 

\author{Alicia J. Koll\'{a}r}
\affiliation{Department of Electrical Engineering, Princeton University, Princeton, NJ 08540, USA}
\affiliation{Princeton Center for Complex Materials, Princeton University, Princeton, NJ 08540, USA}
\author{Mattias Fitzpatrick}
\affiliation{Department of Electrical Engineering, Princeton University, Princeton, NJ 08540, USA}
\author{Peter Sarnak}
\affiliation{Department of Mathematics, Princeton University, Princeton, NJ 08540, USA}
\author{Andrew A.\ Houck}
\affiliation{Department of Electrical Engineering, Princeton University, Princeton, NJ 08540, USA}

\date{\today}

\begin{abstract}
Materials science and the study of the electronic properties of solids are a major field of interest in both physics and engineering.
The starting point for all such calculations is single-electron, or non-interacting, band structure calculations, and in the limit of strong on-site confinement this can be reduced to graph-like tight-binding models.
In this context, both mathematicians and physicists have developed largely independent methods for solving these models.
In this paper we will combine and present results from both fields. In particular, we will discuss a class of lattices which can be realized as line graphs of other lattices, both in Euclidean and hyperbolic space. 
These lattices display highly unusual features including flat bands and localized eigenstates of compact support. We will use the methods of both fields to show how these properties arise and systems for classifying the phenomenology of these lattices,
as well as criteria for maximizing the gaps.
Furthermore, we will present a particular hardware implementation using superconducting coplanar waveguide resonators that can realize a wide variety of these lattices in both non-interacting and interacting form.
\end{abstract}
\maketitle 
\end{CJK*}

\tableofcontents

\section{Introduction}\label{sec:intro}
The study of the electronic properties of materials is one of the major fields of physics, and an extensive toolbox of techniques has been built up for computing them. Within the contexts of analysis, algebra, and combinatorics, mathematicians have developed complementary methods which can be applied to these systems. In this paper we will combine the known methods from physics with theorems from graph theory describing the properties of line graphs in order to analyze a class of lattices which exhibit unusual band structures with spectrally-isolated flat bands. We will present the existing frameworks and results, and explicitly prove relevant corollaries obtained from combining the results of both fields. We will then show how these results can be applied to both Euclidean and non-Euclidean tight-binding solids and review how such models can be realized experimentally using the techniques of circuit Quantum Electrodynamics (cQED) and resonators made from distributed element waveguides such as the microwave coplanar waveguide (CPW).

The structure and dynamics of non-relativistic quantum systems are generally described using the Hamiltonian operator which encodes the total energy of the system. For interacting electrons, the Hamiltonian is a non-linear operator which is generally impossible to solve exactly. However, in many cases, the properties of the corresponding non-interacting Hamiltonian provide a very useful starting point from which to study the fully-interacting model. In this simplified limit, the Hamiltonian of a crystalline solid takes the following form:
\begin{equation}\label{HamiltonianEQ}
\mathbb{H} = -\Delta+ \mathcal{V},
\end{equation}
where $\Delta$ is the Laplacian operator which describes kinetic energy, and $\mathcal{V}$ is a potential that is periodic in space arising due to the ionic cores of atoms in the lattice:
\begin{equation}
\mathcal{V}(\vec{x}) = \sum_{\mbox{sites \it{n}}}{V_0 (\vec{x} - \vec{x}_n)}.
\end{equation}

There are two limits in which there exist well-known techniques for calculating the eigenenergies and eigenvectors of this operator. 
The first is when the periodic potential $\mathcal{V}$ is weak compared to the kinetic energy, the so-called nearly-free-electron limit. 
Here, the lattice produces small perturbations on the the free-electron ($V_0=0$) solution when the momentum is equal to one of the Fourier components of the lattice potential. 

In the opposite limit, called the tight-binding limit, the kinetic energy is weak compared to the potential created by the atomic lattice. Calculations in this limit start from the single-electron bound states for each ionic core, and typically provide accurate solutions for the low-energy portion of the spectrum. If the lattice sites are well-separated compared to the width of the bound states, then solutions to Eqn. \ref{HamiltonianEQ} are well-approximated by linear combinations of these bound states. The tight-binding approximation consists of restricting the Hilbert space to consider only solutions of this form. Within this approximation, the continuum problem can be replaced by a discrete model. (For simplicity assume $V_0$ is centered about the origin and only consider one bound state $\beta (\vec{x})$. An extension to more is straightforward.\cite{Ashcroft:1976ud}) In the discrete model each state $\psi$ is given by a vector in $C_N$, where $N$ is the number of sites in the lattice, which will eventually be taken to infinity.  Each element of $\psi$ encodes the complex amplitude of the bound state $\beta$ on the corresponding site
\begin{equation}\label{TBStateDef}
\psi= (\psi_1, \psi_2, \cdots)  = \sum_{n} {\psi_n \beta(\vec{x} - \vec{x}_n)}.
\end{equation}

By construction, each state in this space is a good approximation to an eigenstate of the potential $\mathcal{V}$ with eigenvalue $E(\beta)$ because each on-site wavefunction is an eigenstate of $-\Delta + V_0 (\vec{x} - \vec{x}_n)$ with this same eigenvalue, and the lattice sites are tightly confined and well-separated in space. The central result of the tight-binding approximation is to systematically compute corrections to this eigenvalue.
At lowest order there are two dominant effects.
The first is a renormalization of the bound state energy $E(\beta)$, which is constant global offset and can be set to zero in the single-band approximation we have taken. 
The second effect is a coupling between neighboring lattice sites due to the fact that they are not infinitely far apart and the tails of the on-site wavefunctions overlap.
It is this coupling that allows particles to move within the lattice by hopping from one site to another. Since it falls of very rapidly with distance, the problem is dominated by the configuration of nearest neighbors.
Assuming there is a single dominant nearest-neighbor distance, the restricted tight-binding Hamiltonian $H$ becomes a particularly simple bilinear form acting on the restricted Hilbert space:
\begin{equation}\label{Hij}
H_{i,j} = -t,
\end{equation}
if sites $i$ and $j$ are nearest neighbors and zero otherwise, where $t^2$ is hopping rate between nearest-neighbor sites. The magnitude of the hopping rate is a property of the bound state $\beta$ and the distance between the lattice sites, which is generally computed numerically.\cite{Jones:2015ic} Defining the creation operator  $a^\dagger_i$ which projects any input state onto the state $\psi_n = \delta_{n,i}$, and the annihilation operator $a_i$ which is its conjugate transpose, $H$ can then be written in its second quantized (and typical physics-notation) form:
\begin{equation}\label{TBHam}
H =  - t \sum_{\left<i,j \right>}{(a_i^{\dagger}a_j + a_j^{\dagger}a_i)},
\end{equation}
where the symbol $\left<i,j \right>$ denotes all nearest-neighbor pairs in the lattice. When the sites of the lattice form a periodic tiling in Euclidean space, there exist well-known methods for obtaining all eigenvalues and eigenvectors of $H$ by exploiting the discrete subgroups of Eudclidean space. (See Ref.\cite{Ashcroft:1976ud} for a physicist's treatment, and Ref.\cite{Kotani:2003jq} for a translation of these methods into the language of abstract algebra.) However, $H$ can also be understood as the transition matrix of a graph, and is very closely related to the graph Laplacian.

The purpose of this work is to collect and combine complementary results from both solid-state physics and mathematics and thereby gain new insight into the properties of $H$. In particular, we will apply these insights to the circuit QED lattices introduced in Refs.\cite{Houck:2012iq, Underwood:2012hx, Kollar:2018vc}.
To this end, the remainder of paper is organized as follows. We will begin by introducing the physics of circuit QED lattices and the tight-binding models that can be realized with them, namely s-wave and p-wave tight binding models on lattices which are the medial lattice or line graph of another.  
We will then introduce the physics reader to relevant theorems from graph theory which relate the spectrum of $H$ on an underlying graph to the effective tight-binding operators on the line graph, guaranteeing that every such lattice has a flat band. To illustrate the consequences of these theorems, we will apply them to two sets of examples. First, we will examine a set of Euclidean lattices which can be treated with both traditional solid state methods and graph-theoretic analysis, contrasting the two sets of techniques and the different types of information that are readily obtained from each. 
Second we will examine a set of hyperbolic examples which can only be treated with graph theoretic methods, and show that despite the absence of an applicable Bloch band theory, these models also display infinite multiplicity degenerate eigenvalues and eigenstates of compact support.

Such flat-band lattices are of particular interest in the fields of many-body physics and quantum simulation because their properties are very sensitive to the presence of interactions that lift the degeneracy of the flat-band.\cite{Leykam:2018vd,Bergman:2008es} Implementing them at the hardware level in superconducting circuits provides a new opportunity to study non-linear and interacting quantum mechanical models on these lattices \cite{Houck:2012iq,Kollar:2018vc,Annunziata:2010dg} and observe many-body physics with photons.
Therefore, we present a study of the conditions under which flat bands arise and are isolated from the rest of the spectrum both in the Euclidean and non-Euclidean cases. We will derive criteria for when gaps can occur and sharp bounds on their maximum size, and identify examples which attain these bounds. 
We will show that frustrated hopping and non-bipartite graphs are essential to the creation of these gaps. Such lattices cannot be divided into two sublattices such that all the neighbors of any site are in the opposite sublattice and they break so-called particle-hole symmetry.
Additionally, we will examine the effects of finite, hard-wall, boundary conditions on the graph spectra and gaps.

\section{The Tight-Binding Hamiltonian and the Graph Laplacian}\label{sec:defs}

Since this paper combines results from both mathematics and physics, and is intended for readers from either field, we will attempt to define and translate the terminology of both solid-state physics and graph theory. The following section is a rapid review of common definitions and conventions in both fields in sufficient detail to allow discussion of the results in the body of the paper. The main results in this particular section will not be proved, only motivated, and we refer the reader to standard texts in both fields for thorough derivations and proofs.\cite{Ashcroft:1976ud, Kittel, GraphTheoryBook, HararyGraphTheory, Kotani:2003jq}

We will define a set of lattice points $\mathcal{P}$ as a set of points periodically spaced in a metric space. For the purposes of this paper, we will restrict ourselves to two dimensional examples. The theoretical results generalize readily to other dimensions, but the circuit QED  hardware implementation is inherently limited to two dimensions, so we will not burden the reader with the notation necessary to keep track of more. The simplest example of a two-dimensional lattice is all integer linear combinations of two linearly independent vectors in $\mathbb{R}_2$. (See for example the vertices of the two-dimensional square lattice in Fig. \ref{fig:G_LG_Examples} \textbf{b}.) However, we will also consider more complicated examples, such as the hexagonal lattice in Fig. \ref{fig:G_LG_Examples} \textbf{a}, which are periodic translations of two or more points. (See Refs.\cite{Ashcroft:1976ud,Kotani:2003jq}).

A lattice $\mathcal{L}$ is defined to be a set of lattice points $\mathcal{P}$ and a set of hopping matrix elements $t_{i,j}$ between them, so that we can consider $\mathcal{L}$ as the tight-binding description of a solid. For now, we will restrict ourselves to lattices where all non-zero $t_{i,j}$ assume the same value, $t$. The choice of these non-zero $t_{i,j}$ therefore defines the notion of a nearest-neighbor in the lattice. Among naturally occurring materials, those that exhibit only a single value of $t$ are a special, highly symmetric class. However, for the artificial materials made from CPW resonators\cite{Kollar:2018vc}, the class of such examples is much broader since uniform hopping can be achieved in the absence of a high-symmetry realization in Euclidean space.\cite{Kollar:2018vc}

With each such lattice $\mathcal{L}$ we will associate a graph $X$, with a vertex set $V(X)$ containing exactly one vertex associated to each of the lattice points in $\mathcal{P}$, and an edge set $\mathcal{E}(X)$ which contains the pair $(xy)$ if and only if $t_{x,y}$ is non-zero.\cite{GraphTheoryBook,HararyGraphTheory}
In that case,
\begin{equation}\label{HT}
H_X=  -t  A_X,
\end{equation}
where $A_X$ is the adjacency (transition) matrix of $X$.
The most common realization (or drawing) of $X$ is to place each vertex at the corresponding point in $\mathcal{P}$, but the properties of $A_X$, and by extension $H$, are independent of the precise realization of $X$. Thus, while this is often a convenient choice, it is a matter of convention.

We define a neighborhood set of a vertex $x$ as 
\begin{equation}\label{neighborhood}
\mathcal{N}_x = \{ y  \in V(X) : xy \in  \mathcal{E}(X)\}.
\end{equation}
The degree, or coordination number, $\mathbb{d}(x)$ of a vertex is defined as the cardinality of its neighbor set, i.e. the number of nearest neighbors. A graph is regular if all its vertices have the same degree.

The physics of particle motion on the graph is governed by the tight-binding (or ``hopping'') Hamiltonian $H_X$ which acts on the components of a state $\psi$  by
\begin{equation}\label{Hpsi}
(H_X\psi)_x = - t \sum_{y \in \mathcal{N}_x}{\psi_y},
\end{equation}
and the eigenvalues of $H_X$ are the allowed eigenenergies.
Mathematical convention is to use the closely related graph Laplacian
\begin{equation}\label{DeltaG}
(\Delta_X \psi)_x = \mathbb{d}(x) \psi_x- \sum_{y \in \mathcal{N}_x}{\psi_y} .
\end{equation}
(The sign of this operator and whether or not it is normalized by  $\mathbb{d}(x)$ is convention and may vary from reference to reference.)
In the case of a regular graph of degree $d$, these two operators are very simply related by a constant offset and a multiplicative factor:
\begin{equation}\label{HandDelta}
\Delta_X =  d I + \frac{H_X}{t},
\end{equation}
where $I$ is the identity.
To simplify the notation, we will assume that $t=-1$ for the remainder of the paper. (This is an unusual convention since most naturally occurring materials exhibit $t>0$; however, the realizations of these models in superconducting circuits have $t<0$ by default.\cite{Schmidt:2013hg,Kollar:2018vc})
For any regular graph, the spectrum of $H_X$ is contained in the interval $[-d,d]$ and the spectrum of $\Delta_X$ in the interval $[0,2d]$.

\section{Circuit QED Lattices}\label{sec:implementation}

\subsection{Tight-Binding Solids in Circuit QED}
 
We now consider the implementation of lattices using distributed elements, most commonly microwave CPW resonators in superconducting circuits.\cite{Houck:2012iq, Schmidt:2013hg, Underwood:2012hx, Kollar:2018vc} Coplanar waveguides are effectively a two-dimensional analog of cylindrical coaxial cables fabricated from metallic films, whose modes are described by the transmission line Lagrangian.\cite{Pozar,GirvinNotes,Schmidt:2013hg} The classical literature \cite{Pozar} gives the mode functions in terms of the voltage along the transmission line, whereas the quantum circuits literature \cite{GirvinNotes,Schmidt:2013hg} typically uses the closely related generalized flux $\Phi(x)$. In the case of CPWs, the two are proportional to one another up to unitful constants. The modes obey a one-dimensional wave equation, and resonators are formed by engineering gaps in the center pin, resulting in capacitive termination.

\begin{figure}[h]
	\begin{center}
		\includegraphics[width=0.9\textwidth]{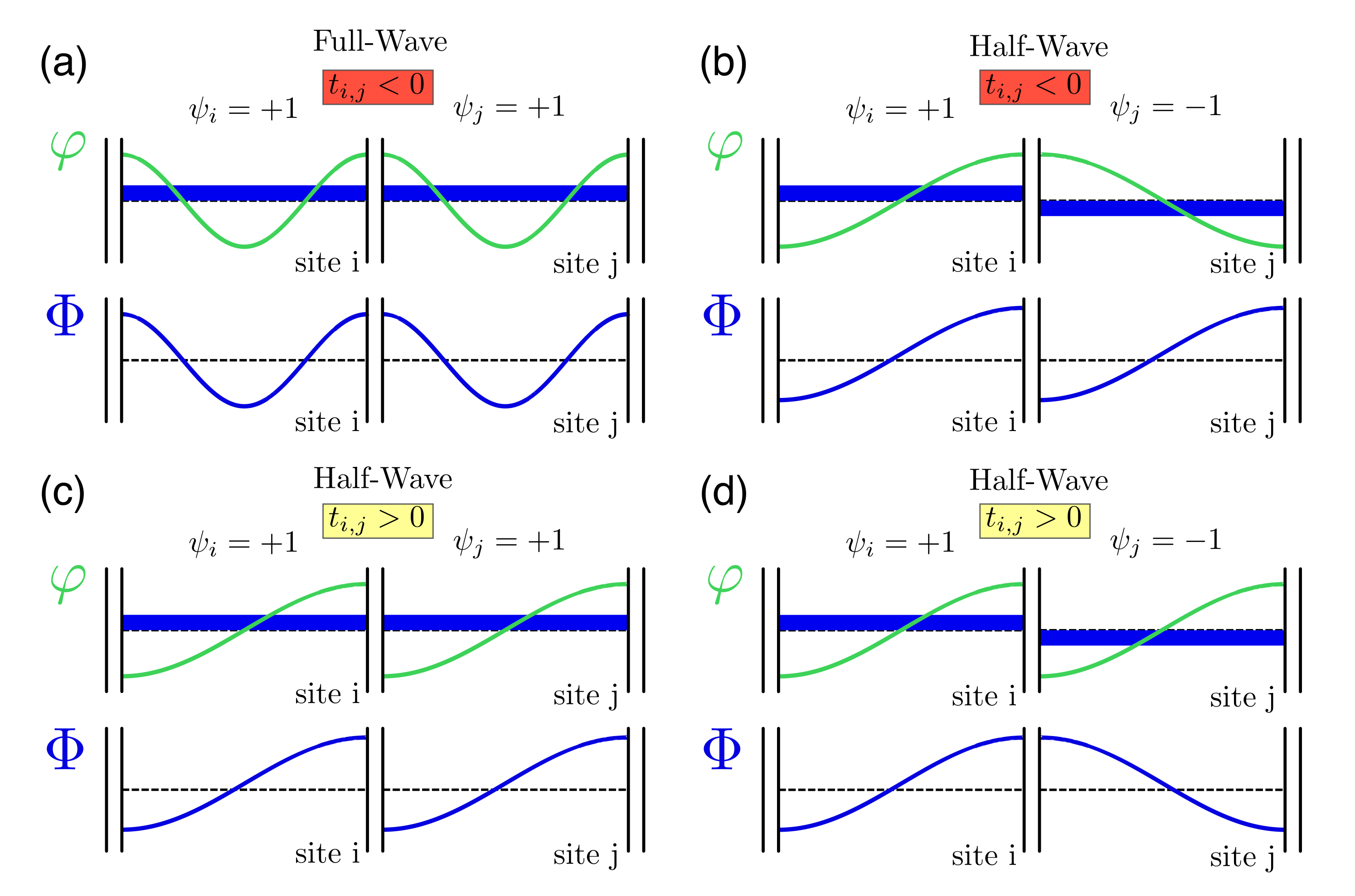}
	\end{center}
	\vspace{-0.6cm}
	\caption{\label{fig:FWHWTB} 
    \textbf{Generalized fluxes and tight-binding wavefunctions.}
    Examples of four possible configurations of $\Phi(x)$ for two coupled cavities and their resulting tight-binding descriptions. For each configuration two copies of the resonator pair are shown. In the upper plot, the on-site wavefunction $\varphi(x)$ is plotted in green for each cavity. The sign of the tight-binding matrix element $t$ is determined by whether or not the $\varphi$ have the same or opposite sign at the coupling capacitor.
    The sign of the tight-binding wavefunction is written above each resonator and indicated by the light blue bar along the length of each cavity.
    The corresponding generalized flux $\Phi(x) = \psi_i \varphi(x,i)$ is plotted in dark blue in the lower pair of cavities.
    \textbf{a} The natural choice of $\varphi(x,i)$ for full-wave modes. The on-site wavefunction is positive at both ends of \textit{all} cavities and $t_{i,j}$ is always negative.
    \textbf{b}-\textbf{d} Possible orientation choices for half-wave modes.
    \textbf{b} shows the choice most similar to the full-wave, where the $\varphi$ are chosen such that they are both positive at the coupling capacitor and $t<0$. However, in non-bipartite graphs, it will not be possible to choose the $\varphi$ such that all coupling capacitors follow this case.
    The choice of $\varphi$'s and $\psi$'s in \textbf{c} corresponds to the same $\Phi(x)$ as in \textbf{b}, but $t>0$ because the $\varphi$ are opposite. Correspondingly, the tight-binding wavefunction is $1,1$ instead of $1,-1$.
    Due to the change in sign of $\varphi(x,j)$ in this choice of orientation, the tight-binding wavefunction $1,-1$ corresponds to a different $\Phi(x)$, shown in \textbf{d}.	
    } 
\end{figure}

Let $x$ be the position along a resonator of length $l$. The boundary condition due to the capacitive termination tends to $d\Phi/dx =0$ in the limit of infinite gap size and vanishing input/output capacitance. The lowest two resonant eigenmodes, shown in Fig. \ref{fig:FWHWTB}, are $\lambda/2 = l$ (half-wave) and $\lambda = l$ (full-wave) standing waves with antinodes at the ends of the resonator and on-site wavefunctions $\varphi(x)  =  \sqrt{2/ c l} \cos(\pi x / l)$ and $\varphi(x)  =  \sqrt{2/ c l} \cos(2\pi x /l)$, where $c$ is the capacitance per unit length of the waveguide.\cite{Schmidt:2013hg}

A single resonator is a one-dimensional object, but the individual eigenmodes can be described as simple harmonic oscillators
\begin{equation}
H_{cav} = \omega_0\, a^\dagger a,
\end{equation}
where $\omega_0$ is the resonant frequency, $a^\dagger$ is a creation operator which adds one photon to the resonator and produces a wavefunction $\varphi$ from the vacuum. The annihilation operator $a$ is its conjugate transpose which removes one photon. Multiple resonators can be coupled capacitively when their ends come into close proximity. Networks of coupled CPW resonators can be regarded as artificial materials in the tight-binding approximation where the individual resonators replace the on-site potential $V_0(\vec{x})$, the single resonator eigenmodes take the place of the bound state wavefunction, and microwave photons replace carrier electrons.\cite{Houck:2012iq,Schmidt:2013hg,Kollar:2018vc}

The appropriate description of kinetic-energy-like terms due to movement of photons between resonators can be derived through careful analysis of the transmission line Lagrangian, and gives rise to a new term in the total Hamiltonian
\begin{equation}\label{HinV1}
\mathcal{H} = \sum_{\mbox{resonators}} \omega_0 a^\dagger_n a_n + \sum_{\mbox{resonator ends}} {\frac{1}{2} \omega_0 C_c \Phi^- \Phi^+},
\end{equation}
where $C_c$ is the coupling capacitance between resonators; and $\Phi^-$, $\Phi^+$, are the values of the generalized fluxes on either side of the coupling capacitor.\cite{Schmidt:2013hg} Eqn. \ref{HinV1} is in a hybrid form involving both creation and annihilation operators and $\Phi$'s, which makes it very difficult to compute with. It is convenient to eliminate one of these pairs. 

Using the half-wave and full-wave mode functions it is possible to convert between $a_n$ and $a^\dagger_n$ and  $\Phi^{(1)}_n$, $\Phi^{(2)}_n$, the voltages at the ends of a resonator,  to produce a new Hamiltonian operator that has the same eigenstates and eigenvalues as Eqn. \ref{HinV1}, but depends only on the value of $\Phi$ at the ends of the resonators:
\begin{equation}\label{HinV2}
\mathcal{H}_\Phi = \sum_{\mbox{resonators}} \frac{cl}{4}\omega_0 (|\Phi^{(1)}_i|^2+ |\Phi^{(2)}_i|^2)\ \  +  \sum_{\mbox{resonator ends}} {\frac{1}{2} \omega_0 C_c \Phi^- \Phi^+},
\end{equation}
with an additional set of constraints $\Phi^{(1)}_i = \beta \Phi^{(2)}_i$, where $\beta = 1$ for the symmetric full-wave modes and $\beta = -1$ for the antisymmetric half-wave modes. Eqn. \ref{HinV2} is extremely useful for gaining physical intuition about voltages and generalized fluxes in the device and interference effects which shape the spectrum and lead to flat bands, but due to the large number of Lagrange multipliers and the need to carefully associate $\Phi^\pm$ with $\Phi^{(1)}_i$ and  $\Phi^{(2)}_i$, it is cumbersome to compute with. A computationally much more convenient form is obtained using the constraints to eliminate all of the generalized fluxes in favor of the creation and annihilation operators:
Eqn. \ref{HinV2} can now be rewritten as a sum over resonators $i$ and pairs $\left<i,j \right>$ of neighboring resonators which share ends points, yielding 
 \begin{eqnarray}\label{HcpwLattice}
\mathcal{H}_{eff}  & =& \sum_{i} \omega_0 a^\dagger_i a_i  - \sum_{\left< i,j \right>}{t_{i,j} (a^\dagger_i a_j + a^\dagger_j a_i)},\\
H_{eff}  & =&  - \sum_{\left< i,j \right>}{t_{i,j} (a^\dagger_i a_j + a^\dagger_j a_i)}, \nonumber
 \end{eqnarray}
where $\omega_0$ has been set to zero in the second equation.
The effective hopping rate is given by
\begin{equation}\label{tij}
t_{i,j} = -\frac{1}{2}\omega_0 C_c \times \varphi(i, x^{(i)}_{end}) \times \varphi(j, x^{(j)}_{end}),
\end{equation}
where $x^{(i)}_{end}$ is the coordinate of the relevant coupling capacitor along the length of the $i$th resonator.

The Hamiltonian in Eqn. \ref{HcpwLattice} is an effective tight-binding model on a lattice which has one site per resonator, and non-zero hopping matrix elements if and only if the respective resonators share end points. 
Therefore, we can associate two graphs with each device, one related to the physical layout of the device and one related to the effective graph which describes the Hamiltonian structure. The first graph $X = (V, \mathcal{E})$ has an element in $V(X)$ for every coupling capacitor in the network, and an element in $\mathcal{E}(X)$ for every resonator. We will refer to this graph as the layout graph since its realizations closely resemble the physical hardware layout. However, $H_X$ and $\Delta_X$ are not the correct operators for describing particle motion in the device. For that purpose a second graph $X_{eff}$ is required, which has a vertex for every resonator and edges connecting two such vertices if and only if the corresponding resonators touch, and whose edges are weighted by the hopping matrix elements $t_{i,j}$. We will refer to this as the effective graph or effective lattice.
The tight-binding wavefunctions $\psi$ on this effective lattice are vectors in $\mathbb{C}^{|\mathcal{E}|}$ which specify the generalized flux on the chosen end of each resonator and encode its value everywhere via $\Phi(x) = \psi_i \varphi(x,i)$.

Henceforth, we will restrict to the simplest case in which all coupling capacitors are equal. The symmetry of $\varphi$ about the center of the resonator guarantees that there are only two possible values of $ \varphi(i, x^{(i)}_{end})$ which we will take to be $\pm 1$, and, therefore, that there are only two possible values of $t_{i,j}$. It will be negative when both values of $\varphi$ in Eqn. \ref{tij} are equal and positive if the $\varphi$ are equal and opposite. 
Examples of orientation choices and the resulting effective hopping matrix elements are shown in Fig. \ref{fig:FWHWTB}. Since the orientation of $\varphi(i,x)$ was chosen arbitrarily for each resonator, each hardware device can be described by many different weighted graph models, depending on the particular choice of gauge.\cite{Schmidt:2013hg,Kollar:2018vc} However, some of these choices are simpler or more illustrative than others.

\subsection{Full-Wave Versus Half-Wave Models: Line-Graph Effective Lattices} \label{subsec:FWHWsetup}

\begin{figure}[h]
	\begin{center}
		\includegraphics[width=0.9\textwidth]{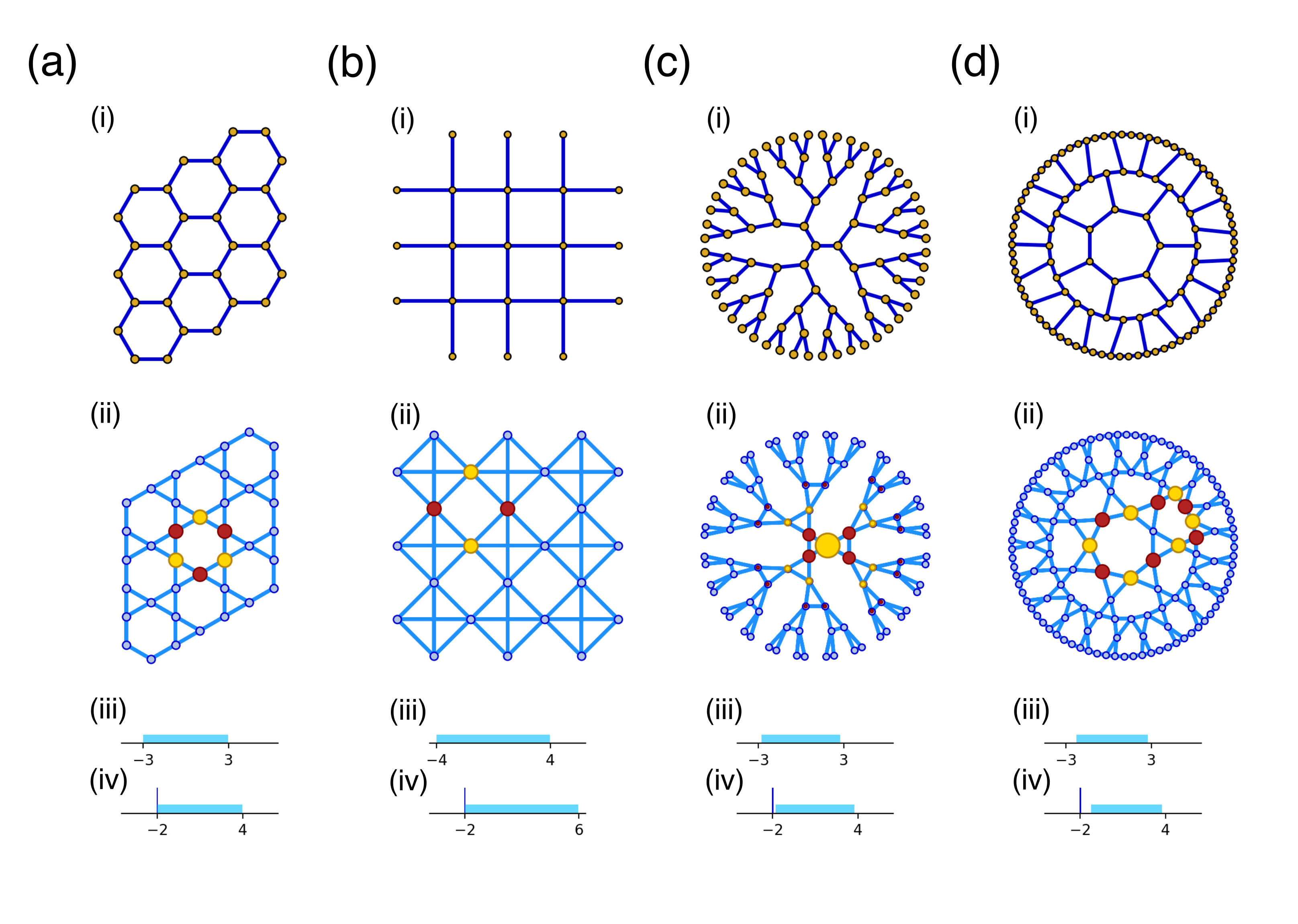}
	\end{center}
	\vspace{-0.6cm}
	\caption{\label{fig:G_LG_Examples} 
    \textbf{Homogeneous infinite line graphs and their energy spectra.}
    \textbf{a}\textit{i} A finite section of the graph of a hexagonal (graphene) lattice. Blue lines indicate hopping matrix elements, and brown circles indicate lattice sites. \textbf{a}\textit{ii} Its line graph, the kagome lattice, where light blue lines indicate hopping matrix elements and white circles indicate lattice sites in the line graph. \textbf{a}\textit{iii} Schematic of the regions in which the $\ell^2$ density of states (DOS) is nonzero for the infinite hexagonal lattice with $t=-1$. This lattice is Euclidean and $3$-regular, and its DOS is supported on the entire interval $[-3,3]$. The corresponding DOS for its line graph is shown in \textbf{a}\textit{iv} and consists of a translated copy of that of the hexagonal lattice plus a flat band at $-2$. An example of one of the hexagonal localized flat-band states is plotted in \textbf{a}\textit{ii}, where the size of the circles indicates the magnitude of the state, red indicates positive sign, and yellow negative.
    \textbf{b}\textit{i-iv} The corresponding plots for the square lattice. Its line graph is a $6$-regular non-planar graph where some of the square plaquettes have additional edges corresponding to next-nearest-neighbor-like hopping. (Note that despite the apparent intersection of edges in the middle of these plaquettes, there is no lattice site there.) The flat band state encloses one of the conventional square plaquettes and is localized due to destructive interference between hopping around the neighboring squares and hopping across the diagonals. 
    \textbf{c}\textit{i-iv} The $3$-regular tree and its line graph. Trees exhibit exponential growth of the number of sites with distance from the origin. Therefore, the $\ell^2$ spectrum of $H$ is gapped away from $3$ and $-3$.\cite{Kesten:1959us,ChavelEigenvalues} Correspondingly the flat band in the spectrum of the line graph is spectrally isolated. Unlike the other cases shown here, the 3-regular tree has no cycles, so that flat-band states are exponentially localized rather than of compact support. 
    \textbf{d}\textit{i-iv} The graph of the heptagon-graphene lattice in hyperbolic space and its line graph, the heptagon-kagome lattice. Unlike the other graphs in this figure, there exists no known method for calculating the exact DOS, or even its support. However, it is known from $C^*$ algebras that there can be at most finitely many gaps\cite{Sunada:1992tv}, and since this graph is hyperbolic its $\ell^2$ spectrum is gapped away from $3$ and $-3$. Motivated by numerical simulations, we therefore sketch the support of the DOS as a single interval, but the presence of additional gaps cannot be ruled out.
    Since heptagon graphene is non-bipartite it exhibits a larger gap at the bottom of the spectrum than at the top, which carries over to its kagome-like line graph. The smallest cycles in the graph are odd and do not support localized flat band states like the one in \textbf{a}\textit{ii}. The smallest localized states therefore form on 14-sided cycles and cover two plaquettes instead.
    } 
\end{figure}

The simplest case occurs when restricting to the second harmonic (full-wave) mode. This mode has the same sign of $\varphi$ at both ends of the cavity; thus, by far the simplest choice is $\varphi(i,x_{end}) = 1$ for all ends in the device. In this case, $t_{i,j}$ is constant and everywhere negative. 
The resulting effective tight-binding Hamiltonian is that of an inverted single-band s-wave model on an effective lattice 
 whose sites are at the midpoints of the edges of $X$, and in which all non-zero hopping matrix elements are equal, regardless of variations in nearest-neighbor distance. $X_{eff}$ is therefore a new graph whose vertices are the edge set $\mathcal{E}(X)$, known as the line graph of $X$, and defined by 
 \begin{eqnarray}\label{LGdef}
V(L(X)) & =& \mathcal{E}(X),\\
\mathcal{E}(L(X)) & =&  \{ (vy)(yz);  vy \in \mathcal{E}(X) \ \mbox{and} \ yz \in \mathcal{E}(X),  v \neq z\}. \nonumber
 \end{eqnarray}
The most intuitive realization of $L(X)$ is to place each vertex at the midpoints between coupling capacitors, in which case its vertices coincide with the medial lattice of the layout. 
Since it arises from symmetric (or s-wave-type) modes on the edges of the graph $X$, we will denote the effective full-wave tight-binding Hamiltonian by $\bar{H}_s (X)$
It maps the space of normalizable wavefunctions in $\mathcal{E}(X)$ to itself, and we find that
\begin{equation}\label{eqn:HbarSubS}
\bar{H}_s (X) = H_{L(X)} = -t A_{L(X)}.
\end{equation}

For each plaquette in a layout lattice, the process of taking the line graph produces a new plaquette of the same shape, but it also adds additional features surrounding each of the original vertices. The prototypical example of this in Euclidean lattices is the hexagonal honeycomb in Fig. \ref{fig:G_LG_Examples} \textbf{a}\textit{i} and its line graph the kagome lattice, shown in Fig. \ref{fig:G_LG_Examples} \textbf{a}\textit{ii}. The line graph of a hyperbolic $3$-regular graph, such as the heptagonal one shown in Fig. \ref{fig:G_LG_Examples} \textbf{d}\textit{i}, will display the same triangular plaquettes around each of the layout vertices, so they will be collectively referred to as kagome-like.
Taking the line graph of a $4$-regular graph produces a non-planar feature which is a square plaquette with diagonal edges of equal amplitude, as seen in Fig. \ref{fig:G_LG_Examples} \textbf{b}. 
As will be shown in Sec. \ref{sec:gengraphs}, the Hamiltonian for any line-graph lattice has an infinite multiplicity eigenvalue of $-2$. In analogy to the Euclidean case, we will refer to these eigenstates as a flat band, regardless of the type of lattice. Sample flat-band eigenstates for both Euclidean and non-Euclidean examples are shown in Fig. \ref{fig:G_LG_Examples} \textbf{a}\textit{ii}-\textbf{d}\textit{ii}.

The situation involving the fundamental (half-wave) modes is more complicated because $\varphi(i,x)$ is positive at one end of the resonator, but negative at the other, and there are multiple ways to write the resulting effective tight-binding Hamiltonian.
As in the full-wave case, it is an operator on a lattice whose sites are $\mathcal{E}(X) = V(L(X))$. 
Its nonzero hopping matrix elements are in exactly the same places as those of $H_{L(X)}$, but their sign will now vary depending on the sign of $\varphi$.
We therefore denote the effective tight-binding operator by $\bar{H}_a (X)$ to indicate that it is the result of antisymmetric modes on the edges of $X$.
If $X$ is bipartite, then its vertices can be split into two non-neighboring groups $V_a$ and $V_b$. It is then possible to chose all of the $\varphi$ such that $\varphi(i,x_{end}) = +1$ at all the vertices in $V_a$ and $\varphi(i,x_{end}) = -1$ at all the vertices in $V_b$. This guarantees that $t_{i,j}$ is once again constant and everywhere negative. Thus $\bar{H}_a (X) = \bar{H}_s (X) =  H_{L(X)}$, and the tight-binding models derived from the full-wave and  half-wave modes are identical. 

However, if $X$ is not bipartite, then not all of the additional minus signs can removed via a judicious choice of the $\varphi$. All possible choices of resonator orientation will result in a combination of positive and negative $t_{i,j}$.
The resulting effective tight-binding Hamiltonian is that of an inverted single-band p-wave model, where the p-wave orbitals are aligned along the edges of the layout graph, and where the magnitudes of all non-zero hopping matrix elements are equal, regardless of variations in nearest-neighbor distance. 
Each choice of orientation for the resonator mode profiles of $X$ can give rise to different matrices for $\bar{H}_a(X)$. However, they are all simply different ways to rewrite the same Hamiltonian $\mathcal{H}_\Phi$. Therefore, they will always have the same eigenvalues. Different orientations correspond to keeping track of different ends of the resonators, so eigenvectors from one orientation can be converted to any other by multiplying the corresponding element of $\psi$ by $-1$. We will therefore abuse notation and refer to $\bar{H}_a$ without specifying the choice of orientation.

\begin{figure}[h]
	\begin{center}
		\includegraphics[width=0.9\textwidth]{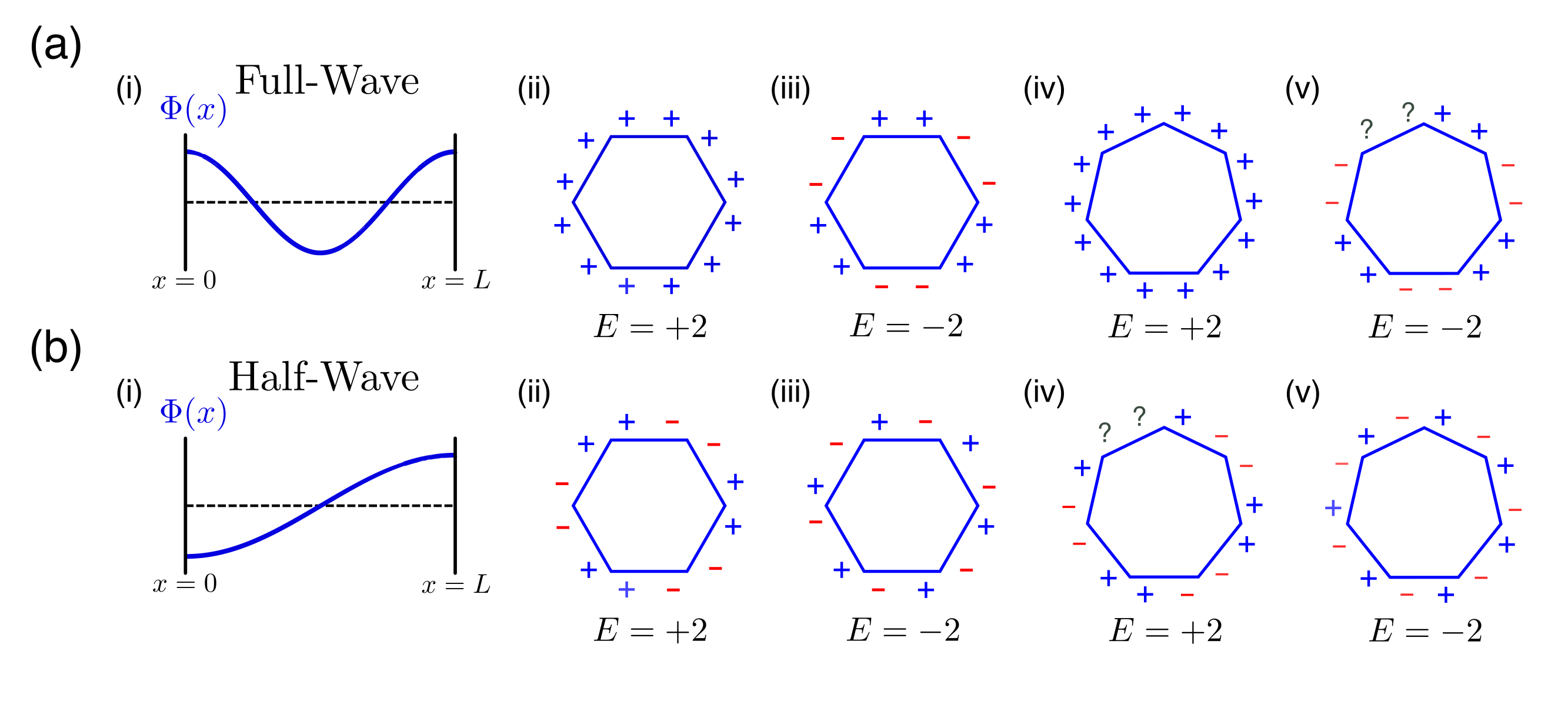}
	\end{center}
	\vspace{-0.6cm}
	\caption{\label{fig:Voltage_Configs} 
    \textbf{Resonator and ring eigenmodes.}
    \textbf{a}\textit{i} and \textbf{b}\textit{i} The intra-resonator profiles of $\Phi(x)$ for the second harmonic full-wave and fundamental half-wave modes, respectively. Both modes exhibit antinodes at the ends of the resonator, but the half-wave mode is antisymmetric about the cavity center.
    \textbf{a}\textit{ii}-\textit{iii} Eigenmodes of a 6-sided cycle of full-wave resonators (shown as straight edges) with eigenvalues $+2$ and $-2$. The $+$ and $-$ signs indicate the sign of $\Phi$ on the end of the resonator they are drawn next to. The full-wave modes are symmetric so both ends of the resonator have the same sign. The maximal eigenvalue of $2$ is obtained by having the same sign of $\Phi(x)$ on both sides of all coupling capacitors, which is achieved when $\Phi(x_{end})$ has the same sign on all resonators. The minimal eigenvalue of $-2$ is obtained by alternating the sign on neighboring resonators to obtain opposite values of $\Phi^\pm$ on all coupling capacitors. (Recall that $t<0$.) Since this layout graph is bipartite, both states are possible.
    \textbf{b}\textit{ii}-\textit{iii} The corresponding state for half-wave modes on a hexagonal cycle. The half-wave modes are antisymmetric so the sign of $\Phi(x)$ changes from one end of the resonator to the other. The maximal eigenvalue of $2$ is obtained by choosing the orientation of $\Phi(x)$ on each site such that all coupling capacitors have equal values of $\Phi^\pm$ on both sides. This is achieved by reversing the orientation on neighboring resonators. Keeping the orientation the same on each site gives the configuration with eigenvalue $-2$.
    \textbf{a}\textit{iv}-\textit{v} Attempts at realizing the same states using full-wave modes on a 7-sided cycle. The state at $2$ is analogous to the 6-sided case, but  because this graph is non-bipartite, the alternating pattern required to achieve $-2$ is geometrically frustrated and this state can no longer be defined, indicated by question marks.
    \textbf{b}\textit{iv}-\textit{v} Corresponding half-wave states on a 7-sided cycle. Because this graph is non-bipartite, the possible eigenvalues for full-wave and half-wave modes are not identical. The frustration is transferred from the low energy end of the spectrum to the upper, and the half wave modes can form a state at $-2$, but not one at $2$.
    } 
\end{figure}

The operator $\bar{H}_a$ is somewhat cumbersome because of the need to chose a particular orientation of $X$ in order write it as a matrix acting on vectors on $\mathcal{E}(X)$. However, it shares many features with $\bar{H}_s$. As long as there exists a finite $d_{max} = \max_{L(X)} d_x$, both are bounded self-adjoint operators from  the space of normalizable wavefunctions $\ell^2(\mathcal{E}(X)) = \ell^2(V(L(X)))$ to itself and their spectra are contained in the interval $[-d_{max}, d_{max}]$. There are, however, some striking differences between them which can already be seen by considering the underlying configurations of $\Phi$ on a simple graph, such as a single cycle $c_k$. This graph is 2-regular, and as long as $k$ is even, the maximum and minimum eigenvalues of both $\bar{H}_s$ and $\bar{H}_a$ are $\pm 2$ respectively. The corresponding $\Phi$ patterns in both full-wave and half-wave cases are shown in Fig. \ref{fig:Voltage_Configs} \textbf{a}\textit{ii}-\textit{iii} and \textbf{b}\textit{ii}-\textit{iii}. 
The corresponding tight-binding wavefunctions are $1$ everywhere or alternating $1$ and $-1$. 
If, however, $k$ is odd, then $c_k$ is not bipartite and the spectra are asymmetric. In the full-wave case, there exists a state with eigenvalue $2$, which has the same tight-binding wavefunctions as in the even-$k$ case, and whose $\Phi$ pattern is shown in Fig. \ref{fig:Voltage_Configs} \textbf{a}\textit{iv}.
However, the state at $-2$ no longer exists because the alternating sign of $\Phi$ and tight-binding wavefunctions cannot be consistently maintained. 
The odd-$k$ half-wave case, shown in Fig. \ref{fig:Voltage_Configs} \textbf{b}\textit{iii}-\textit{iv} is reversed. The patterns of $\Phi$ clearly show that the state at $-2$ exists while the one at $2$ cannot. To understand the same result in the tight-binding picture requires a choice of gauge. The simplest choice is to orient each resonator such that $\varphi$ goes from negative to positive going around the cycle in a clockwise direction. In this case, all $t_{i,j}$ are positive. The state $1,1,1,\cdots$ can easily be formed, but unlike the full-wave or bipartite cases, it has eigenvalue $-2$, and the alternating state $1,-1,1,-1,\cdots$, which would correspond to eigenvalue $2$ cannot be consistently defined.

\subsection{Flexible Resonators and Non-Euclidean Graphs}
Distributed-element waveguide resonators, like CPW resonators, have two unique properties which make them particularly versatile for realizing different layout graphs. First, the frequency of the resonator depends only on its total arc length, not on its shape. Therefore, straight resonators, and those fabricated with turns or meanders can produce effective photonic lattice sites with identical on-site energies. Second, the effective hopping matrix elements between resonators are set by the geometry of the coupling capacitor regions at the end of the resonators. Thus, hopping rates do not depend on center-of-mass distance as they normally would in an atomic lattice.\cite{Schmidt:2013hg,Kollar:2018vc} Therefore, CPW lattices can realize a planar layout graph $X$ even if a two-dimensional realization of $X$ or $L(X)$ with equidistant nearest-neighbor vertices is impossible, and opens the door to two new classes of lattice models: first, Euclidean lattices with unusual unit cells, and second, lattices in non-Euclidean spaces. Concrete examples of each of these kinds will be discussed in Secs. \ref{sec:euclidean} and \ref{sec:hyperbolic}, respectively.

For non-Euclidean graphs like those described in Ref.\cite{Kollar:2018vc}, the traditional Bloch theory-based methods of solid state physics fail, and there is no known method for computing the complete spectrum of an arbitrary non-Euclidean lattice. However, the more general methods of graph theory still provide considerable insight into the spectra and states of the effective models produced by lattices of CPW resonators, even for Euclidean lattices which are amenable to traditional Bloch-theory methods.
The remainder of this paper is therefore devoted to applying these methods and analyzing their physical consequences.
Due to the nature of CPW fabrication, non-planar layout graphs cannot be realized without multilayer fabrication. Furthermore, layout graphs with degree-3 vertices are by far the easiest to realize and the most robust to fabrication errors. Higher coordination numbers in the layout graph typically result in asymmetry of the coupling capacitors and unequal $t_{i,j}$ for different pairs of resonators incident on a vertex.\cite{Houck:2012iq,Schmidt:2013hg, Underwood:2012hx,Kollar:2018vc} We will therefore concentrate on planar layout graphs with coordination numbers less than or equal to three, although many of these arguments generalize readily to higher coordination numbers and non-planar graphs.

\subsection{Sample Sizes}
Experiments on infinite lattices are not possible, so we must instead work with only a finite set of vertices of order a few hundred.
A typical way to produce such a set is a hard-wall truncation: removal of all vertices and edges outside of a finite region. The resulting truncated graph $S(X)$ is an induced subgraph of the infinite lattice $X$, and is no longer d-regular.
As we will show in Sec. \ref{sec:gengraphs},
there is a close correspondence between $H_{S(X)}$ and the effective tight-binding operators $\bar{H}_s(S(X))$ and $\bar{H}_a(S(X))$.
However, the irregularity of $S(X)$ can produce additional states not found in the spectrum of $\bar{H}_s(X)$ or $\bar{H}_a(X)$, and care must be taken to account for these boundary effects. The Euclidean case is well known from solid state physics, but the hyperbolic case, where the boundary constitutes a finite fraction of the total volume, is more subtle. Both will be examined in detail in Secs. \ref{sec:euclidean} and \ref{sec:hyperbolic}, respectively.

\section{Spectra of Tight-Binding Hamiltonians on Graphs}\label{sec:gengraphs}
\subsection{Finite Layouts}\label{subsec:finitelayouts}

We examine the spectra of the Hamiltonians $\bar{H}_s(X)$ and $\bar{H}_a(X)$ for layouts $X$. While our main interest is in large planar $X$'s which are induced subgraphs of homogeneous cubic graphs, understanding the location and possible spectral gaps of general layouts is instructive and of independent interest. We restrict ourselves to finite layouts $X$, which are connected loopless graphs whose vertices have degree at most 3. Denote by $n$ the cardinality of the vertex set $V(X)$ and $m$ that of the edge set $\mathcal{E}(X)$.

Let $w_s:\mathcal{E} \times \mathcal{E} \rightarrow \{0,1\}$ be given by
\begin{equation}\label{eqn:w0}
w_s (e, e')= \begin{cases} 1, & \mbox{if } e  \mbox{ and }  e' \mbox{ share a vertex}\\
		 0  &  \mbox{otherwise}.
		  \end{cases} 
\end{equation}
Given an orientation of the edges of $X$ denote by $e^+$ and $e^-$ the head and foot in $V$ of an edge $e$.
Define $w_a:\mathcal{E} \times \mathcal{E} \rightarrow \{0,1, -1\}$ by 
\begin{equation}\label{eqn:w1}
w_a (f, g)= \begin{cases} 1, & \mbox{if } f^+ = g^+ \mbox{ or }  f^- = g^- \\
		 -1 & \mbox{if } f^+ = g^- \mbox{ or }  f^- = g^+,\\
		 0  &  \mbox{otherwise}.
		  \end{cases} 
\end{equation}
Note that $w_\alpha(e,e')$ is symmetric in $e$, $e'$ for $\alpha = a$ or $\alpha = s$.

The vector space of functions $f: \mathcal{E} \rightarrow \mathbb{C}$ comes with an inner product
\begin{equation}\label{eqn:dotprod}
\left< f,g\right> = \sum_{e \in \mathcal{E}}{f(e) g^*(e)},
\end{equation}
and we denote the inner product space by $\ell^2(\mathcal{E})$. The effective tight-binding Hamiltonians $\bar{H}_\alpha(X)$ for $\alpha = s \mbox{ or } a$ on $\ell^2(\mathcal{E})$ that were introduced in Sec. \ref{sec:implementation} are given in terms of $w_\alpha$ by 
\begin{equation}\label{eqn:HasW}
\bar{H}_\alpha  f(e) = \sum_{e' \in \mathcal{E}}{w_\alpha (e, e') f(e')}.
\end{equation}
$\bar{H}_\alpha$ is self-adjoint on $\ell^2(\mathcal{E})$ and we denote its spectrum by $\sigma(\bar{H}_\alpha)$. Since $w_\alpha(e,e')$ is not zero for at most four $e'$ for each $e$, it follows that $||\bar{H}_\alpha|| \leq 4$ and that $\sigma(\bar{H}_\alpha) \subset [-4,4]$. We will see below that $\bar{H}_\alpha$ can be factorized, from which it will follow that $\sigma(\bar{H}_\alpha) \subset [-2,4]$. Our aim is to study these spectra and their gaps at the bottom when $n\rightarrow \infty$. Something that we will exploit repeatedly and which follows from the definitions (see Sec. \ref{subsec:FWHWsetup}) is that the adjacency operator $A_{L(Y)}$ of the line graph of $Y$ is equal to the s-Hamiltonian $\bar{H}_s(Y)$.

The key factorizations involve incidence matrices. For $\alpha = s$ let $M$ be the $m \times n$ matrix:
\begin{equation}\label{eqn:fwincidence}
M(e,v)= \begin{cases} 1, & \mbox{if } e \mbox{ and } v \mbox{ are incident}, \\
		 0  &  \mbox{otherwise}.
		  \end{cases} 
\end{equation}
The following is well known\cite{BiggsGraphTheory, Cvetkovic:1980} and easy to check:
\begin{eqnarray}\label{eqn:fwincidence2}
M^t M & =& D_X + A_X, \mbox{ while} \\
MM^t & =&  2I + \bar{H}_s, \nonumber
 \end{eqnarray}
where $D_X = \mbox{diag}(\mathbb{d}(v))_{v\in V}$ .
The kernel of $D_X + A_X$ has dimension $\eta = \eta(X)$ which is $1$ or $0$ depending on whether $X$ is bipartite or not. Hence, $\mbox{rank } M = m-\eta$ and $\mbox{ker } M^t$ has dimension $m-n+\eta$ (we are assuming $m\geq n$). It follows that 
\begin{equation}\label{eqn:simgaH1}
\sigma(\bar{H}_s) = \{-2\}^{m - n +\eta} \cup \{ -2 + \sigma^*(D_X + A_X)    \},
\end{equation}
where $\{ \gamma\}^\nu$ means $\gamma$ with multiplicity $\nu$, and $\sigma^*(D_X + A_X)  $ consists of the nonzero (in fact positive) eigenvalues.\cite{BiggsGraphTheory, Cvetkovic:1980}

For $\alpha = a$ let $N$ be the $m \times n$ incidence matrix given by
\begin{equation}\label{eqn:hwincidence}
N(e,v)= \begin{cases} 1, & \mbox{if } e^+ =  v, \\
		-1 &  \mbox{if } e^- =  v,\\
		 0  &  \mbox{otherwise}.
		  \end{cases} 
\end{equation}
A calculation similar to Eqn. \ref{eqn:fwincidence2} (detailed in Appendix \ref{app:HWshirai}) yields 
\begin{eqnarray}\label{eqn:hwincidence2}
N^t N & =& D_X - A_X,  \mbox{ while} \\
NN^t & =&  2I + \bar{H}_a. \nonumber
 \end{eqnarray}
 Note that $N^tN$ is the Laplacian $\Delta_X$ on functions defined in Eqn. \ref{DeltaG}, or equivalently, the combinatorial Laplacian on $0$-chains\cite{Ray:1971va}, 
 and its kernel is $1$-dimensional, corresponding to the constant function on $V$. Hence the rank of $N$ is $n-1$ and the kernel of $N^tN$, which is the combinatorial Laplacian on $1$-chains, has dimension $m-n +1$. It follows that 
 \begin{equation}\label{eqn:simgaH2}
\sigma(\bar{H}_a) = \{-2\}^{m-n +1} \cup \{ -2 + \sigma^*(D_X - A_X)    \}.
\end{equation}
Note that $X$ is bipartite if and only if $\sigma(D_X - A_X) = \sigma(D_X + A_X)$, and in this case, as was observed in Sec. \ref{subsec:FWHWsetup}, $\sigma(\bar{H}_s(X)) = \sigma(\bar{H}_a(X))$. Furthermore, $\sigma(\bar{H}_a(X))$ does not depend on the choice of orientation of $\mathcal{E}$ in the definition of $\bar{H}_a(X)$.
 For a $3$-regular layout Eqns. \ref{eqn:simgaH1} and \ref{eqn:simgaH2} simplify and give $\sigma(\bar{H}_s)$ and $\sigma(\bar{H}_a)$ in terms of $\sigma(A_X)$:
 \begin{eqnarray}\label{eqn:finiteregularspectra}
\sigma(\bar{H}_a) = \{-2\}^{m - n + 1} \cup \{ 1 - \sigma(A_X) \},\\
\sigma(\bar{H}_s) = \{-2\}^{m - n +\eta} \cup \{ 1 + \sigma(A_X) \}. \nonumber
 \end{eqnarray}
 
Equations. \ref{eqn:simgaH1} and \ref{eqn:simgaH2} show that $\sigma(\bar{H}_\alpha) \subset [-2,4]$ and give the exact (high) multiplicity of the eigenvalue $-2$. The question that we address in what follows is whether the bottom of the spectrum is gapped as $n(X) \rightarrow \infty$. Let $\lambda(\bar{H}_\alpha(X))$ be the smallest eigenvalue of $\bar{H}_\alpha(X)$ which is larger than $-2$. For $\alpha = a$ it follows from Eqn. \ref{eqn:hwincidence2} that 
\begin{equation}\label{eqn:lambdaa}
\lambda(\bar{H}_a(X)) = -2 + \lambda_1(\Delta_X),
\end{equation}
where $\lambda_1(\Delta_X)$ is the smallest positive eigenvalue of the Laplacian. Whether $\lambda(\bar{H}_a(X))$ is bounded below by a positive constant along a sequence of such layout graphs has been studied extensively,\cite{Alon:1985wi} and it is equivalent to the sequence being an expander.
 The separator theorem\cite{Lipton:1980wt} shows that a sequence of planar graphs is never an expander, and hence $\bar{H}_a(X)$ cannot be gapped at $-2$ for planar layouts:
 \begin{equation}\label{eqn:Halim}
 \limsup_{n(x) \rightarrow \infty, X \mbox{ planar}} {\lambda(\bar{H}_a(X)) = -2}
 \end{equation}
 
\begin{figure}[h]
	\begin{center}
		\includegraphics[width=0.9\textwidth]{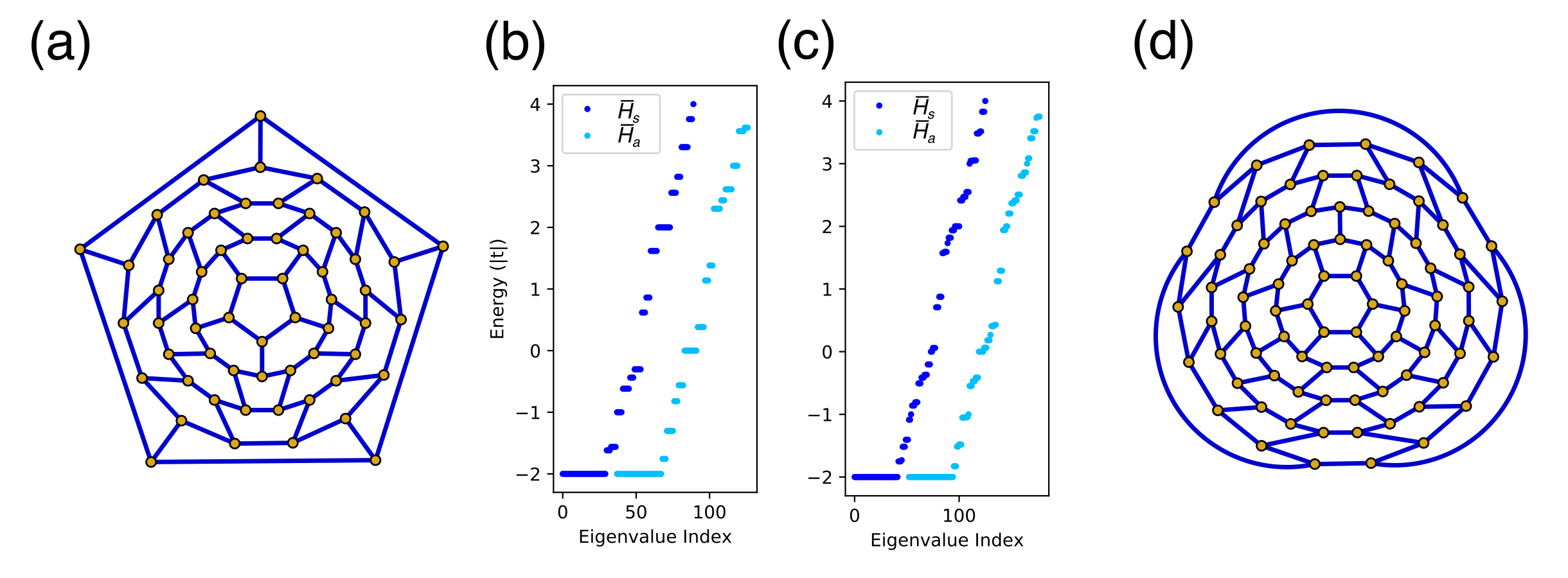}
	\end{center}
	\vspace{-0.6cm}
	\caption{\label{fig:Fullerenes} 
    \textbf{Fullerenes.}
    \textbf{a} The graph for the well known spherical tiling found in soccer balls and $C_{60}$ Buckminsterfullerene.\cite{C60}
    \textbf{b} Corresponding spectrum of $\bar{H}_s$ and $\bar{H}_a$. The spectrum of $\bar{H}_a$ has been offset horizontally for clarity. Because this layout is finite and small, $\bar{H}_a$ has a gap above $-2$ despite being planar. Additionally, this gap is bigger than the $3-2\sqrt{2}$ limit for non-planar expander graphs. This $60$-site graph is one of the largest planar graphs that violates this bound.
    \textbf{c}-\textbf{d} Corresponding plots for one version of the  larger fullerene $C_{84}$.\cite{Raghavachari:1991ey, RamanujanFullerenes} This graph is one of largest known planar Ramanujan graphs and $\bar{H}_a$ has a gap of almost exactly $3-2\sqrt{2}$. (The other $7$-fold symmetric realization of $C_{84}$ is not Ramanujan.)
    } 
\end{figure}
 
 Expanders exist (this is by no means obvious) and one can ask about the maximal gap at $-2$ for $\bar{H}_a(X)$. The Alon-Boppana theorem in the form established in Ref.\cite{Nilli:1991wd} asserts that 
 \begin{equation}\label{eqn:expanderbound}
 \limsup_{n(X) \rightarrow \infty} {\lambda_1(\Delta_X) \leq 3 - 2\sqrt{2}}.
 \end{equation}
 Hence for our general layouts we have 
 \begin{equation}\label{eqn:Hagapbound}
 \limsup_{n(X) \rightarrow \infty} {\lambda(\bar{H}_a(X))} \leq 1 - 2\sqrt{2} = -1.828\cdots
 \end{equation}
We will call a layout $X$ with $\lambda(\bar{H}_a(X)) \geq 1-2\sqrt{2}$ a Ramanujan layout. These exist (with $n(X)$ arbitrarily large) given explicitly as $3$-regular Ramanujan graphs.\cite{Lubotzky:1987ub,RamanujanGraphs}
The largest planar Ramanujan layout that we know of is depicted in Fig. \ref{fig:Fullerenes}.
Thus, as far as the optimal gap at $-2$ for $\bar{H}_a(X)$ we have 
\begin{equation}\label{eqn:Hagapbound2}
\limsup_{n(X) \rightarrow \infty} {\lambda (\bar{H}_a(X))} = 1 -2 \sqrt{2}.
\end{equation}

The story of the gap at the bottom for $\bar{H}_s(X)$ is quite different, at least for non-bipartite $X$ (the bipartite cases reduce to the discussion above since $\sigma(\bar{H}_a(X))) = \sigma(\bar{H}_s(X))$ for these). $X$ is not bipartite if and only if $X$ carries a nontrivial cycle of odd length. The gap $2 + \lambda(\bar{H}_s(X))$ is a quantitative measure of $X$ not being bipartite. However, the existence of a short odd cycle is neither necessary nor sufficient for the gap to be positive as $n(X) \rightarrow \infty$. One can construct large $X$'s with a $3$-cycle and for which $\lambda(\bar{H}_s(X))$ tends to $-2$ as $n(X) \rightarrow \infty$, while the large girth and non-bipartite Ramanujan graphs\cite{Lubotzky:1987ub} give (using Eqn. \ref{eqn:finiteregularspectra}) examples of $X$'s with $\lambda(\bar{H}_s(X)) \geq 1 - 2\sqrt{2}$ which have no short odd cycles. The following local condition on short odd cycles in $X$ ensures that the bottom of $\sigma(\bar{H}_s(X))$ is gapped and it applies to all our layouts of interest.
\begin{gather}
\mbox{If for a fixed } r \geq 2 \mbox{ the induced subgraphs on } \nonumber \\ 
B_r(x) = \{y\in X : d_X(x,y) \leq r \} \subset X
\mbox{ are nonbipartite for every } x \in X, \mbox{ then}\nonumber \\
\lambda(\bar{H}_s(X)) \geq -2 + (48(3.2^{2r-1} - 1)^2)^{-1}. \label{thm:cycledensity}
\end{gather}
We postpone the proof of the above to Appendix \ref{app:Hsgap}.
This criterion yields many planar $X$'s for which $-2$ is uniformly gapped for $\bar{H}_s(X)$ as $n(X) \rightarrow \infty$.

\begin{figure}[h]
	\begin{center}
		\includegraphics[width=0.98\textwidth]{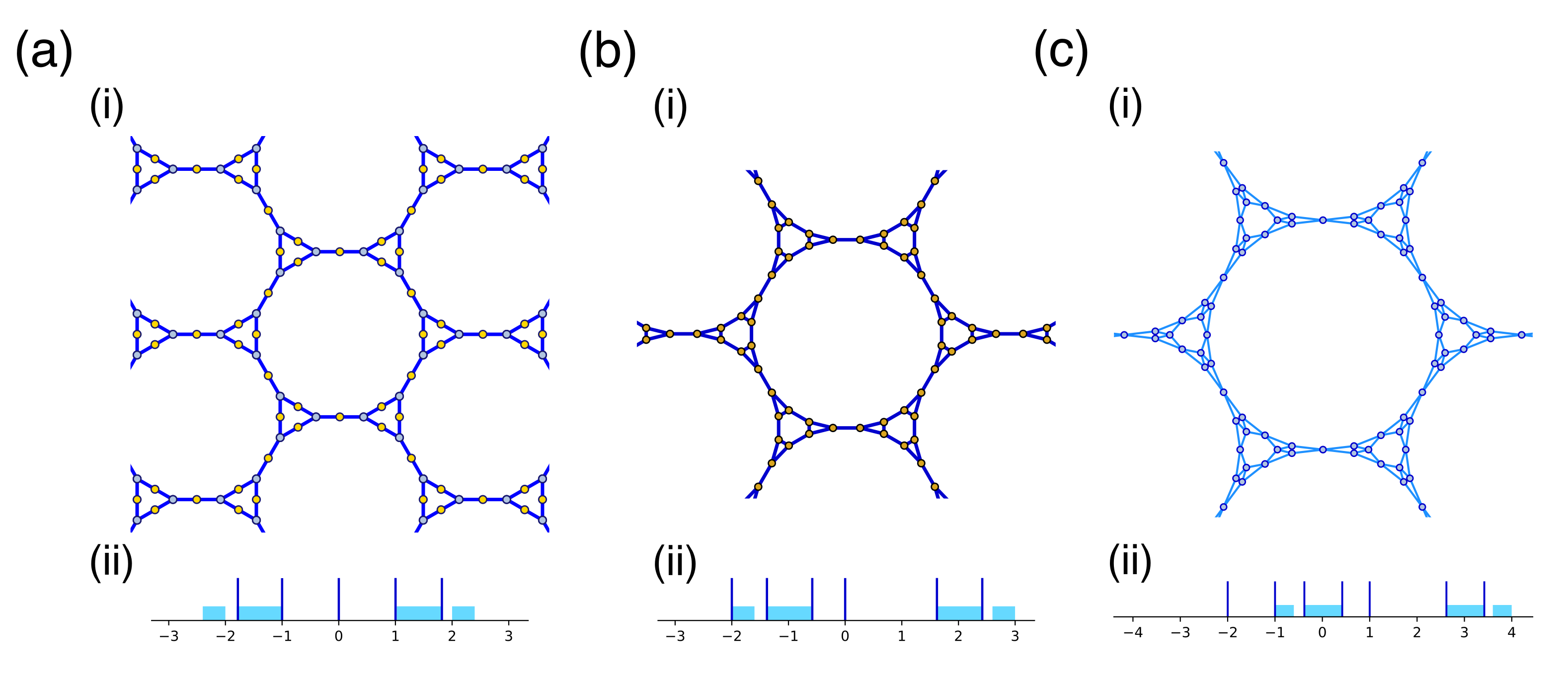}
	\end{center}
	\vspace{-0.6cm}
	\caption{\label{fig:AliciumDOS} 
    \textbf{Maximally gapped flat bands.}
    Example of the construction of a $3$-regular layout $\mathfrak{X}$ for which $\bar{H}_s(\mathfrak{X})$ realizes the locally maximal gap interval in Eqn. \ref{eqn:CharonGaps}, and has the maximum possible gaps around its flat bands at $-2$ and $1$.
    \textbf{a}\textit{i} The $3,2$-biregular graph  $Y$ obtained by successive subdivision and line graph operations on the hexagonal honeycomb $T_6$ (graphene): $Y = \mathbb{S}(\mathcal{X})$, where $\mathcal{X} = L(\mathbb{S}(T_6))$. 
    The density of states (DOS) of $A_Y$ is sketched in \textbf{a}\textit{ii}. The support of the DOS is shown in light blue, and the flat bands are highlighted in dark blue.
    The locations of the flat bands and the support of the density of states (DOS) can be derived from the properties of $T_6$ via Eqns. \ref{eqn:XSX} and \ref{eqn:XLSX}. 
    The flat band at $0$ with gaps of $\pm 1$ on either side arises because $Y$ is the subdivision graph of a line graph. The additional flat bands at $\pm 1$ and $\pm \sqrt{3}$ are present because $\mathcal{X}$ itself is the line graph of a subdivision graph. 
    \textbf{b}\textit{i}-\textit{ii} The $3$-regular graph $\mathfrak{X} = L(Y)$ which has flat bands at $-2$, $(1\pm \sqrt{13})/2$, $(1\pm \sqrt{5})/2$. This graph realizes the locally maximal gap interval given in Eqn. \ref{eqn:CharonGaps}.
    \textbf{c}\textit{i}-\textit{ii} The $4$-regular line graph $\bar{\mathfrak{X}}$ obtained as the effective lattice of $\mathfrak{X}$. The spectrum of this graph is equal to that of $\bar{H}_s(\mathfrak{X})$, and has optimally gapped flat bands at both $-2$ and $1$.
    } 
\end{figure}

We turn to the optimal gap at $-2$ for $\bar{H}_s(X)$. In Appendix \ref{app:hoffman} we apply the classification initiated by Hoffman\cite{Cameron:1975vu}
of graphs $Y$ for which all the eigenvalues of $A_Y$ are at least $-2$ to show that
\begin{equation}\label{eqn:Haextremalgap}
\limsup_{n(X) \rightarrow \infty}\lambda(\bar{H}_s(X)) \leq -1.
\end{equation}
This puts a limit on the gap and we call layouts $X$ which achieve equality in Eqn. \ref{eqn:Haextremalgap} Hoffman layouts. If $Y$ is a bipartite biregular layout of degrees $2$ and $3$ respectively, then its line graph $X = L(Y)$ is $3$-regular. Using the equivalence between $\bar{H}_s(Y)$ and $A_X$ along with Eqn. \ref{eqn:simgaH1} yields that the smallest eigenvalue of $A_X$ is $-2$. Hence from Eqn. \ref{eqn:finiteregularspectra} it follows that for such an $X$ 
\begin{equation}\label{eqn:Hsgapbound}
\lambda(\bar{H}_s(X)) = -1,
\end{equation}
that is $X$ is a Hoffman graph. There is nothing preventing us from choosing $Y$ and also $X$ in this construction to be planar. 
See Fig. \ref{fig:AliciumDOS} for a planar example.
Hence
\begin{equation}\label{eqn:Hsgapbound2}
\limsup_{\substack{n(X) \rightarrow\infty,\\ X \mbox{ planar}}} {\lambda(\bar{H}_s(X))} = -1.
\end{equation}
This concludes our analysis of the gap at $-2$ for the  $\bar{H}_\alpha$'s of finite graphs.

In examining the possible gaps in these spectra we restrict our layouts to be $3$-regular. In Appendix \ref{app:hoffman} we show that for these a Hoffman layout is achieved by the process leading to Eqn. \ref{eqn:Hsgapbound}. For regular layouts it follows from Eqn. \ref{eqn:finiteregularspectra} that it suffices to analyze $\sigma(A_X)$.
Given a disjoint union $I$ of open intervals in $[-3,3]$, $X$ is $I$-gapped if $\sigma(A_X) \cap I = \emptyset$. We say that $I$ is a gap (resp planar) interval if there is a sequence of $3$-regular $X$'s (resp planar) with $n(X)\rightarrow \infty$ and which are $I$-gapped. 
$I$ is a locally maximal gap interval if by increasing any of its component intervals the resulting interval is no longer a gap interval. $I$ is a globally maximal gap interval if for any $J \supsetneqq I$, $J$ is not a gap interval. Understanding these gap intervals is the question of what gaps can be achieved by regular layouts, and in particular in the spectra of $\bar{H}_\alpha(X)$.

In this notation Eqn. \ref{eqn:Hagapbound2} is equivalent to $(2\sqrt{2},3]$ being a locally maximal gap interval while Eqn. \ref{eqn:Hsgapbound2} is equivalent to $[-3,-2)$ being a locally maximal planar gap interval. Non-bipartite cubic Ramanujan graphs are $I = [-3, -2\sqrt{2}) \cup (2 \sqrt{2}, 3]$-gapped, and the recent result in Ref.\cite{Abert:2016dg} shows that this $I$ is globally maximal.

In Appendix \ref{app:hoffman} we extend our analysis to give further examples of locally and globally maximal gap intervals. We record a couple of these here:
\begin{equation}\label{eqn:McLaughlinGaps}
I = [-3,-2) \cup (-2, b) \cup (c', 0) \cup (0,c) \cup (b',3)
\end{equation}
 is a globally maximal gap interval with, $b = \frac{1 - \sqrt{1+4(3 +2 \sqrt{3})}}{2} = -1.965\cdots$, $b' = \frac{1 + \sqrt{1+4(3 +2 \sqrt{3})}}{2} = 2.965\cdots$, $c = \frac{1 + \sqrt{1+4(3 -2 \sqrt{3})}}{2} = 1.149\cdots$,  and  $c' = \frac{1 + \sqrt{1+4(3 -2 \sqrt{3})}}{2} = -0.149\cdots$. As was shown in Ref.\cite{McLaughlin:1986vb}, one graph which realizes this gap interval is the Cayley graph of $\mathbb{Z}_2 *\mathbb{Z}_3$. This graph and its line graph are shown in Fig. \ref{fig:McLaughlin}.
\begin{equation}\label{eqn:CharonGaps}
I = [-3,-2)\cup \left( \frac{1-\sqrt{5}}{2}, 0 \right) \cup    \left(0, \frac{1+\sqrt{5}}{2} \right) 
\end{equation}
is a locally maximal planar gap interval. An example of a graph that realizes this is shown in Fig. \ref{fig:AliciumDOS}.

\subsection{Infinite Regular Layouts}\label{subsec:infinitelayouts}

When $X$ is an infinite layout we will take it to be $3$-regular. The vector spaces of functions from $V(X)$ and $\mathcal{E}(X)$ to $\mathbb{C}$ are infinite dimensional and come with the inner products in Eqn. \ref{eqn:dotprod}, making them into Hilbert spaces $\ell^2(V)$ and $\ell^2(E)$. The linear operators $A_X$ on $\ell^2(V)$ and the $\bar{H}_\alpha(X)$ on $\ell^2 (\mathcal{E})$ are self-adjoint and bounded. We denote their $\ell^2$ spectra by $\sigma(A_X)$ and $\sigma(\bar{H}_\alpha (X))$. The relation in Eqn. \ref{eqn:finiteregularspectra} extends to this $\ell^2$ setting:\cite{Shirai:1999wm}
\begin{eqnarray}\label{eqn:shiraispectrum}
\sigma(\bar{H}_s (X))  & =& \{ -2\}^\infty \cup \{  1 + \sigma(A_x)  \},\\
& \mbox{and} &  \nonumber \\
\sigma(\bar{H}_a (X))  & =& \{ -2\}^\infty \cup \{  1 - \sigma(A_x)  \}. \nonumber
 \end{eqnarray}
 In particular, $-2$ is a point eigenvalue of $\bar{H}_\alpha (X)$ of infinite multiplicity. Moreover, $\bar{H}_s(X)$ has a compactly supported eigenfunction with eigenvalue $-2$ if and only if $X$ has a non-backtracking circuit of even length, while $\bar{H}_a(X)$ has such an eigenfunction if and only if $X$ is not the $3$-regular tree. The existence of the point eigenvalue is proved in Ref.\cite{Shirai:1999wm} and we give the construction of the eigenstates in Appendix \ref{app:HWshirai}.

From Eqn. \ref{eqn:shiraispectrum} the primary spectrum to be understood is $\sigma(A_X)$. The end-points $\lambda_{min}(A_X)$ and $\lambda_{max}(A_X)$ of $\sigma(A_X)$ have variational characterizations:
\begin{equation}\label{eqn:variational1}
\lambda_{min}(A_X)   = \inf_{f: V\rightarrow \mathbb{R}}   \frac{\sum\limits_{v \sim w}{f(v) f(w)}}{\sum\limits_{v} {f(v)^2}} ,
\end{equation}
\begin{equation}\label{eqn:variational2}
\lambda_{max}(A_X)    = \sup_{f: V\rightarrow \mathbb{R}}   \frac{\sum\limits_{v \sim w}{f(v) f(w)}}{\sum\limits_{v} {f(v)^2}}.
 \end{equation}
If follows that 
\begin{equation}\label{eqn:lminlmax3}
|\lambda_{min}(A_X)| \leq \lambda_{max}(A_X) \leq 3.
\end{equation}
In the $\inf$ and $\sup$ in Eqns. \ref{eqn:variational1} and \ref{eqn:variational2} one can restrict to $f$'s of compact support and achieve the same extrema for this $\ell^2$-spectrum. If $W\subset V$ is a finite subset, then the $\inf$ and $\sup$ in Eqns. \ref{eqn:variational1} and \ref{eqn:variational2} for $f$'s supported on W are the bottom and top, $\lambda_{min}(A_Y)$ and $\lambda_{max}(A_Y)$, of the spectrum of the adjacency matrix $A_Y$ of the $W$-induced subgraph $Y$ of $X$ (which we denote by $Y\subset X$). Hence
\begin{eqnarray}\label{eqn:finiteextrema}
\lambda_{min} (A_X) & =& \inf_{\substack{Y\subset X,\\ Y \mbox{ finite}}} {\lambda_{min} (A_Y)},\\
& \mbox{and} &  \nonumber \\
\lambda_{max} (A_X) & =& \sup_{\substack{Y\subset X, \\ Y \mbox{ finite}}} {\lambda_{max} (A_Y)}.
 \end{eqnarray}
Note that $3$ itself is not an eigenvalue of $A_X$ since if it were the corresponding eigenfunction would have to be constant (X is connected) and hence will not be in $\ell^2(V)$. Thus if $3\in \sigma(A_X)$, then $3$ must be an accumulation point of $\sigma(A_X)$ and the same applies for $-3$. We conclude that $-2$ is gapped in $\bar{H}_a(X)$ if and only if $\lambda_{max}(A_X) <3$, and in $\bar{H}_s(X)$ if and only if $\lambda_{min}(A_X) > -3$.

In order to analyze the extrema of $\sigma(A_X)$, as well as other of its properties we assume $X$ is homogeneous, or at least almost homogenous: that there is a finitely generated subgroup $G = G(X)$ of automorphisms of $X$ which acts transitively on $V(X)$, or, in the almost homogeneous case, that the orbit set $G\backslash V(X)$, is finite. In this case, $G$ acts on $\ell^2(V)$ as unitary operators by the representation $\mathcal{R}(g), g \in G$
\begin{equation}\label{eqn:Gaction}
\mathcal{R}(g)f(v) = f(g^{-1}v),
\end{equation}
and this action commutes with $A_X$. Decomposing $\ell^2(V)$ according to the $G$-action brings $G$ and its representation theory into the analysis.

There is a clean answer as to whether $\lambda_{max} = 3$ in terms of $G$. This is due to Kesten in Ref.\cite{Kesten:1959us} and for our setting of almost homogeneous $X$'s to Brooks in Ref.\cite{Brooks:1982vi}, and it asserts that 
\begin{equation}\label{eqn:3iff}
\lambda_{max}(A_X) = 3 \mbox{ if and only if } G \mbox{ is amenable}.
\end{equation}
Examples of amenable groups are ones which have finite index subgroups which are Abelian, while non-Abelian free groups and the hyperbolic tessellation groups (discussed in Sec. \ref{sec:hyperbolic}) are examples of non-amenable groups. From Eqn. \ref{eqn:lminlmax3} it follows that if $G$ is not amenable, then $\lambda_{min}(A_X) > -3$ and $\lambda_{max}(A_X) < 3$, and hence;
\begin{equation}\label{eqn:nonamenable}
\text{\parbox{.85\textwidth}{  If $G$ is not amenable $-2$ is gapped for $\bar{H}_a(X)$ and $\bar{H}_s(X)$.   }}
\end{equation}
\begin{equation}\label{eqn:amenable}
\text{\parbox{.85\textwidth}{
If $G$ is amenable then $-2$ is not gapped for $\bar{H}_a(X)$ and gapped for $\bar{H}_s(X)$ if and only if $X$ is not bipartite.
}}
\end{equation}

In Eqn. \ref{eqn:amenable} all that needs clarification is that if $X$ is not bipartite, then $\lambda_{min}(A_X) > -3$. This and a bit more will follow from Eqn. \ref{thm:cycledensity}. Firstly, since $X$ is not bipartite, it has an odd $k$-cycle and, moreover, since it is homogeneous every vertex is at most distance $r$ (for some finite $r$) from such a $k$-cycle. Fix $v_0 \in V(X)$ and for $\rho$ a large integer let $Y_\rho$ be the subgraph of $X$ induced on $B_\rho(v_0) = \{ v \in V(X): d_X(v,v_0) \leq \rho  \}$.  $Y_\rho$ satisfies the conditions in Eqn. \ref{thm:cycledensity} and hence
\begin{equation}\label{eqn:lminnonzero}
\lambda_{min}(D_{Y_\rho} + A_{Y_\rho}) \geq \epsilon_0 > 0 \mbox{ independent of }\rho.
\end{equation}
Now $3I \geq D_{Y_\rho}$ (as quadratic forms) hence 
$$
\lambda_{min}(3I + A_{Y_\rho}) \geq \epsilon_0, \mbox{ or} $$
$$
\lambda_{min}(A_{Y_\rho}) \geq -3 + \epsilon_0. $$
As $\rho\rightarrow \infty$, $B_\rho(V_0)$ exhausts $X$ and hence from Eqn. \ref{eqn:finiteextrema} we see that 
$$\lambda_{min}(A_X) \geq -3 +\epsilon_0, $$
which proves Eqn. \ref{eqn:amenable} in the stronger form that the induced (finite) $Y_\rho$'s of $X$ have $-2$ gapped in $\sigma(\bar{H}_s(X))$.

This completes the qualitative description of the gap at $-2$ for $\bar{H}_\alpha(X)$ when $X$ is (almost) homogeneous. We turn to the quantitative study of $\sigma(\bar{H}_\alpha(X))$ for a given homogeneous $X$. If $G$ is Abelian, then all its irreducible representations are 1-dimensional and one can decompose $\ell^2(V)$ accordingly. This reduces the problem to finite dimensions. For example, if $G = \mathbb{Z}\times \mathbb{Z}$ as is the case for planar Euclidean crystallographic groups, the (unitary) dual group $\hat{G}$ of $G$ is the $2$-dimensional torus $\mathcal{T} = \mathbb{R}^2/\mathbb{Z}^2$. $A_X$ leaves invariant the subspaces $\ell^2(V, \chi)$ for $\chi\in \hat{G}$ given by the functions $f:V\rightarrow \mathbb{C}$ satisfying
\begin{equation}\label{eqn:chieigen}
f(gv) = \chi(g)f(v), g \in G.
\end{equation}
Denote the spectrum of $A_X$ on this (say) $l$-dimensional space by $\lambda_1(\chi), \ldots, \lambda_l(\chi), \chi \in \mathcal{T}$ and these continuous functions of $\chi$ give the bands and gaps in the spectrum of $A_X$ on $\ell^2(V)$.
This analysis is a well-developed theory in this planar Euclidean setting and is known as Bloch Wave Theory. We review and exploit it in Sec. \ref{sec:euclidean}, leading to explicit computations of these spectra. In this Bloch-Wave setting $\lambda$ is an eigenvalue of $A_X$ (that is it has a corresponding $\ell^2(V)$ bound state) if and only if $\lambda(\chi)$ is a constant function of $\chi$, which is called a `flat band' (and in this case $\lambda$ has infinite multiplicity). Although there is no apparent Bloch Wave Theory for general $G$ (or for our finite layouts) we continue to use this suggestive terminology of ``flat-bands" for eigenvalues of infinite multiplicity in the homogeneous (infinite) setting and for very large multiplicity eigenvalues in the finite layout setting.

When $G$ is not amenable there are few examples for which $\sigma(A_X)$ can be computed explicitly. The (unitary) dual $\hat{G}$ is no longer a friendly object,  specifically, the groups that we encounter are not of type I.\cite{Glimm:1961ta}
There is a qualitative and quite general theorem which asserts that for any $G$ satisfying the conditions in Ref.\cite{Puschnigg:2002ju} (and these include all of the groups we consider and in particular the tessellation groups $G_k$ in Sec. \ref{sec:hyperbolic}), $\sigma(A_X)$ consists of finitely many closed intervals or bands (see Ref.\cite{Sunada:1992tv}). There are some special examples for which $\sigma(A_X)$ can be computed and which are significant. The first is the $3$-regular tree $X_3$ which was computed by Kesten in Ref.\cite{Kesten:1959us}:
\begin{equation}\label{eqn:kesten}
\sigma(A_{X_3}) =[-2\sqrt{2}, 2\sqrt{2}], 
\end{equation}
and the spectral measure is absolutely continuous on this interval. 
$X_3$ can be realized as the Cayley graph of 
$G = \mathbb{Z}_2\star \mathbb{Z}$; w.r.t. the symmetric generating set $\{ Q, R, R^{-1} \}$, with $Q^2 = 1$.
$X_3$ is the universal cover for any $3$-regular layout and the $3$-regular Ramanujan Graphs are exactly those which have their non-constant spectrum contained in $\sigma(A_{X_3})$. 
 
\begin{figure}[h]
	\begin{center}
		\includegraphics[width=0.9\textwidth]{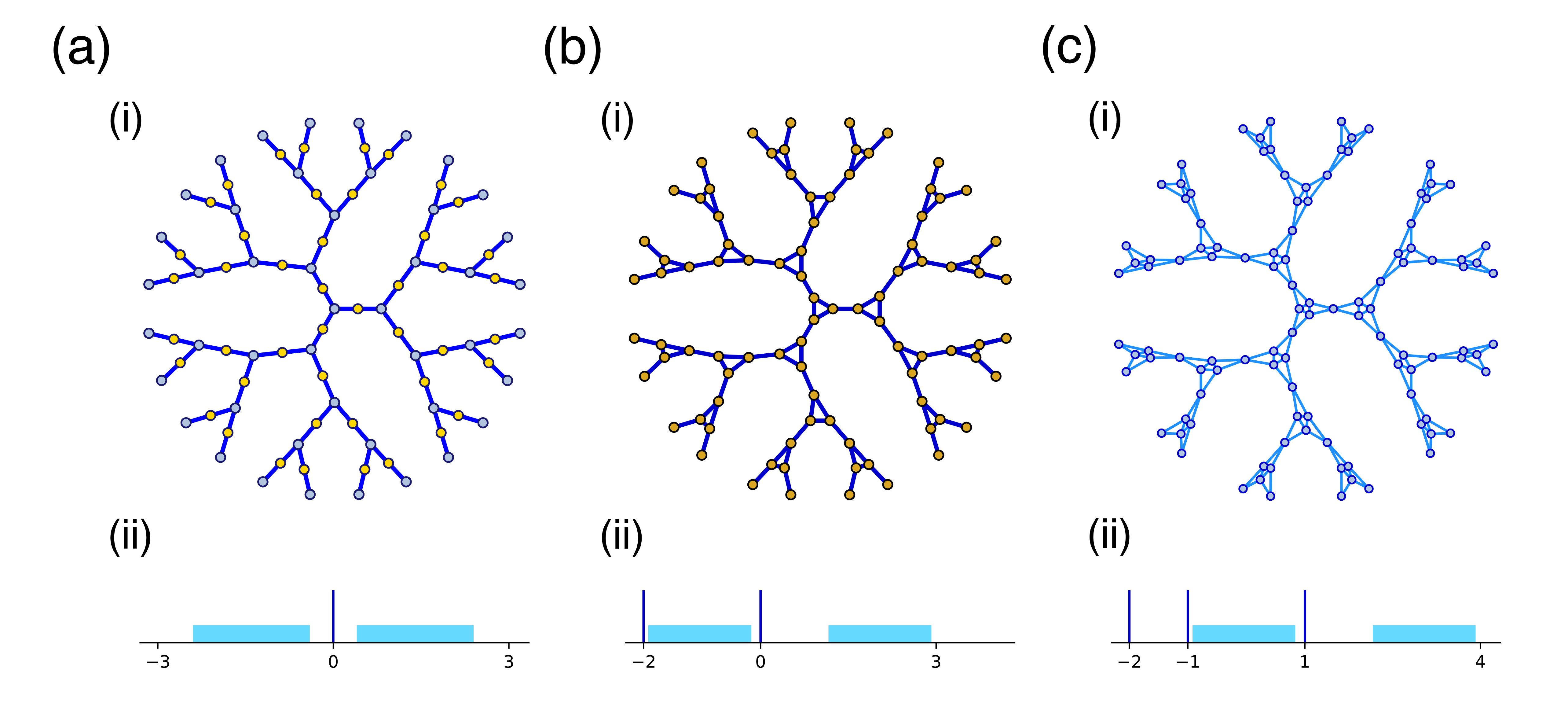}
	\end{center}
	\vspace{-0.6cm}
	\caption{\label{fig:McLaughlin} 
    \textbf{The Cayley graph of $\mathbb{Z}_2 \star \mathbb{Z}_3$ and its line graph.}
    \textbf{a}\textit{i} The biregular graph obtained from the 3-regular tree by adding new vertices, shown in yellow, in the middle of each edge, and splitting each edge in two.
    Its spectrum, shown schematically in \textbf{a}\textit{ii}, can be derived from that of the $3$-regular tree and consists of a flat band at $0$, and the two intervals $(\pm \sqrt{3-2\sqrt{2}}, \pm \sqrt{3+2\sqrt{2}}) = (\pm 0.414, \pm 2.414)$.
    \textbf{b}\textit{i} The resulting 3-regular line graph $\mathbb{M}$, which is also the Cayley graph of $\mathbb{Z}_2 \star \mathbb{Z}_3$. Its spectrum, shown in \textbf{b}\textit{ii}, was derived in Ref.\cite{McLaughlin:1986vb} It consists of two flat bands at $-2$ and $0$ with exponentially localized eigenstates and two absolutely continuous intervals $(-1.965\cdots, -0.149\cdots)$ and $(1.149\cdots, 2.965\cdots)$, whose end points are $(1 \pm \sqrt{1+4(3 \pm2 \sqrt{3})})/2$. This graph realizes the globally maximal gap interval in Eqn. \ref{eqn:McLaughlinGaps}.
    \textbf{c}\textit{i} The 4-regular line graph of $\mathbb{M}$.
    \textbf{c}\textit{ii} Its spectrum exhibits a gap of $1$ above the flat band, which is the maximum possible among 4-regular graphs.
    The graph and line graph in \textbf{b} and \textbf{c} are universal covers for all 3-regular examples in this paper.
    } 
\end{figure}
\begin{figure}[h]
	\begin{center}
		\includegraphics[width=0.9\textwidth]{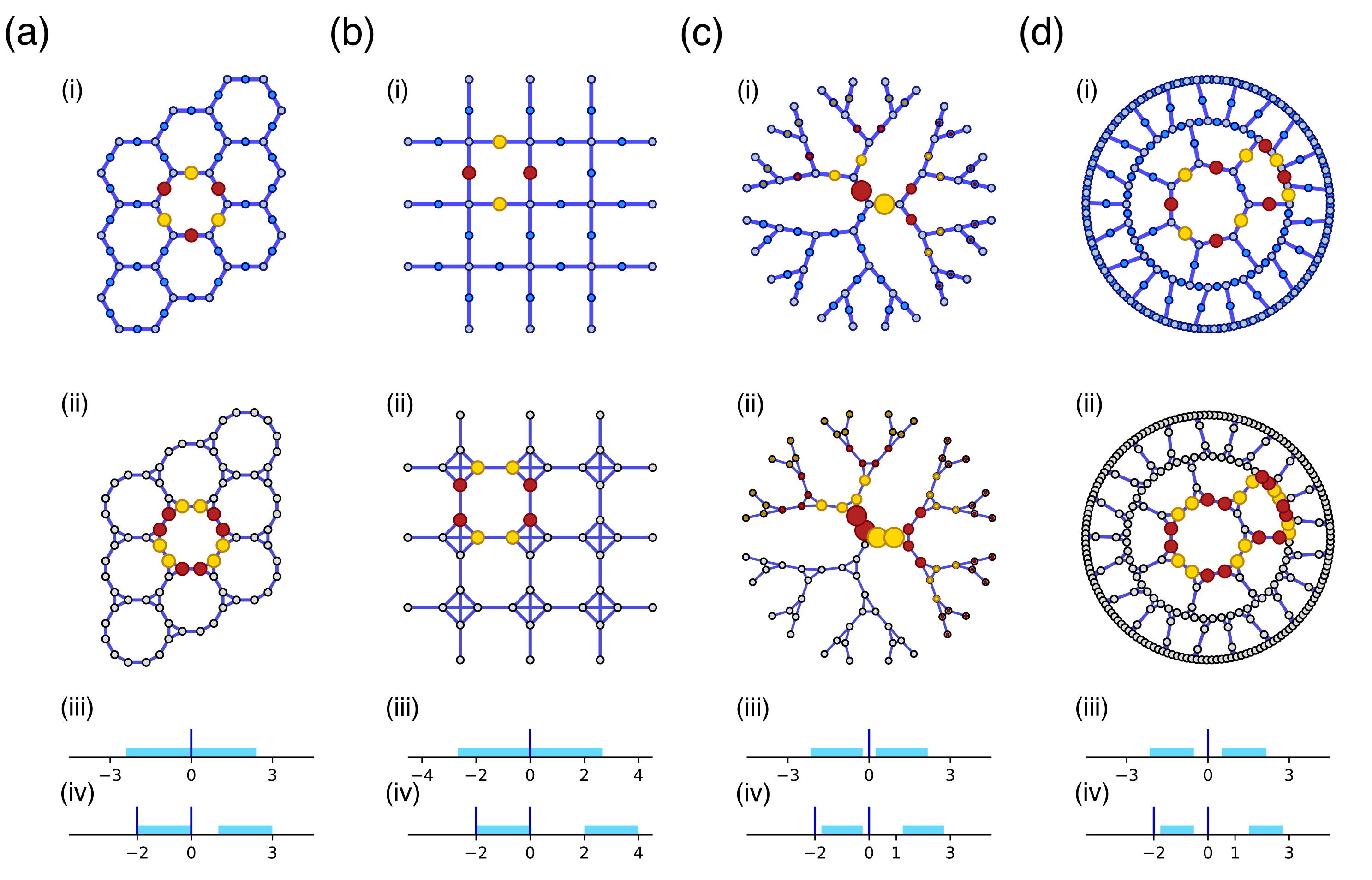}
	\end{center}
	\vspace{-0.6cm}
	\caption{\label{fig:SplitGraphsAndLineGraphs} 
    \textbf{Line graphs of subdivided graphs.}
    \textbf{a}\textit{i} The $3,2$-biregular graph $\mathbb{S}(T_6)$, obtained by adding a new vertex in the middle of each edge of $T_6$. Blue lines indicate edges, light blue circles the original vertices of $T_6$, and dark blue circles the additional vertices of $\mathbb{S}(T_6)$. 
    \textbf{a}\textit{ii} Its $3$-regular line graph $L(\mathbb{S}(T_6))$.
    \textbf{a}\textit{iii}-\textit{iv} Schematics of the regions where the $\ell^2$ DOS is nonzero for $\mathbb{S}(T_6)$ and $L(\mathbb{S}(T_6))$, respectively. In addition to the flat band at $-2$ expected from Eqn. \ref{eqn:fwincidence2} for $H_L(\mathbb{S}(T_6))$, both graphs exhibit a flat band at zero. Compactly supported eigenstates from both of these bands are plotted in \textbf{a}\textit{i}-\textit{ii}, where the size of the circle indicates the amplitude of the state, red indicates positive sign, and yellow negative. The remainder of the spectra consist of two intervals given by the eigenenergies of $T_6$ and Eqns. \ref{eqn:XSX} and \ref{eqn:XLSX}.
    \textbf{b}\textit{i}-\textit{iv} Corresponding plots starting from the $4$-regular square lattice, rather than $T_6$. The resulting $4,2$-biregular subdivided graph is well known in solid-state physics where it was introduced by Lieb\cite{Lieb:1988ba} as a thought example for studying the ground state properties of the Hubbard model.  Since both $T_6$ and the square lattice are bipartite and Euclidean, none of the flat bands in \textbf{a} or \textbf{b} are gapped.
    \textbf{c}\textit{i}-\textit{iv} Equivalent plots starting from the $3$-regular tree. These graphs, also shown in Fig. \ref{fig:McLaughlin}, exhibit gapped flat bands because the $\ell^2$ spectrum of the $3$-regular tree is the interval $(-2\sqrt{2}, 2\sqrt{2})$.\cite{Kesten:1959us} Since the tree has no cycles, the flat bands at $0$ consist of exponentially localized states.
    \textbf{d}\textit{i}-\textit{iv} Corresponding plots starting from the hyperbolic tessellation $T_7$ discussed in Sec. \ref{sec:hyperbolic} . This tiling is non-bipartite and has an asymmetric spectrum, therefore the minimum gap near $0$ is larger than that near $-2$ for $H_{L(\mathbb{S}(T_7))}$.
    } 
\end{figure}

 The subdivision graph of $X_3$ is the universal $3,2$-biregular bipartite graph and its line graph $\mathbb{M} = L(\mathbb{S}(X_3))$ is the McLaughlin graph depicted in Fig. \ref{fig:McLaughlin}. $\sigma(A_\mathbb{M})$ was computed in Ref.\cite{McLaughlin:1986vb} and it consists of two isolated flat bands at $-2$ and $0$ and is otherwise supported on the two indicated intervals, with absolutely continuous spectrum.  
$\mathbb{M}$ is the Cayley graph of $G =\mathbb{Z}_2 \star \mathbb{Z}_3$, w.r.t. the generators $\{ Q, R, R^{-1} \}$, with $Q^2 = R^3 = 1$. 
 $\mathbb{M}$ is a Hoffman graph and it covers all large finite $3$-regular Hoffman graphs. This follows from the classification of the latter (see Appendix \ref{app:hoffman}) as being line graphs of the $3,2$-biregular bipartite graphs. The $3$-regular Hoffman graphs in Eqn. \ref{eqn:McLaughlinGaps} have all their non-constant eigenvalues contained in $\sigma(A_\mathbb{M})$. 
 
The explicit computation of $\sigma(A_{X_3})$ and $\sigma(A_\mathbb{M})$ are special cases of the computation in terms of algebraic functions of the spectra of homogeneous $X$'s for which $G(X)$ has a free subgroup of finite index (see Ref.\cite{Woess:1987wc}).

The same operation $L(\mathbb{S}(X))$ which generates $\mathbb{M}$ from $X_3$ can also be applied to general regular graphs, and it always produces flat bands at both $-2$ and $0$. A series of examples is shown in Fig. \ref{fig:SplitGraphsAndLineGraphs}. Proofs of the existence of these flat bands and formulas for determining $\sigma(\mathbb{S}(X))$ and $\sigma(L(\mathbb{S}(X)))$ from $\sigma(A_X)$ are modified from Ref.\cite{Cvetkovic:1980} and is given in Appendix \ref{app:subdivisionGraphs}. A discussion of Euclidean examples which realize the locally maximal gap interval in Eqn. \ref{eqn:CharonGaps} is carried out it in Sec. \ref{subsec:EuclideanMaxGaps}.

\section{Euclidean Lattices}\label{sec:euclidean}
\subsection{Bloch Theory}
In the special case relevant to conventional solid-state physics, where the graph corresponds to a lattice which is a regular periodic tiling of Eucliean space, stronger statements can be made by exploiting the symmetries of the space. For every such lattice, there exists a smallest fundamental domain, or unit cell, $U_c$ such that $U_c$ contains finitely many lattice  points and translating $U_c$ by all integer linear combinations of two linearly independent vectors produces all the points in the lattice. These vectors are known as the lattice vectors or lattice generators $\vec{a}_1$ and $\vec{a}_2$. Translation by all integer linear combinations of these two vectors produces an Abelian group $\mathcal{A}$ which is isomorphic to $\mathbb{Z} \times \mathbb{Z}$.

Because this problem is periodic in space, the Hamiltonan $H$ is invariant under special translations, and the group $\mathcal{A}$ is precisely the largest possible group of such translations. By Bloch's theorem\cite{Ashcroft:1976ud} $H$ and $\mathcal{A}$ can be simultaneously diagonalized. Since $\mathcal{A}$ encodes a large fraction of the structure of $H$, it is sensible to classify lattices by the structure of $\mathcal{A}$. Physics literature, however, does not typically refer to the group structure of $\mathcal{A}$ explicitly. Instead it is standard to consider the set of points obtained by acting on a single point with $\mathcal{A}$. This set is known as the Bravais lattice $B_r$, and the field of crystallography classifies lattices in terms of the geometry of $B_r$.

If the fundamental domain contains only one point, then the set of lattice points $\mathcal{P}$ is itself a Bravais lattice, and the eigenfunctions of $H$ will also be eigenfunctions of $\mathcal{A}$. The states can therefore be parametrized in terms of their momentum and written as particularly simple Bloch waves
\begin{equation}\label{BlochWave}
\psi_{\vec{k}}(l) = e^{i \vec{k} \cdot \vec{x_l}},
\end{equation}
where $x_l$ is the location of the $l$th lattice site in $\mathcal{P}$.
This wavefunction $\psi_{\vec{k}}$ is an eigenfunction of $H$ which obeys the following relations
\begin{equation}\label{EigenEnergy}
H\psi_{\vec{k}} = E(\vec{k}) \psi_{\vec{k}}, 
\end{equation}
\begin{equation}\label{translate1}
T_{\vec{a}_1} \psi_{\vec{k}} =e^{-i \vec{k} \cdot \vec{a}_1} \psi_{\vec{k}},
\end{equation}
\begin{equation}\label{translate2}
T_{\vec{a}_2} \psi_{\vec{k}} =e^{-i \vec{k} \cdot \vec{a}_2} \psi_{\vec{k}},
\end{equation}
where $T_{\vec{v}}$ is an operator which translates the state by the vector $\vec{v}$.\cite{Ashcroft:1976ud}
The eigenenergy $E(\vec{k})$ varies continuously, and usually smoothly, with $\vec{k}$.
Since this is a discretized model with a restricted wavefunction that only exists on the lattice sites, multiple values of $\vec{k}$ produce identical wavefunctions. Therefore, both $E(\vec{k})$ and $\vec{k}$ are restricted to a fundamental domain  known as the first Brillouin zone. 
If the unit cell contains more than one site, then the Bravais lattice is smaller than $\mathcal{P}$; however, the solution above can be generalized to a vector-valued Bloch wave with $J$ entries, one for each site in the unit cell. 
See Ref.\cite{Ashcroft:1976ud} for a full description and proof of this procedure in physics notation and Ref.\cite{Kotani:2003jq} for a translation of it into the terminology of Cayley graphs and abstract algebra.
Once this is done, there are $j = 1,..., J$ solutions, one for each lattice point in the first Brillouin zone. Each of these solutions can be parametrized continuously as a function of $\vec{k}$, yielding energy bands $E^{(j)}(\vec{k})$ and corresponding eigenstates $\psi^{(j)}_{\vec{k}}$. The decomposition of these solutions into $J$ bands parametrized by $\vec{k}$ is known as a band structure. 

When thinking of a tight-binding lattice as a graph, the notion of $\vec{k}$ as we have used it here becomes ill-defined because it depends on the precise realization of the graph. However, the eigenvalue under translation by the generators of $\mathcal{A}$ is an equivalent, realization-independent, quantity. In a slight abuse of notation we will refer to the band structure for a given lattice as $B(X)$, where $X$ is the corresponding graph, and chose the parametrization that is most convenient for each case.

\begin{figure}[h]
	\begin{center}
		\includegraphics[width=0.9\textwidth]{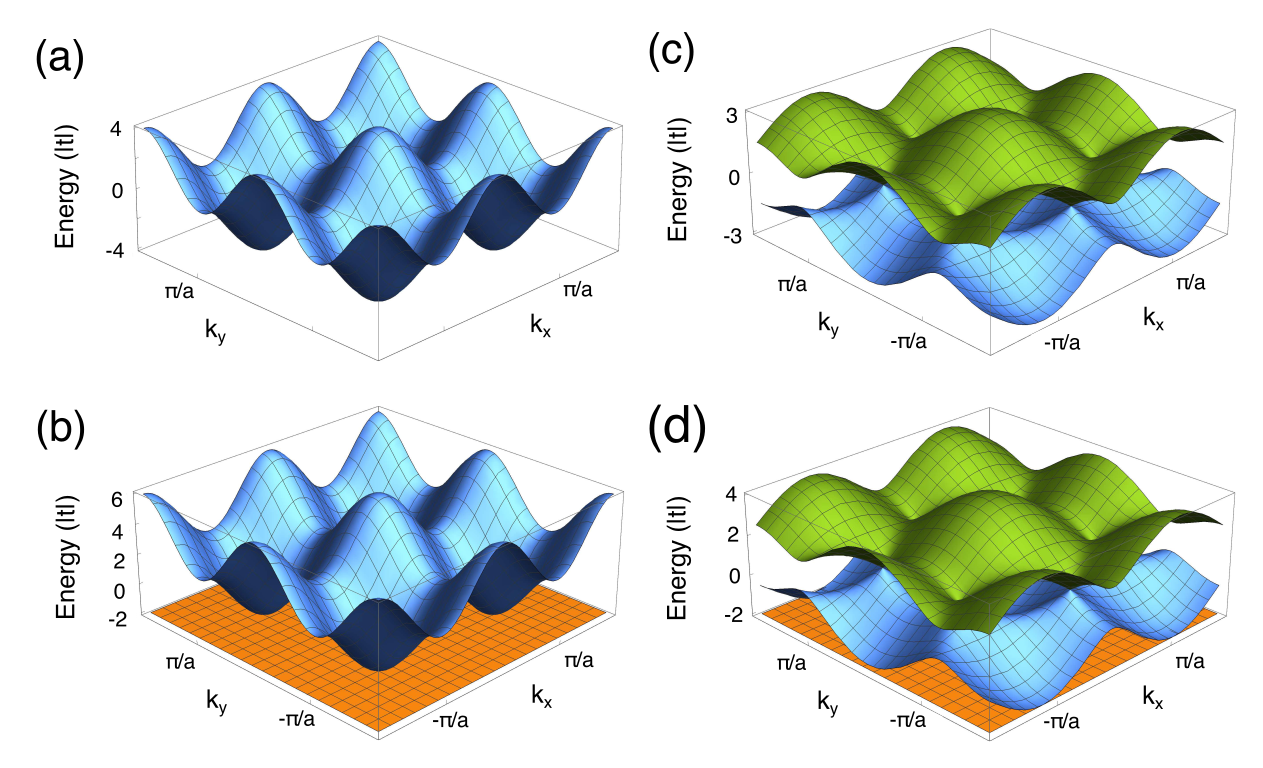}
	\end{center}
	\vspace{-0.6cm}
	\caption{\label{fig:G_LG_BandStructure} 
    \textbf{Band structure of line graphs.}
    \textbf{a} The band structure of the square lattice (shown in Fig. \ref{fig:G_LG_Examples} \textbf{b}\textit{i}) with $t=-1$.
    \textbf{b} Band structure of the line graph of the square lattice (shown in Fig. \ref{fig:G_LG_Examples} \textbf{b}\textit{ii}). This band structure consists of a flat band at $-2$ and copy of the square lattice band structure shifted up by $2$.
    \textbf{c} The band structure of graphene (shown in Fig. \ref{fig:G_LG_Examples} \textbf{a}\textit{i}) with $t=-1$. 
    \textbf{d} The band structure of the kagome lattice, the line graph of graphene (shown in Fig. \ref{fig:G_LG_Examples} \textbf{a}\textit{ii}). This band structure consists of a flat band at $-2$ and copy of the graphene band structure shifted up by $1$.
    } 
\end{figure}

Let $X$ be a $d$-regular Euclidean lattice with $d\geq 3$ and all nearest-neighbor hopping matrix elements equal.
It can be converted into a circuit QED lattice by placing one resonator on each edge in $X$. The effective graph $L(X)$ is then a $2d-2 > 3$ regular Euclidean lattice, which is realized with its vertices at the midpoints of the edges of a realization of $X$. In this particular realization both $X$ and $L(X)$ have the same Bravais lattice. Other realizations yield the same results, but are more cumbersome to compute with, so for the remainder of this section we will always assume this medial lattice construction. We showed previously that $\sigma(\bar{H}_s (X)) = d- 2 + \sigma(H_X)$ and that  $\sigma(\bar{H}_a (X)) = d- 2 - \sigma(H_X)$. However, combining Bloch theory with the results from Sec. \ref{sec:gengraphs}, we can make stronger statements about the band structures and not just the spectra. Note that many of these results were known in the mathematical physics community studying ferromagnetic ground states of the Hubbard model in flat bands, see for examples Refs.\cite{Mielke:1991bands, Mielke:1991multiplicity}. However, that body of literature focuses almost entirely on bipartite layouts, and pays little attention to the momentum structure of the higher bands.

If $X$ is an infinite $d$-regular graph corresponding to a Euclidean lattice, 
then not only the spectra, but also the band structures of $\bar{H}_s(X)$ and $\bar{H}_a(X)$ are completely determined by $H_X$. We denote the energy bands of $\bar{H}_s(X)$ by $E^{(j)}_{s(X)}$, and those of $\bar{H}_a(X)$ by $E^{(j)}_{a(X)}$. It then follows that
for each band in the spectrum of $H_X$, we have
\begin{equation}\label{Brelation}
E^{(j)}_{s(X)}(\vec{k}) = d-2 + E^{(j)}_{X}(\vec{k}),
\end{equation}
and
\begin{equation}\label{BrelationHW}
E^{(j)}_{a(X)}(\vec{k}) = d-2 - E^{(j)}_{X}(\vec{k}).
\end{equation}
The remaining bands in $B(\bar{H}_s(X))$ and $B(\bar{H}_a(X))$ are flat bands of the form
$$E^{(j)}_{s(X)}(\vec{k}) = -2,$$
and 
$$E^{(j)}_{a(X)}(\vec{k}) = -2,$$
 consisting of localized eigenstates of compact support.

\textit{Proof:}
First, consider a state $\psi$ which is an eigenstate of $H_X$ with eigenvalue $E_X$. The incidence matrices $M$ and $N$ are operators which map states in $\ell^2(X)$ to states in $\ell^2(L(X))$. They produce two new states $\Psi_s = M\psi$ and $\Psi_a = N \psi$.
Using Eqn. \ref{eqn:fwincidence2}, it follows that 
\begin{eqnarray}
M M^t M \psi  &=& M (M^t M) \psi = M (D + H_X) \psi \\
              &=& M (d + E_X) \psi = (d+ E_X) M \psi \nonumber \\
              &=& (d + E_X) \Psi_s, \nonumber
\end{eqnarray}
and 
\begin{eqnarray}
M M^t M \psi  &=& (M M^t) M \psi = (2I + \bar{H}_s) M \psi \\
              &=& (2 + \bar{H}_s) \Psi_s. \nonumber
\end{eqnarray}
Combining these two relations we obtain 
\begin{equation}
(2 + \bar{H}_s)\Psi_s = (d + E_X)\Psi_s,
\end{equation}
which can be rearranged to 
\begin{equation}
\bar{H}_s(X) \Psi_s = (d-2 + E_X)\Psi_s.
\end{equation}

The corresponding half-wave relations are very similar and yield.
\begin{eqnarray}
N N^t N \psi  &=& N (N^t N) \psi = N (D -H_X) \psi \\
              &=& N (d - E_X) \psi = (d- E_X) N \psi \nonumber\\
              &=& (d - E_X) \Psi_a, \nonumber
\end{eqnarray}
\begin{eqnarray}
N N^t N \psi  &=& (N N^t) N \psi = (2I +\bar{H}_a) N \psi \\
              &=& (2 + \bar{H}_a) \Psi_a, \nonumber
\end{eqnarray}
\begin{equation}
(2 + \bar{H}_a)\Psi_a = (d - E_X)\Psi_a,
\end{equation}
and 
\begin{equation}
\bar{H}_a(X) \Psi_a = (d-2 - E_X)\Psi_a.
\end{equation}
 Therefore, for all eigenstates of $H_X$ there exist corresponding eigenstates of $\bar{H}_s(X)$ and $\bar{H}_a(X)$. The remaining states on $L(X)$ are the kernel of $MM^t$ or $NN^t$ and give rise to the flat band(s) at $-2$ for $\bar{H}_s(X)$ and $\bar{H}_a(X)$. As shown in Fig. \ref{fig:FWHW_FB_States}, these flat bands will consist of localized eigenstates of compact support arising from destructive interference of hopping amplitudes and voltages.

In order to show the relation between the band structures, it suffices to show that if $\psi$ is a Bloch wave with momentum $\vec{k}$, then $\Psi_s$ and $\Psi_a$ are also Bloch waves with momentum $\vec{k}$. We proceed by choosing a  specific realization of $X$. The location of a vertex $x$ is given by a vector $\vec{x}$. The vertex $(xy)$ in $L(X)$ will be drawn at the midpoint of the bond, $\vec{z} := (\vec{x} + \vec{y})/2$. Since $\psi_{\vec{k}}$ is a Bloch wave, we know that 
\begin{equation}
\psi_{\vec{k}}(\vec{x}) = u_{\gamma_x} (\vec{k}) e^{i \vec{k} \cdot \vec{x}},
\end{equation}
where $\gamma$ indexes the different sites in the unit cell of $X$, and $u_{\gamma_x} (\vec{k})$ is a function which depends only on the parameter $\vec{k}$. Using the definitions of $M$ and $N$, we find that 
\begin{eqnarray}
\Psi_s (\vec{z})  &= &  \left[ M(\vec{z}, \vec{x}) u_{\gamma_x} (\vec{k}) e^{i \vec{k} \cdot \vec{x}}    + M(\vec{z}, \vec{y})\, u_{\gamma_y} (\vec{k}) e^{i \vec{k} \cdot \vec{y}}    \right],\\
  &= &  \left[  M(\vec{z}, \vec{x}) u_{\gamma_x} (\vec{k}) e^{i (\vec{k} \cdot \vec{x} -  \vec{k} \cdot \vec{y})/2}    + M(\vec{z}, \vec{y})\, u_{\gamma_y} (\vec{k}) e^{i (\vec{k} \cdot \vec{y} -  \vec{k} \cdot \vec{x})/2}    \right] \ e^{i (\vec{k} \cdot \vec{x} +  \vec{k} \cdot \vec{y})/2} \nonumber\\
  & = & \tilde{u}_s(\vec{k}, \vec{z}) e^{i \vec{k} \cdot \vec{z}}, \nonumber
\end{eqnarray}
where $M(\vec{z}, \vec{y})$ is the entry of the incidence matrix $M$ for the edge whose center point is $\vec{z}$ and for the vertex drawn at $\vec{y}$. The state $\Psi_a$ obeys a similar relation with the incidence matrix $N$:
\begin{eqnarray}
\Psi_a (\vec{z})  &= &  \left[ N(\vec{z}, \vec{x}) u_{\gamma_x} (\vec{k}) e^{i \vec{k} \cdot \vec{x}}    + N(\vec{z}, \vec{y})\, u_{\gamma_y} (\vec{k}) e^{i \vec{k} \cdot \vec{y}}    \right],\\
  &= &  \left[N(\vec{z}, \vec{x})   u_{\gamma_x} (\vec{k}) e^{i (\vec{k} \cdot \vec{x} -  \vec{k} \cdot \vec{y})/2}    + N(\vec{z}, \vec{y})\, u_{\gamma_y} (\vec{k}) e^{i (\vec{k} \cdot \vec{y} -  \vec{k} \cdot \vec{x})/2}    \right] \ e^{i (\vec{k} \cdot \vec{x} +  \vec{k} \cdot \vec{y})/2} \nonumber\\
  & = & \tilde{u}_a(\vec{k}, \vec{z}) e^{i \vec{k} \cdot \vec{z}}, \nonumber
\end{eqnarray}
These two new wavefunctions $\Psi_\alpha$ are Bloch waves on $L(X)$ with momentum $k$ if they are proportional to $e^{i \vec{k} \cdot \vec{z}}$ and if $\tilde{u}_\alpha(\vec{k}, \vec{z})$ depends only on which site of unit cell $\vec{z}$ is equivalent to. The correct complex exponential has been explicitly factored out, so all that remains is to show that $\tilde{u}_\alpha(\vec{k}, \vec{z})$ depends only on position within the unit cell, and not on the absolute position of $\vec{z}$. Let $\beta$ index the sites in the unit cell of $L(X)$, i.e. the bonds of $X$. Each such bond will always have the same orientation, so $\vec{x} - \vec{y}$ will be the same for all such sites. For a given $\beta$, its endpoints will always be equivalent to the same two sites in the unit cell of the layout lattice, $\gamma_1$ and $\gamma_2$. Therefore, the pair $u_{\gamma_x}$ and $u_{\gamma_y}$ will always be the same. If we choose an orientation of the layout resonators that respects the Bravais lattice symmetry, then the pair $M(\vec{z},\vec{x})$ and $M(\vec{z}, \vec{y})$ will also be the same for all instances of the bond $\beta$, and same will be true of $N$. As a result, the functions $\tilde{u}_s$ and $\tilde{u}_a$ depends only on $\beta$ and $\vec{k}$, not on the specific value of $\vec{x}$, $\vec{y}$, or $\vec{z}$; and $\Psi_\alpha$ is a Bloch wave on $L(X)$ with momentum $\vec{k}$. Examples of the band structures of $H_X$ and $\bar{H}_s$ are shown in Fig. \ref{fig:G_LG_BandStructure} for Euclidean 3-regular and 4-regular cases. Those for $\bar{H}_s (X)$ are clearly shifted copies of that of $H_X$ plus a flat band at $-2$. For these particular cases $\bar{H}_a(X)$ has the same band structure as $\bar{H}_s$ and is not shown separately. More subtle non-bipartite examples will be shown later in Fig. \ref{fig:HPKTopology}.


We are interested in understanding when the flat bands in the spectra of $\bar{H}_s(X)$ and $\bar{H}_a(X)$ are gapped. From the correspondence between their dispersive bands and those of $H_X$, it is clear that if $X$ is $d$-regular, the existence and magnitude of this gap is completely determined by the density of states of $H_X$ near $-d$ in the case of $\bar{H}_s(x)$, and near $d$ in the case of $\bar{H}_a(X)$, and two general statements can be made. First, in Euclidean lattices, the flat bands of $\bar{H}_a(X)$ are never gapped, and second, the flat bands of $\bar{H}_s(X)$ are gapped if and only if $X$ is non-bipartite.

Consider first the simpler case of $\bar{H}_a(X)$. Its flat bands are gapped if and only if there exists a non-zero positive $\epsilon$ and an interval $(d-\epsilon, d]$ in which $H_X$ has no eigenstates. Otherwise the states of $X$ in this interval will give rise to a set of states on $L(X)$ with eigenvalues in the interval $[-2, -2+\epsilon]$ which touch the flat band. The constant function on $X$ is an eigenfunction of $H_X$ with eigenvalue $d$, but it is not strictly speaking $\ell^2$-normalizable. However, in this section we are considering only Euclidean lattices, so it is in the $\ell^2$ closure and no such non-zero $\epsilon$ exists.

The case of $\bar{H}_s(X)$ is slightly more complicated. In this case, the flat band is gapped if and only if there is a nonzero $\epsilon$ such that $H_X$ has no states in the interval $[-d, -d +\epsilon]$.
If $X$ is bipartite, then its vertices can be divided into two sublattices $\mathcal{V}_A$ and $\mathcal{V}_B$ such that if $x \in \mathcal{V}_A$, then $\mathcal{N}_x \subset \mathcal{V}_B$ and vice versa. Each state on $X$ can therefore be decomposed into a state on each sublattice $\psi = (\psi_A, \psi_B)$. If $\psi$ is an eigenstate of $H_X$ with eigenvalue $E$, consider the state $\psi' = (\psi_A, -\psi_B)$. Since all the neighbors of a given lattice site are in the opposite sublattice, this state reverses the sign of every term in $H \psi$, and is an eigenstate with eigenvalue $-E$. The spectrum of $H_X$ is therefore symmetric about $E = 0$.
As before, the state $\psi(x) = {1 \ \forall\  x \in V(G)}$ has eigenvalue $d$ and is in the closure of $\ell^2(X)$. Therefore, the state $M\psi'$ is a state in the closure of $\ell^2(L(X))$ with eigenvalue $-2$ and touches the flat band. 

The non-bipartite case can be proved by contradiction. Assume that $X$ is not bipartite, and that the flat band is ungapped. Then there exists a state $\zeta$ in the $\ell^2$ closure such that $H_X \zeta = -d$. The Hamiltonian $H_G$ commutes with the symmetry transformations of the lattice, and these two sets of operators can be simultaneously diagonalized. As a result, there will exist a $\zeta$ which will return to itself under all translations in the Bravais lattice. If the magnitude of the eigenvalue with respect to translations is greater than one, $\zeta$ will grow exponentially and be severely non-normalizable, so the value of $\zeta$ in each unit cell must return to itself up to a phase factor. The only remaining possibility is amplitude variations within the unit cell. If $\zeta$ has a local maximum within the unit cell, then at that point $H_X\zeta < d \zeta$. If it does not have a local maximum, then it must grow in one direction, which contradicts the normalizability requirement. Therefore, $|\zeta(x)|$ must be constant, in which case  $\zeta$ will have the property that
$$ \zeta(y) = -\zeta(x)\  \forall\  y \in \mathcal{N}_x. $$
As a result, every vertex in $X$ can be labeled as either type-A or type-B according to the sign of $\zeta$, which contradicts the assumption that $X$ is non-bipartite.

\subsection{Real-Space Topology and Gapped Flat Bands at -2}

\begin{figure}[h]
	\begin{center}
		\includegraphics[width=0.9\textwidth]{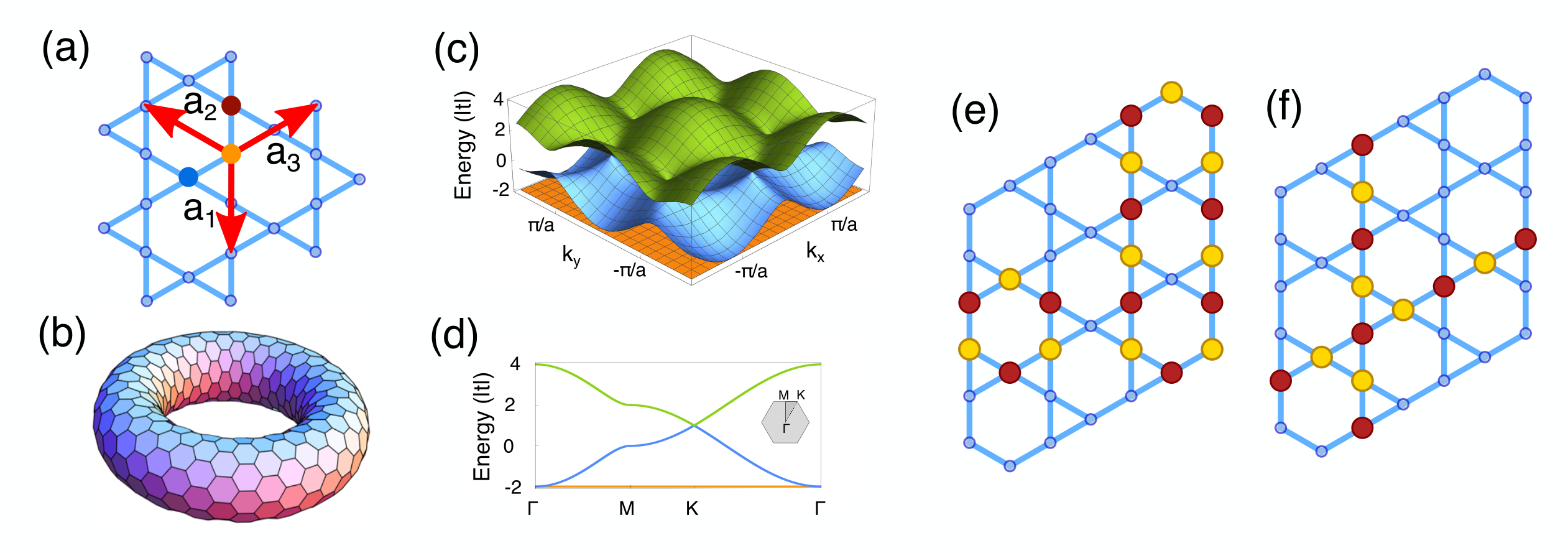}
	\end{center}
	\vspace{-0.6cm}
	\caption{\label{fig:Kagome_Topology} 
    \textbf{Kagome real space topology.} \textbf{a} The three-site unit cell and smallest lattice generators of the kagome lattice, only two of which are linearly independent.
    \textbf{b} Schematic hexagonal unit cells under periodic boundary conditions.
\textbf{c} The band structure of the kagome lattice for $t = -1$. The flat band at $-2$ is shown in orange, and the two dispersive graphene-like bands above it are shown in blue and green. The middle band touches the flat band at $k=0$.
\textbf{d} Line cut through the band structure along a path through the first Brillouin zone shown in the inset.
\textbf{e} Two localized flat band states. Each hexagonal plaquette supports one 6-site compactly supported eigenstate. Sums of these minimum states produce larger loops, such as the three plaquette one shown here. Since band structure is computed on the torus, which is without boundary, summing up the hexagonal states on all plaquettes results in a vanishing linear combination. Thus, $\mathcal{M}$ plaquettes on the torus produce only $\mathcal{M}-1$ linearly independent localized states of this type, one less than in a Bloch band.
\textbf{f} Noncontractible-loop flat-band states. These two states wrap fully around the torus, and are linearly independent of the contractible loop states in \textbf{e}, giving a total of $\mathcal{M}+1$ states with energy $-2$. The presence of this extra state requires a band touch between the flat and dispersive bands and forbids the presence of a gapped flat band.\cite{Bergman:2008es}
    } 
\end{figure}

In the case of Euclidean lattices, in addition to the graph theory results presented above, there is a topology argument due to Bergman \textit{et al.}\cite{Bergman:2008es} which is conventionally used to understand when flat bands can be gapped. It was originally derived for the simplest case of the kagome lattice, which arises when the layout graph is graphene (a hexagonal honeycomb). We will sketch their argument for this case before applying it to more unusual examples. Step one, determine the unit cell and Bravais lattice of the lattice under consideration. The three-site unit cell and the generators of the triangular Bravais lattice of the kagome lattice are shown in Fig. \ref{fig:Kagome_Topology} \textbf{a}. 

Step two, consider a paralellogram of $\mathcal{N} \times \mathcal{N}$ unit cells and apply periodic boundary conditions by wrapping it onto a torus, as sketched in Fig. \ref{fig:Kagome_Topology} \textbf{b}. From standard Bloch theory \cite{Ashcroft:1976ud} it is known that there will be $J$ energy bands, where $J$ is the number of sites in the unit cell. The three bands of the kagome lattice are plotted in Fig. \ref{fig:Kagome_Topology} \textbf{c}. Since this is a finite-sized sample with periodic boundary conditions, we will not obtain the full continuum surfaces. Instead, we will obtain a uniform-mesh sampling of them with a discreteness set by the periodicity of the torus and $\mathcal{M} = \mathcal{N}^2$ total points per band.\cite{Ashcroft:1976ud}

Step three is to consider the eigenstates and count the total number with eigenvalue $-2$ that are linearly independent. One of the three bands is completely flat, so we expect to find $\mathcal{M}$ states at this energy. We know that there will be a localized eigenstate of compact support on every hexagonal plaquette of the lattice, shown in Fig. \ref{fig:Kagome_Topology} \textbf{e}. Summing these states together will result in linearly dependent configurations which are analogous states on larger and larger cycles in the lattice. Since the torus is without boundary, as we sum up more and more of the hexagonal localized states, the resulting loop will grow until it meets itself and annihilates. 
This indicates the presence of a linearly-dependent single-plaquette state, and therefore, there exist only $\mathcal{M}-1$ independent hexagonal localized states, which is one less than expected. The missing state is a noncontractible loop which wraps around the torus. In fact, there are two such states, shown in Fig. \ref{fig:Kagome_Topology} \textbf{f}, giving a total of $\mathcal{M}+1$ states with eigenvalue $-2$. This is one more state than is provided by the flat band. Therefore, one of the other bands must dip down to $-2$ and touch the flat band, guaranteeing that there is no energy gap above it.

\begin{figure}[h!]
	\begin{center}
        \includegraphics[width=0.9\textwidth]{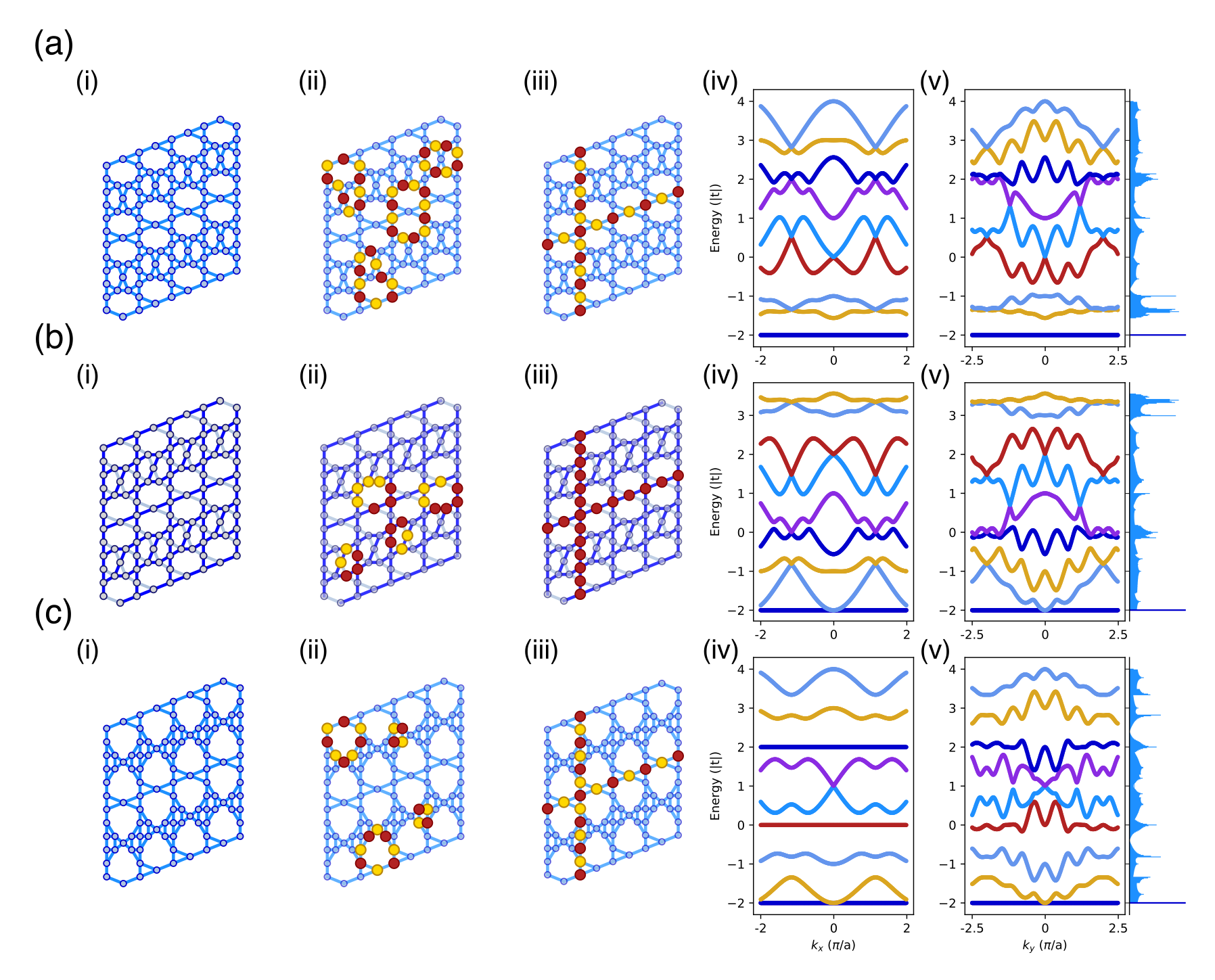}
	\end{center}
	\vspace{-0.6cm}
	\caption{\label{fig:HPKTopology} 
\textbf{Modified kagome lattices.} 
\textbf{a}\textit{i} The heptagon-pentagon-kagome lattice formed by adding interstitials to a traditional kagome lattice and creating two heptagonal and two pentagonal plaquettes in the space of three hexagonal ones. It is a Euclidean lattice and can be treated with Bloch theory, and it is the line graph of the non-bipartite heptagon-pentagon-graphene lattice whose smallest cycles are all odd. 
\textbf{a}\textit{ii} The smallest localized eigenstates, which give rise to four flat bands at $-2$. As in the hyperbolic lattices discussed in Ref.\cite{Kollar:2018vc}, these states all enclose two plaquettes. Because they interlock and overlap, translating these states forms two complete covers of the torus and two vanishing linear combinations for a total of $4\mathcal{M}-2$ linearly independent states.
\textbf{a}\textit{iii} The two noncontractible loop states. These, together with the contractible loop states in \textbf{a}\textit{ii}, are all the $4\mathcal{M}$ flat-band states. Unlike the kagome lattice in Fig. \ref{fig:Kagome_Topology}, there is no extra state in the flat bands, and a band touch is not required. The real space topology argument does not preclude an accidental band touch, but since the layout graph for this lattice is non-bipartite, the flat bands will be gapped.
\textbf{a}\textit{iv}-\textit{v} Cuts through the band structure along $k_y = 0$ and $k_x = 0$ which confirm the presence of the gapped flat bands. A bar plot of the DOS ($0.02t$ resolution) is shown next to the cuts with energy on the vertical axis, with the number of states in each energy bin indicated by the width of the bars. Flat bands are indicated in dark blue and dispersive bands in light blue.
\textbf{c}\textit{i}-\textit{v} Equivalent plots for the half-wave heptagon-pentagon kagome lattice. The smallest flat band states encircle single plaquettes. Combined with the noncontractible loop states they give rise to $4\mathcal{M}+1$ linearly independent states with eigenvalue $-2$ and a dispersive band is required to bend down to $-2$ and supply the additional one. The DOS and cuts through the band structure in \textbf{b}\textit{iv}-\textit{v} confirm that the gap above the flat bands closes at $k=0$.
\textbf{c}\textit{i}-\textit{v} Equivalent plots for the octagon-square-kagome lattice. This lattice is the line graph of a bipartite lattice, and the full-wave and half-wave models are identical. As in \textbf{b}, the smallest flat band states enclose single plaquettes and do not overlap, and the gap above the flat bands closes at $k=0$.
    } 
\end{figure}

This argument generalizes naturally to the line graph of any 3-regular Euclidean lattice with only even cycles. The kagome lattice is the simplest lattice of this type, and the only one that arises as the tight-binding approximation of a lattice of atoms with constant nearest neighbor spacing. CPW lattices, however,  make line graphs directly and easily have uniform hopping rates  even if there is no realization of $L(X)$ in which the nearest-neighbor vertices are all equidistant. They can therefore realize a much broader class of such examples, some examples of which are shown in Fig. \ref{fig:HPKTopology}, along with their band structures.

As was shown by Bergman et al. in their original paper\cite{Bergman:2008es}  the topology of how the localized states cover the torus, and thus the combinatorics of the flat band states, can change drastically if the smallest localized states overlap. This is precisely what happens in the band structure of $\bar{H}_s(X)$ when $X$ is non-bipartite and the smallest cycle is odd. Consider for example the lattice shown in Fig. \ref{fig:HPKTopology} \textbf{a}\textit{i}. It is a variant of the kagome lattice formed by adding interstitials sites. These sites are then connected in a way that transforms three hexagonal plaquettes into two heptagonal and two pentagonal ones. Due to its similarity to the kagome lattice and the hyperbolic kagome analogs presented in Ref.\cite{Kollar:2018vc}, we refer to this lattice as the heptagon-pentagon kagome lattice. The smallest cycles in this lattice are odd, and therefore do not support localized flat-band states with eigenvalue $-2$. However, the lattice does contain even cycles, the smallest of which are realized by encircling two plaquettes, rather than just one. For each unit cell, there are four such states which are linearly independent. One copy of each is shown in Fig. \ref{fig:HPKTopology} \textbf{a}\textit{ii} (spatially separated for clarity). 

Each of these states gives rise to a flat band, so Bloch theory predicts a total of $4\mathcal{M}$ states in the flat bands. Just as in the kagome case, there are two noncontractible loop states, shown in Fig. \ref{fig:HPKTopology} \textbf{a}\textit{iii}. However, because the flat band states overlap, there are now two independent ways to cover the entire torus, and therefore two vanishing linear combinations. As a result and unlike in the kagome case, a dispersive band is not required to touch the flat bands. In principle there could still be an accidental touch between the bands, but since the heptagon-pentagon-graphene layout graph is not bipartite, the results in Sec. \ref{sec:gengraphs} guarantee the presence of a gap between the flat band and the rest of the spectrum. This gap is clearly visible in the band structure calculations shown in Fig. \ref{fig:HPKTopology} \textbf{a}\textit{iv}-\textit{v}.

For $\bar{H}_s(X)$ and bipartite $X$, or for $\bar{H}_a(X)$ and \textit{any} $X$, the smallest localized states cover only a single plaquette and do not interlock. Therefore, the real-space topology argument for these lattices is identical to that for the kagome lattice, and both analysis methods conclude that the flat bands at $-2$ cannot be gapped. Examples of two such models, their flat band states, and band structures are shown in Fig. \ref{fig:HPKTopology} \textbf{b}\textit{i}-\textit{v} and \textbf{c}\textit{i}-\textit{v}.

\subsection{Finite-Size Effects}

Experiments must necessarily occur on finite-sized samples. The simplest such graph is an induced subgraph $S(X)$ in which all vertices outside a finite region have been removed. This graph shares many properties and symmetries with the infinite lattice $X$, but the coordination numbers will vary at the boundary.
As shown in Sec. \ref{sec:gengraphs} the spectra of 
$\bar{H}_s(S(X))$ and $\bar{H}_a(S(X))$ are determined by that of $H(S(X))$, but 
the irregularity of the boundary of $S(X)$ produces additional eigenvalues which do not correspond to eigenvalues of $H_X$. 
For Euclidean lattices, this difficulty can be removed theoretically by applying periodic boundary conditions.
Unfortunately, periodic boundary conditions usually result in a highly non-planar graph which is incompatible with the single-layer fabrication process of circuit QED lattices, and experimentally we must work with hard-wall truncations. Fortunately, however, Euclidean geometry guarantees that as the system size increases, the edge states induced by the truncation constitute a vanishing fraction of the possible states.

\subsection{Maximally Gapped Flat Bands}\label{subsec:EuclideanMaxGaps}

\begin{figure}[h]
	\begin{center}
		\includegraphics[width=0.8\textwidth]{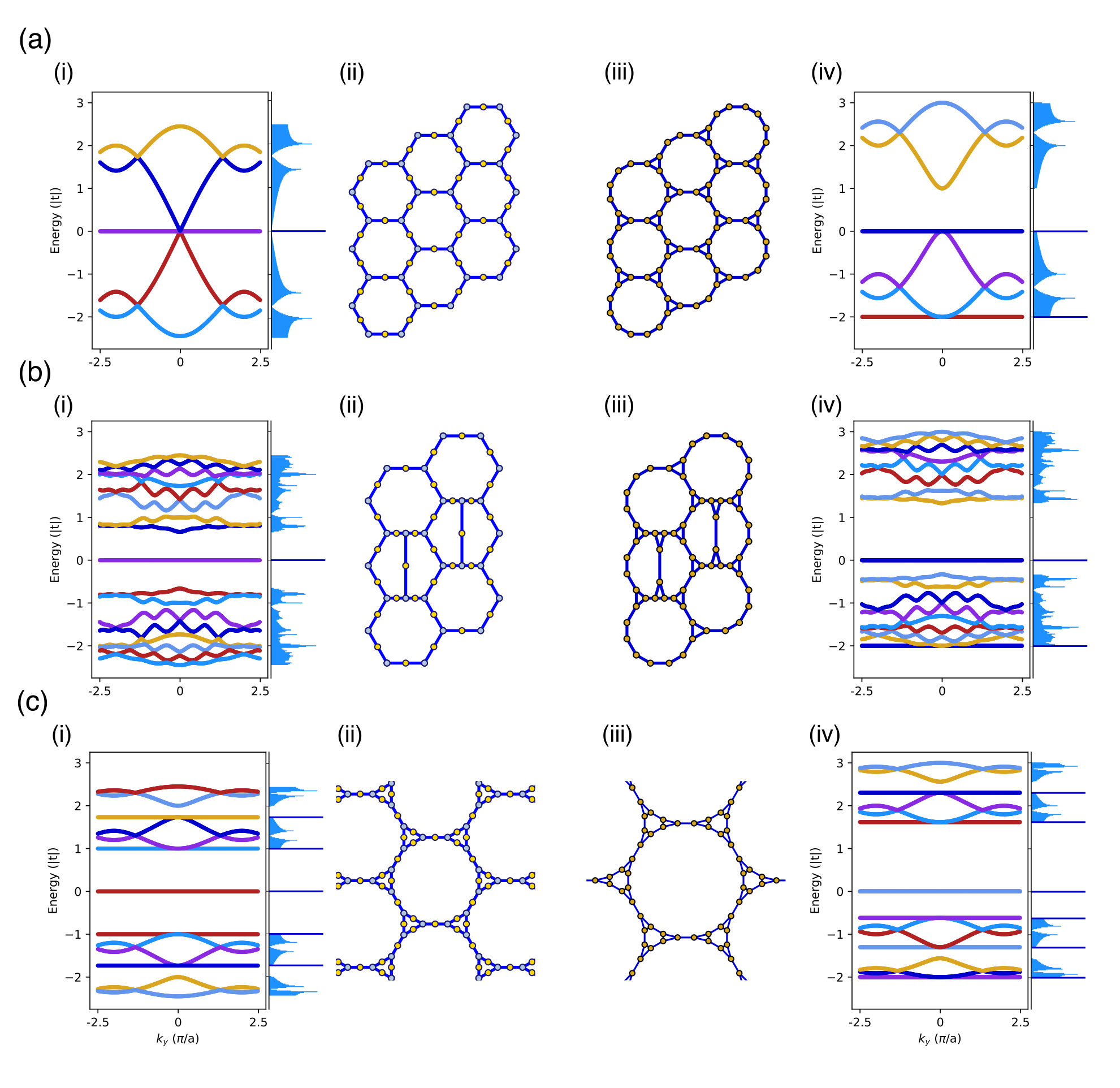}
	\end{center}
	\vspace{-0.6cm}
	\caption{\label{fig:SplitGraphs} 
    \textbf{Spectra of subdivided graphs and their line graphs.}
    Three examples of Hoffman graphs with flat bands at $-2$ and $0$.
    \textbf{a} The sample layout graph $X = \mathbb{S}(T_6)$ which is $3,2$-biregular and its $3$-regular effective lattice $L(\mathbb{S}(T_6))$ are shown in \textbf{a}\textit{ii}-\textit{iii}, and cuts through the resulting band structures for the layout Hamiltonian $H_{\mathbb{S}(T_6)}$ and the effective Hamiltonian $\bar{H}_s(\mathbb{S}(T_6))$ are shown in \textbf{a}\textit{i} and \textbf{a}\textit{iv}, respectively. A bar plot of the DOS ($0.02t$ resolution) is shown next to the cuts with energy on the vertical axis, with the number of states in each energy bin indicated by the width of the bars. Flat bands are indicated in dark blue and dispersive bands in light blue. 
    Both band structures consist of stretched versions of that of $H_{T_6}$ and additional flat bands. Since $T_6$ is bipartite, none of the flat bands are gapped, and the gap above $0$ in \textbf{a}\textit{iv} takes on the minimum possible value for a Hoffman graph: 1.
    \textbf{b}\textit{i}-\textit{iv} Corresponding plots starting from the heptagon-pentagon graphene lattice. Since this graph is non-bipartite, both flat bands at $0$ are gapped. That of $H_{\mathbb{S}(X)}$ in \textbf{b}\textit{i} is gapped symmetricallly, and that of $\bar{H}_s(\mathbb{S}(X))$ is gapped below by $\mathcal{R} >0$ and above by $1+ \mathcal{R}$. However, the least eigenvalue of the heptagon-pentagon graphene lattice is not $-2$ so $\mathcal{R}$ is only half of the maximal value.
    \textbf{c}\textit{i}-\textit{iv} Corresponding plots for a Euclidean construction that does attain the maximal value of $\mathcal{R}$. This graph, shown in \textbf{c}\textit{iii} is obtained as $\mathfrak{X} = L(\mathbb{S}(\mathcal{X}))$, where $\mathcal{X} =L(\mathbb{S}(T_6))$. The graph $\mathcal{X}$ is itself a $3$-regular line graph and Hoffman graph with least eigenvalue $-2$ and is a quotient of the Cayley graph of $\mathbb{Z}_2*\mathbb{Z}_3$. $\mathfrak{X}$ is therefore also a $3$-regular line graph which exhibits the maximal $\mathcal{R} = (1 + \sqrt{5})/2)$.
    If the graph in \textbf{c}\textit{iii} is used as a layout, the resulting line graph, shown in Fig. \ref{fig:AliciumBands} \textbf{c}-\textbf{d}, is $4$-regular and $\bar{H}_s(\mathfrak{X})$ has maximally gapped flat bands at $-2$ and $1$. 
    } 
\end{figure}

As was shown in Sec. \ref{sec:gengraphs} and is illustrated in Fig. \ref{fig:HPKTopology}, producing gapped flat bands at $-2$ in Euclidean lattices requires restricting to $\bar{H}_s(X)$ and non-bipartite layout graphs $X$. In order to ensure that the required all-way couplers are not unphysical, we will restrict to layouts which are $3$-regular. The size of the gap will then be determined by the minimum eigenvalue of $X$, so we are interested in the maximally non-bipartite $X$'s. As was shown in Sec. \ref{sec:gengraphs} and Appendix \ref{app:hoffman}, the largest that this eigenvalue can be is $-2$, which gives rise to a gap of $1$ above the flat band, and is achieved by Hoffman graphs. With finitely many exceptions, this value is achieved only by $3$-regular line graphs. As with the the McLaughlin graph $\mathbb{M}$, such graphs are realized by starting from a $3$-regular graph $X$, taking the $3,2$-biregular graph $\mathbb{S}(X)$, and finally its line graph $L(\mathbb{S}(X))$. Starting from a $3$-regular Euclidean lattice $X$ will produce a Euclidean $L(\mathbb{S}(X))$ which is also $3$-regular, but will have least eigenvalue $-2$. Several examples of such lattices, their band structures, and their densities of states are shown in Fig. \ref{fig:SplitGraphs}.

\begin{figure}[h]
	\begin{center}
		\includegraphics[width=0.98\textwidth]{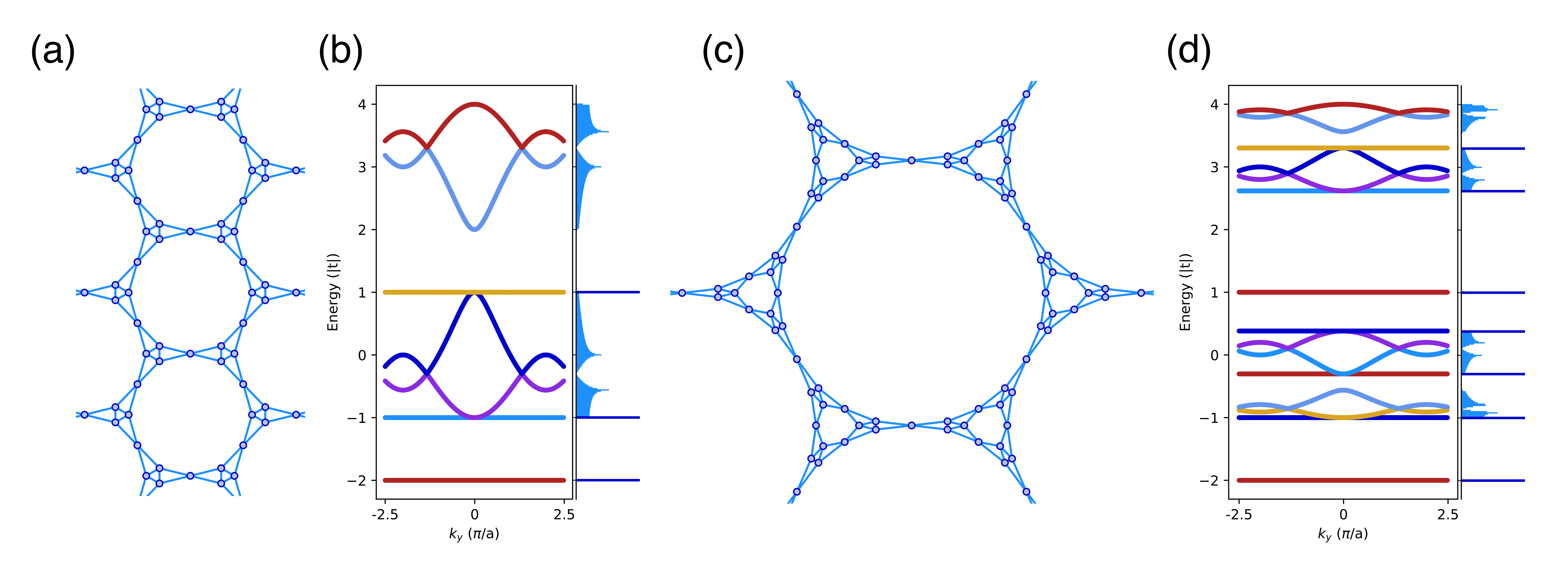}
	\end{center}
	\vspace{-0.6cm}
	\caption{\label{fig:AliciumBands} 
    \textbf{Optimally gapped flat bands.}
    \textbf{a} The effective lattice $\bar{\mathcal{X}}$ which is the line graph of the graph $\mathcal{X} = L(\mathbb{S}(T_6))$ shown in Fig. \ref{fig:SplitGraphs} \textbf{a}\textit{iii}. 
    The band structure and DOS of $\bar{H}_s(\mathcal{X})$ are shown in \textbf{b}.
    They display a maximally large band gap of $1$ above the flat band at $-2$ because this layout is a $3$-regular Hoffman graph and has least eigenvalue $-2$. The flat band at $1$ is ungapped because $T_6$ (graphene) is bipartite.
    \textbf{c} The effective lattice $\bar{\mathfrak{X}}$ which is the line graph of the graph $\mathfrak{X} = L(\mathbb{S}(\mathcal{X}))$ shown in Fig. \ref{fig:SplitGraphs} \textbf{c}\textit{iii}. 
    The band structure and DOS of $\bar{H}_s(\mathfrak{X})$ are shown in \textbf{d}.
    Since $\mathcal{X}$ is a Hoffman graph, $\bar{H}_s(\mathfrak{X})$ displays maximal gaps around the flat bands at both $-2$ and $1$. The flat band at $-2$ is isolated by a gap of $1$, and the flat band at $1$ has band gaps of $\frac{1+\sqrt{5}}{2}$ (above) and $\frac{1-\sqrt{5}}{2}$ (below).
    } 
\end{figure}

Using the results of Ref.\cite{Cvetkovic:1980} (summarized in Appendix \ref{app:subdivisionGraphs}), it can be shown that the eigenenergies of $X$, $\mathbb{S}(X)$ and $ L(\mathbb{S}(X))$ obey the following relations:
\begin{equation}\label{eqn:EsubSX}
E_{\mathbb{S}(X)}= \begin{cases} \pm\sqrt{E_X +3}\\
		 0,
		  \end{cases} 
\end{equation}
and 
\begin{equation}\label{eqn:EsubLSX}
E_{L(\mathbb{S}(X))}= \begin{cases} \frac{1 \pm \sqrt{1 + 4(E_X + 3)}}{2}\\
		0,\\
		 -2.
		  \end{cases} 
\end{equation}
Therefore, as in the case of $\bar{H}_\alpha(X)$, whether or not the flat bands are gapped is determined by the spectrum and band structure of $X$. Since $\mathbb{S}(X)$ is always bipartite, $L(\mathbb{S}(X)$ will never have a gapped flat band at $-2$, but the one at $0$ is gapped if and only if $X$ is non-bipartite, and the magnitude of this gap is determined by how non-bipartite $X$ is. 
The simplest example is to start from graphene, which we will also refer to as $T_6$ in anticipation of Sec. \ref{sec:hyperbolic}. The graphs $\mathbb{S}(T_6)$ and $\mathcal{X} = L(\mathbb{S}(T_6))$ are shown in Fig. \ref{fig:SplitGraphs} \textbf{a}. Since graphene is bipartite, none of these flat bands are gapped. $\bar{H}_s(\mathcal{X})$ has an optimally gapped flat band at $-2$, but no gap at $1$, as shown in Fig. \ref{fig:AliciumBands} \textbf{a} and \textbf{b}. 
Starting from a non-bipartite graph like heptagon-pentagon graphene gives rise to large but non-maximal gaps, as shown in Fig \ref{fig:SplitGraphs}\textbf{b}. 

By induction, the largest gaps are obtained when the initial graph itself is already a Hoffman graph. The simplest such graph is the graph $\mathcal{X}$, and it gives rise to the new layout graph $\mathfrak{X} = L(\mathbb{S}(\mathcal{X}))$ shown in Fig. \ref{fig:SplitGraphs} \textbf{c}\textit{iii}. This graph achieves the locally maximum gap interval in Eqn. \ref{eqn:CharonGaps}, with gaps of $(1+\sqrt{5})/2$ and $(1-\sqrt{5})/2$, above and below the flat band at $0$, respectively. $\bar{H}_s(\mathfrak{X})$ has optimally gapped flat bands at both $-2$ and $1$, as shown in Fig. \ref{fig:AliciumBands} \textbf{c},\textbf{d}.

\section{3-Regular Tessellations of Regular Polygons}\label{sec:hyperbolic}

For $k \geq 3$ denote by $T_k$ the trivalent graphs obtained from tessellations by regular $k$-gons, $P_k$. For $k = 3,4,5$ these are tessellations of the round sphere corresponding to Platonic solids: the tetrahedron, cube, and dodecahedron, respectively. $k = 6$ corresponds to the infinite Euclidean tessellation by hexagons. For $k \geq 7$ the tessellations are infinite hyperbolic. 

\begin{figure}[h]
	\begin{center}
		\includegraphics[width=0.7\textwidth]{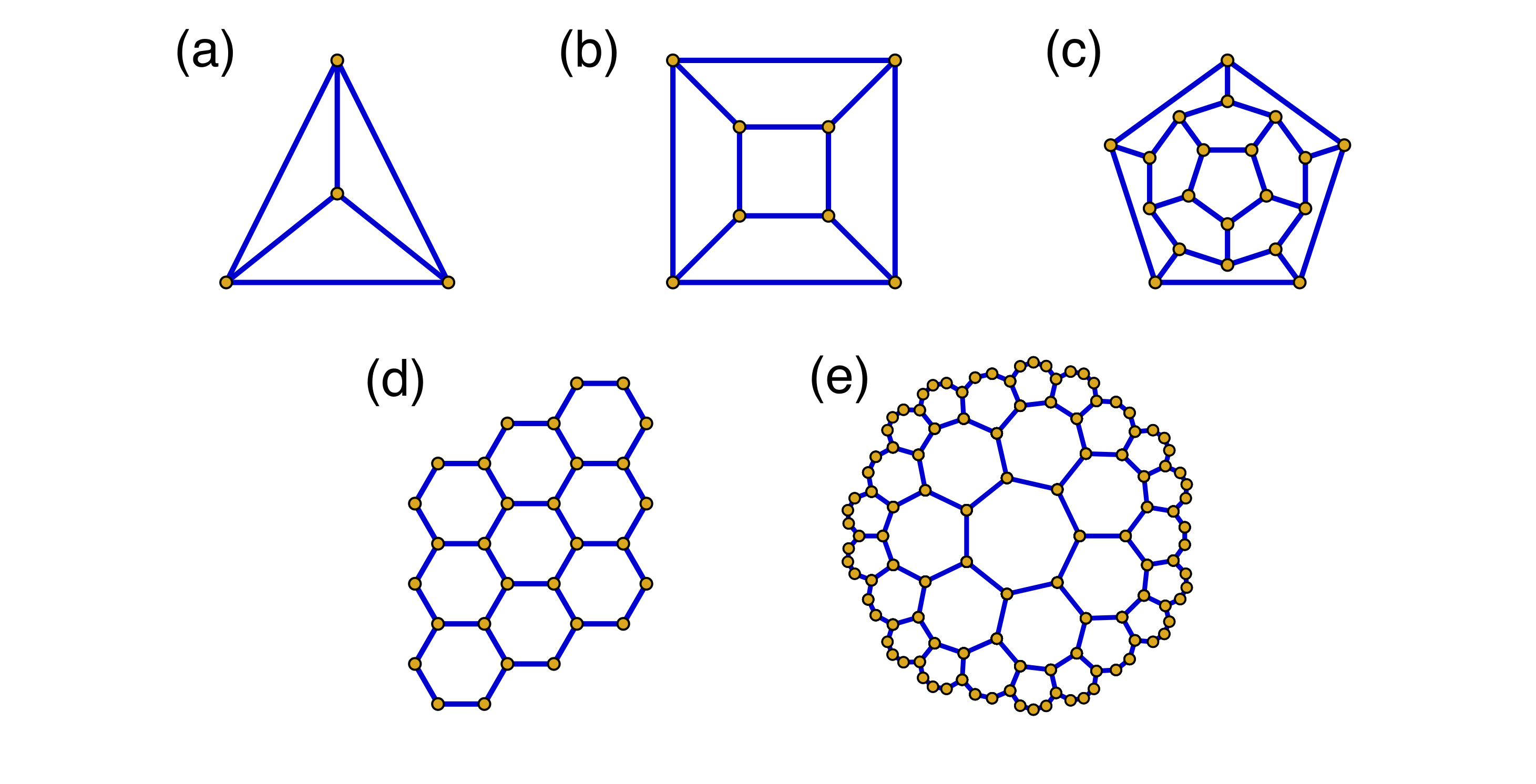}
	\end{center}
	\vspace{-0.6cm}
	\caption{\label{fig:TsubK} 
   \textbf{The trivalent graphs $\bf{T_k}$. \rm{}} 
   \textbf{a} The 3-regular graph with triangular plaquettes, $T_3$. The vertices and edges of this graph can be equated with those of the tetrahedron.
   \textbf{b} The 3-regular graph with square plaquettes, $T_4$, which corresponds to the cube.
   \textbf{c} The dodecahedron graph $T_5$ made using pentagonal plaquettes.
   \textbf{d} A finite section of the hexagonal honeycomb, or graphene lattice $T_6$.
   \textbf{e} A finite section of the heptagonal honeycomb, i.e. the heptagon-graphene lattice, $T_7$. This sample consists of all polygons within a distance $3$ of the origin, and will be denoted by $S_3(7)$.
   } 
\end{figure}

In this section we will consider the spectra of $\bar{H}_s(X)$ and $\bar{H}_a(X)$, where $X = T_k$ for some $k$. To simplify the notation we will denote $\sigma(A_X)$ simply by $\sigma(X)$. As was established in Sec. \ref{sec:gengraphs}, $\sigma(T_k) \subset [-3,3]$ and $\sigma(\bar{H}_\alpha (T_k)) \subset [-2,4]$. Combining the general results in Sec. \ref{sec:gengraphs} with the structure of the $T_k$'s we will examine when $\sigma(\bar{H}_\alpha (T_k))$ is gapped at -2.

Note that $T_k$ is bipartite if and only if $k$ is even.
The graphs $T_k$ are homogeneous, indeed the `Coexter' group $G_k$ generated by reflections in the sides of the base $(2,3,k)$-triangles (see Appendix \ref{app:Cstar}) yields the full symmetry groups of the $T_k$'s. Hence, $G_k$ commutes with $A_{T_k}$ and acts on its eigenspaces.

For the finite $T_k$'s one computes the spectra:
 $$
\begin{array}{ c| c| c}
\sigma(T_3) = \{ (-1)^3, 3\},  &  \sigma(\bar{H}_s(T_3)) = \{ (-2)^{2}, 0^{3}, 4 \},  &  \sigma(\bar{H}_a(T_3))  = \{ (-2)^3, 2^3 \},  \\ 
\sigma(T_4) = \{ -3, (-1)^3,1^3, 3\},  &  \sigma(\bar{H}_s(T_4)) = \{ (-2)^{5}, 0^{3},2^3,  4 \},  &  \sigma(\bar{H}_a(T_4))  = \sigma(\bar{H}_s(T_4)),  \\ 
\end{array}
$$
 $$
\begin{array}{ c}
\sigma(T_5) =  \{ (-\sqrt{5})^3, (-2)^4, 0^4, 1^5, \sqrt{5}^3, 3    \}\\ 
  \sigma(\bar{H}_s(T_5)) = \{ (-2)^{10},(1-\sqrt{5})^{3}, (-1)^{4}, 1^{4}, 2^{5}, (1+\sqrt{5})^{3},4\}    \\
  \sigma(\bar{H}_a(T_5)) =   \{  (-2)^{11}, (1-\sqrt{5})^3, 0^5, 1^4, 3^4, (1+\sqrt{5})^3     \}\\
\end{array}
$$
The multiplicities of the eigenvalues in these cases can be explained by the dimensions of the irreducible representations of the $G_k$'s.

The group $G_6$ has an Abelian subgroup of finite index (which is isomorphic to $\mathbb{Z} \times \mathbb{Z}$) and hence one can use Bloch waves to compute $\sigma (T_6)$, and with it $\sigma(\bar{H}_\alpha(T_6))$, which is the spectrum of the kagome lattice, and is well-known. 
\begin{equation}
  \sigma(\bar{H}_s(T_6)) = {(-2)^{\infty}} \cup \left( -2, 4 \right].
\end{equation}
Here, -2 has an infinite dimensional space of $\ell^2(\mathcal{E}(T_6))$ eigenvectors and the rest of the spectrum on (-2,4) is absolutely continuous.

\subsection{$k$ $\geq$ 7: The Hyperbolic Cases}\label{subsec:PeterHyperbolic}

For $k \geq 7$ the Coxeter groups $G_k$ of symmetries of $T_k$ are infinite and have finite index surface subgroups, see Appendix \ref{app:Cstar}. In particular they are not amenable and are known to satisfy the Kadison property.\cite{Puschnigg:2002ju} Hence, according to the general qualitative results of Sec. \ref{sec:gengraphs} we have that
$$ |\lambda_{min} (T_k)| \leq \lambda_{max}(T_k) < 3,$$
and
$$ \sigma(T_k) \mbox{ consists of finitely many bands}.$$

Further qualitative features of $\sigma(T_k)$ can be inferred from the spectral measure (i.e. the density of states) $\mu_k$ of $A_{T_k}$. As in Kresten\cite{Kesten:1959us} this measure is supported on $[-3,3]$ and is determined by its moments. For $l > 0$
\begin{equation}\label{eqn:moments}
\int_{-3}^{3}{\lambda^l \, d\mu_k (\lambda) } = \varrho_k(l),
\end{equation}
where $\varrho_k(l)$ is the number of walks of length $l$ on $T_k$ starting and ending at a vertex $v$. Note that $T_\infty$ is the $3$-regular tree and that locally $T_k$ converges to $T_\infty$ as $k\rightarrow \infty$. Clearly,
\begin{equation}\label{eqn:rho1}
\varrho_k (l) \geq \varrho_\infty (l) \mbox{ for any } k \mbox{ and } l,
\end{equation}
and 
\begin{equation}\label{eqn:rho2}
\varrho_k (l) = \varrho_\infty (l) \mbox{ as } k \rightarrow \infty  \mbox{ for fixed } l.
\end{equation}
From Eqn. \ref{eqn:rho1} it follows that 
\begin{equation}\label{eqn:lmabdamaxlim}
\lambda_{max}(T_k) \geq \lambda_{max} (T_\infty) = 2 \sqrt{2}.
\end{equation}
In fact, from Kesten\cite{Kesten:1959us} it follows that 
\begin{equation}\label{eqn:lmaxlb}
\lambda_{max} (T_k) > 2 \sqrt{2}, \mbox{ for } 7 \leq k\leq \infty.
\end{equation}
On the other hand, Eqn. \ref{eqn:rho2} implies that $\mu_k\rightarrow \mu_\infty$ as $k \rightarrow \infty$ in the sense that 
\begin{equation}\label{eqn:muconvergence}
\int_{-3}^{3}{f(\lambda) \, d\mu_k (\lambda) } \rightarrow \int_{-3}^{3}{f(\lambda) \, d\mu_\infty (\lambda) },
\end{equation}
for any fixed continuous $f$.
In particular, the support of $\mu_k$, that is $\sigma(T_k)$, converges to $\sigma(T_\infty)$ as $k\rightarrow\infty$. Hence, 
$$ \lim_{k \rightarrow \infty} \lambda_{max} (T_k) = 2 \sqrt{2}, $$
and
\begin{equation}\label{eqn:lminlmaxTree}
\lim_{k \rightarrow \infty} \lambda_{min} (T_k) = -2 \sqrt{2}, 
\end{equation}

From these and Eqn. \ref{eqn:shiraispectrum} we deduce that $-2$ is an isolated flat band at the bottom of both $\sigma(\bar{H}_a(T_k))$ and  $\sigma(\bar{H}_s(T_k))$. Furthermore, both $\bar{H}_a(T_k)$ and $\bar{H}_s(T_k)$ display compactly supported states at $-2$, with slightly different character. A detailed construction of these states is given in Appendix \ref{app:compactSupport}. As far as the spectrum of the induced balls $S_r(k) \subset T_k$ for radius $r$ about a given $v \in V(T_k)$, we have from Eqn. \ref{eqn:lminnonzero} that $\bar{H}_s(S_r(k))$ is gapped at $-2$ if and only if $T_k$ is non-bipartite, i.e. $k$ is odd, while for $\bar{H}_a(S_r(k))$ $-2$ is not gapped (as $r \rightarrow \infty$).

To obtain more quantitative information about these spectra associated with $T_k$ we first examine the analytic arguments above and supplement them with a numerical study leading to a reasonably complete picture. As far as the number of bands for $\sigma(T_k)$, we show in Appendix \ref{app:Cstar} how the solution of the Kaplansky-Kadison conjecture concerning idempotents in the reduced $C^*$-algebra of a torsion-free hyperbolic group may be used to show that the number of bands in $\sigma(T_k)$ is at most $m_k/3$, where $m_k$ is the (essentially linear) arithmetic function of $k$ given in Table \ref{table:numbands} in Appendix \ref{app:Cstar}. In particular for $k = 7,8,9,10,30$ the bounds are $28,16,12,10,10$, respectively.

One can estimate $\lambda_{max}(T_k)$ from above by estimating the Cheeger constant for $T_k$ and applying a combinatorial version of Cheeger's inequality. This is done in Ref.\cite{Higuchi:2003vw} where they show that 
\begin{equation}\label{eqn:sqrtkbound}
\lambda_{max}(T_k) \leq 2 \sqrt{  \frac{2k-3}{k-2  }}.
\end{equation}
On the other hand Pashke\cite{Paschke:1992ti} observes that $T_k$ is covered by the Cayley graph $X_k$ of $J_k =\left< Q \right> *\left< R\right>$ with $Q^2 = 1, \ R^k = 1$ w.r.t the generators $\{ Q, R, R^{-1} \}$. Hence,
\begin{equation}\label{eqn:anotherbound}
\lambda_{max}(T_k) \geq \lambda_{max}(X_k).
\end{equation}
The spectrum of $X_k$ was computed in Ref.\cite{Paschke:1992ti} and 
\begin{eqnarray}\label{eqn:numericalbound}
2 \sqrt{2} < \lambda_{max}(X_k) & = &\min_{s \geq 0} {   \left\{   2 \cosh s  + \mathcal{Q}\left(  \frac{\cosh(ks) + 1}{\sinh(s)\sinh(ks)}  \right)   \right\}}    \nonumber\\
 \mbox{where } \mathcal{Q}(x) &=& \frac{\sqrt{x^2+1} - x}{x} .
 \end{eqnarray}
Equations \ref{eqn:sqrtkbound} and \ref{eqn:anotherbound} give tight bounds for $\lambda_{max}(T_k)$; for example
\begin{eqnarray}\label{eqn:numericallims}
2.862 \leq \ &\  \lambda_{max}(T_7) \ & \ \leq 2.966\ldots   \nonumber\\
2.852 \leq \ &\  \lambda_{max}(T_8) \ &\  \leq 2.943\ldots
 \end{eqnarray}
 
 To end our analytic estimates on these spectra, we turn to an explicit lower bound on the gap at $-2$ for $\sigma( \bar{H}_s (S)$, where $S_r = S_r(k)$ is the induced layout in $T_k$ which is a ball of radius $r$ (and $r$ arbitrarily large). From Eqn. \ref{eqn:simgaH1} this is dictated by $\sigma(A_{S_r} + D_{S_r})$. We have 
\begin{align}\label{equ:lambda}
\lambda_{min} (A_{S_r} + D_{S_r}) &= \inf_{f:V(S_r) \rightarrow \mathbb{R}} \frac{\sum\limits_v \large( \sum\limits_{\omega\sim v} f(\omega) + d (v) f(v) \large) f(v)}{\sum\limits_v f^2 (v)} \\ \nonumber
& = \inf_{f:V(S_r)\rightarrow \mathbb{R}} \frac{\frac{1}{2}\sum\limits_{e \in \mathcal{E}(S_r)} \left[ f(e^{+}) + f(e^{-}) \right]^2}{\sum\limits_{v\in V(S_r)} f^2(v)}
\end{align}
where $e^{+}$, $e^{-}$ are the vertices that $e$ joins.
So $\lambda_{min} (A_{S_r} + D_{S_r})$ is a measure of how close $S_r$ is to being bipartite.

For $k \geq 7$ and even, $T_k$ is bipartite as is $S_r(k)$. Therefore it follows from Sec. \ref{sec:gengraphs} that the gap at $-2$ must vanish as $r\rightarrow \infty$.
For $k$-odd there is a gap isolating the eigenvalue -2 in $\sigma(\bar{H}_s(S_r))$, which is both striking and practically useful.\cite{Kollar:2018vc,Leykam:2018vd}
Bounds on this feature can be seen by estimating the quotient in Eqn. \ref{equ:lambda}
directly. For a $k$-sided polygon (or cycle graph) $e_k$; and $f:V(c_k)\rightarrow \mathbb{R}$
\begin{equation} \label{equ:lambda2}
  \frac{1}{2} \sum_{e \in \mathcal{E}(c_k)} (f(e^+) + f(e^-))^2 = \sum_v \left( \sum_{\omega \sim v} f(\omega) + 2 f(v)\right) f(v)
\end{equation}
The spectrum $\sigma (c_k)$ of $c_k$ is $\{ 2 \cos ( \frac{2 \pi j}{k} ), j = 1,2,...,k \}$,
hence if $k = 2 \nu + 1$, then
\begin{equation}\label{equ:fplusfminus}
  \frac{1}{2} \sum_{e \in \mathcal{E}(c_k)} (f(e^+) + f(e^-))^2 \geq \left( 2 + 2 \cos (\frac{2 \pi \nu}{2\nu + 1} ) \right) \sum_{v \in V(c_k)} f^2(v)
\end{equation}

Applying Eqn. \ref{equ:fplusfminus} to each of the polygons $P$ in $S_r(T_k)$, each of whose boundary $\partial P$ is a $k$-cycle yields:
\begin{equation}
  \sum_{e \in \mathcal{E}(S_r)} \frac{1}{2} \left[f(e^+) + f(e^{-}) \right]^2 = \sum_{P \subset S_{r}} \sum_{e\in \partial P} \frac{\mathsf{w} (e)}{2} \left[ f(e^+) + f(e^-)\right]^2,
\end{equation}
where $\mathsf{w}(e) = 1$ or $1/2$ depending on whether $e$ is an edge which bounds one or two of the $P$'s in $S_r$.

Hence
\begin{align}
\sum_{e\in S_r} \frac{1}{2} \left[ f(e^+) + f(e^-)\right]^2 &\geq \frac{1}{2} \sum_{P \subset S_r} \sum_{e \in \partial P} \frac{(f(e^+) + f(e^-))^2}{2} \nonumber \\
& \geq \frac{1}{2} \sum_{P \subset S_{r}} \left(2 + 2 \cos\left(\frac{2 \pi \nu}{2 \nu + 1} \right) \right) \sum_{v \in \partial P} f^2 (v)  \nonumber \\
& \geq  \left[ 1 + \cos \left(\frac{2 \pi \nu}{2\nu +1} \right) \right] \sum_{v \in V(S_r)} {f^2 (v)}
\end{align}
It follows that 
\begin{equation}\label{eqn:ksquaredbound}
  \lambda_{min} (A_{S_r} + D_{S_r}) \geq 1 + \cos\left(\frac{2 \pi \nu}{2 \nu + 1} \right)\geq \frac{1}{k^2}
\end{equation}
Hence -2 is an eigenvalue of $\bar{H}_s(S_r)$ with multiplicity $|\mathcal{E}(S_r)| - |V(S_r)|$ and $\sigma (\bar{H}_s (S_r))$ has no points in $(-2, -2 + \frac{1}{k^2})$.

\subsection{Numerics}

\begin{figure}[h]
	\begin{center}
		\includegraphics[width=0.9\textwidth]{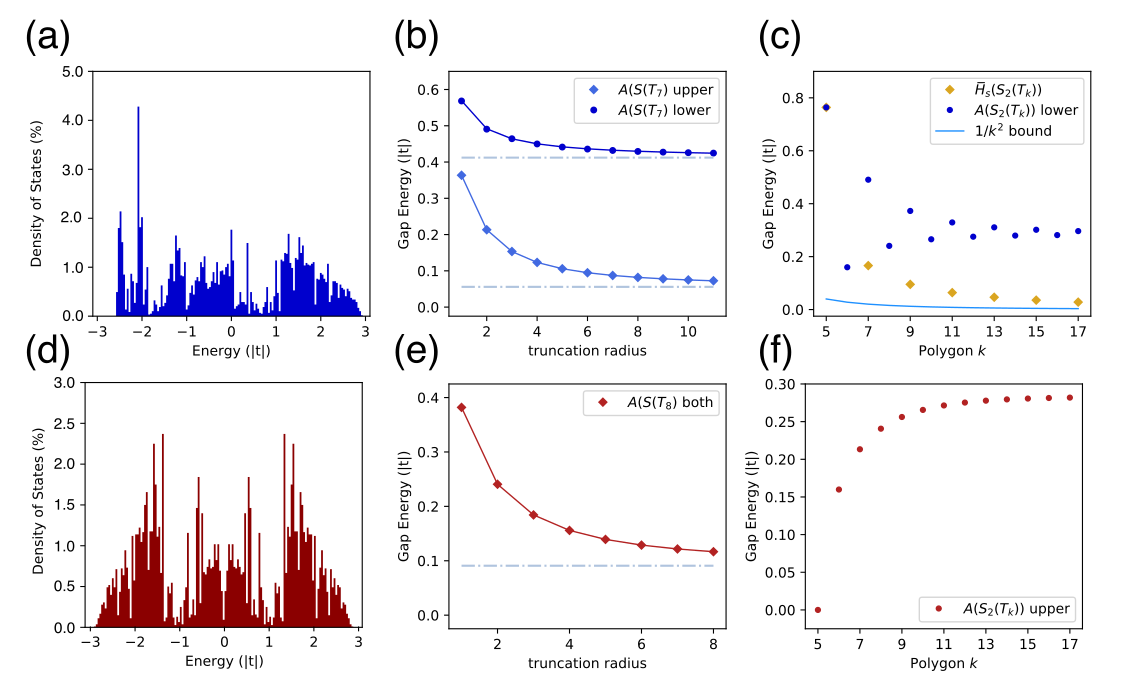}
	\end{center}
	\vspace{-0.6cm}
	\caption{\label{fig:layoutNumerics} 
    \textbf{Truncated hyperbolic layouts.}
     \textbf{a} Numerical DOS for the adjacency matrix $A_{S_6(T_7)}$, where the vertical axis indicates the percentage of the total states found in each energy bin (bin width $= 0.04 |t|$). The maximum eigenvalue is below $3$, reflecting the $\ell^2$ gap in $\sigma(A_{T_7})$. Because $T_7$ is non-bipartite, the gap near $-3$ is considerably larger. 
     As the system size increases, these values asymptote to those for the infinite regular graph $T_7$ (See Eqn. \ref{eqn:finiteextrema}).
     The magnitude of both of the upper and lower gaps versus radius of the induced subgraph is shown in \textbf{b}, and by fitting these curves, we extrapolate to asymptotic values of $\sim 0.06$ (upper) and  $\sim 0.41$ (lower). Symbols are the numerical data, and solid lines the fits. The fitted asymptotic values are indicated by horizontal dot-dashed lines.
     \textbf{c} Plots of the lower gap versus $k$, the number of sides in the layout polygon. The gaps for $A(S_2(T_k))$ are denoted as blue circles and shown for all values of $k$ between $5$ and $17$. They oscillate depending on whether $k$ is even or odd (and thus whether $T_k$ is bipartite or not) and converge to a value of $\sim 0.3$ corresponding to a truncated $3$-regular tree. Note that since these graphs are quite small, these gaps are quite far from their asymptotic infinite-system values. Gap energies for $\bar{H}_s(S_2(T_k))$, denoted by gold diamonds, are plotted only for odd $k$ since they vanish otherwise. In contrast to the infinite-layout gaps which converge to a finite value as $k$ increases, these gaps decrease with increasing $k$. However, they are always markedly larger than the $1/k^2$ bound in Eqn. \ref{eqn:ksquaredbound}. Note that $T_5$ is a spherical tiling and that $S_2(T_5)$ is already the complete tiling. Therefore the gaps for $A$ and $\bar{H}_s$ are identical, and the two markers coincide. In all other cases, the gap for $\bar{H}_s(S(T_k))$ is systematically smaller than that for $A(S(T_k))$ and $A(T_k)$.
      \textbf{d}-\textbf{e} The corresponding plots for $A_{S_5(T_8)}$. Since this layout is bipartite, the spectra and the gaps of $A_{T_8}$ are symmetric about zero, and both gaps are relatively small with fitted asymptotic values of $\sim 0.09$. 
      \textbf{f} Plots of the upper gap of $A(S_2(T_k))$ versus $k$. Unlike the lower gap, it is not strongly effected by whether the layout is bipartite of not, so it converges monotonically to the value for a truncated $3$-regular tree.
     } 
\end{figure}

To gain further insight into the spectra, we conducted numerical diagonalization studies of $A_{S_r(T_k)}$, $\bar{H}_s(S_r(T_k))$, and $\bar{H}_a(S_r(T_k))$ for a series of polygons $k$ and system sizes $r$. From Eqn. \ref{eqn:finiteextrema} and Sec. \ref{sec:gengraphs}, it is clear that $\sigma(A_{S_r(T_k)})$ is much more readily extrapolated to $r= \infty$ than the effective Hamiltonians $\bar{H}_\alpha$. We therefore start by examining the layout for both bipartite and non-bipartite cases, in particular $\sigma(A_{S_r(T_7)})$ and $\sigma(A_{S_r(T_8)})$. Using numerical diagonalization we compute the density of states up to $r=6$ for $T_7$ and $r = 5$ for $T_8$, shown in Fig. \ref{fig:layoutNumerics} \textbf{a} and \textbf{d}, respectively. $A_{S_r(T_7)}$ shows a large gap near $-3$ due to the non-bipartite nature of this tiling, and a significantly smaller gap near $3$. Since $T_8$ is bipartite it does not display the larger, asymmetric frustration gap. In this case, the DOS is completely symmetric and shows only relatively small gaps at $\pm 3$. For larger values of $r$ the system sizes begin to exceed $\sim 10,000$ vertices and we are unable to perform full exact diagonalization.

However, the adjacency matrix is highly sparse, so we are able to use sparse matrix methods to determine the largest and smallest eigenvalues for much larger system sizes. This allows us to determine $\lambda_{max}(r)$ and $\lambda_{min}(r)$ up to system sizes of $\sim 1,000,000$. The resulting values for $T_7$ are plotted in Fig. \ref{fig:layoutNumerics} \textbf{b} up to $r = 11$. $T_8$ grows more rapidly with $r$, so we were able to compute only up to $r=8$, as shown in Fig. \ref{fig:layoutNumerics} \textbf{e}. As a result of the variational characterization of the $\ell^2$ spectrum given in Sec. \ref{sec:gengraphs}, the gaps $3- \lambda_{max}(r)$ and $3+ \lambda_{min}(r)$ give rigorous upper bounds on the $\ell^2$ gaps at $\pm 3$ for $r = \infty$. In all cases the computed gaps decrease monotonically with $r$ and appear to asymptote to a non-zero value. In order to determine the $\ell^2$ gap for the infinite lattice, we fit the computed gaps to an empirical Lorentzian-like fit function
\begin{equation}\label{eqn:lorishff}
gap(r) = A \pm \frac{1}{w + (r/s)^p},
\end{equation}
where all four parameters ($A$, $w$, $s$, and $p$) are allowed to vary. The resulting fits are very good and shown as solid lines in Fig. \ref{fig:layoutNumerics} \textbf{b} and \textbf{e}. The fitted asymptotic values $3\pm A$ are $\sim 0.41$ for the lower gap of $T_7$, $\sim ~0.06$ for the upper gap of $T_7$, and $\sim 0.09$ for both gaps of $T_8$. The corresponding fit parameters are shown in Table \ref{table:fitparams}.

\begin{table}[ht]
\caption[A]{Fit Parameters}
\centering
\begin{tabular}{ c| c| c c c c}
k & end & A & w & s & p \\ 
  \hline
7 & lower & -2.59 & 1.30 & 0.24 & 1.26 \\  
7 & upper & 2.94 & 1.48 & 0.68 & 1.46 \\ 
8 & lower & -2.91 & 1.52 & 0.64 & 1.43 \\ 
8 & upper & 2.91 & 1.52 & 0.64 & 1.43 \\ 
$\infty$ & lower & -2.83 & 0.52 & 1.81 & 1.45 \\ 
$\infty$ & upper & 2.83 & 0.52 & 1.81 & 1.45 \\ 
\end{tabular}
\label{table:fitparams}
\end{table}

In order to validate our empirical fit function and asymptotic values, we benchmark against the $3$-regular tree $T_\infty$. We find that the fits are excellent here as well, and that the numerical asymptotic value agrees with the theoretical value of $2 \sqrt{2}$ to better than $0.01$. This level of agreement constitutes an estimate of the error in this numerical method, and we therefore present all numerical asymptotic values rounded to this precision.

For $k$ large and finite, the system size grows very rapidly with $r$, and we are unable to compute for large $r$. We therefore study the $k$ dependence of the gaps for $r = 2$ (which allows us to also include the spherical tiling $k=5$). The gaps for $A_{S_2(T_k)}$ are plotted in Fig. \ref{fig:layoutNumerics} \textbf{c} and \textbf{d} for $k = 5, \ldots 17$. The upper gap is not influenced by non-bipartite frustration effects, so it increases monotonically with $k$ and tends to an asymptotic value near $\sim 0.28$. This value is significantly larger than the infinite size limit of $\sim 0.17$ because these are relatively small induced subgraphs. The lower gap however depends strongly on how non-bipartite the graph is. This gap therefore oscillates with $k$ and is always larger than the asymptotic limit if $k$ is odd.

Additionally, we have computed the lower gap of $\bar{H}_s(S_2(T_k))$ at $-2$ for $k = 5,7, \ldots 15,17$. The resulting values are plotted in Fig. \ref{fig:layoutNumerics} \textbf{c} alongside the corresponding results for $A_{S_2(T_k)}$. The tiling $T_5$ is a complete spherical tiling and, in fact, $S_2(T_5) = T_5$. Therefore, for $k=5$, the gaps in $A_{S_2(T_5)}$ and $\bar{H}_s(S_2(T_5))$ are identical, as expected for a regular graph. For $k > 5$, however, the observed gap for $\bar{H}_s(S_2(T_k))$ is much smaller than that for $A_{S_2(T_k)}$, and decreases monotonically with increasing $k$. The $1/k^2$ bound from Eqn. \ref{eqn:ksquaredbound} is also plotted in Fig. \ref{fig:layoutNumerics} \textbf{c}, and is visibly not sharp.

\begin{figure}[h]
	\begin{center}
		\includegraphics[width=0.9\textwidth]{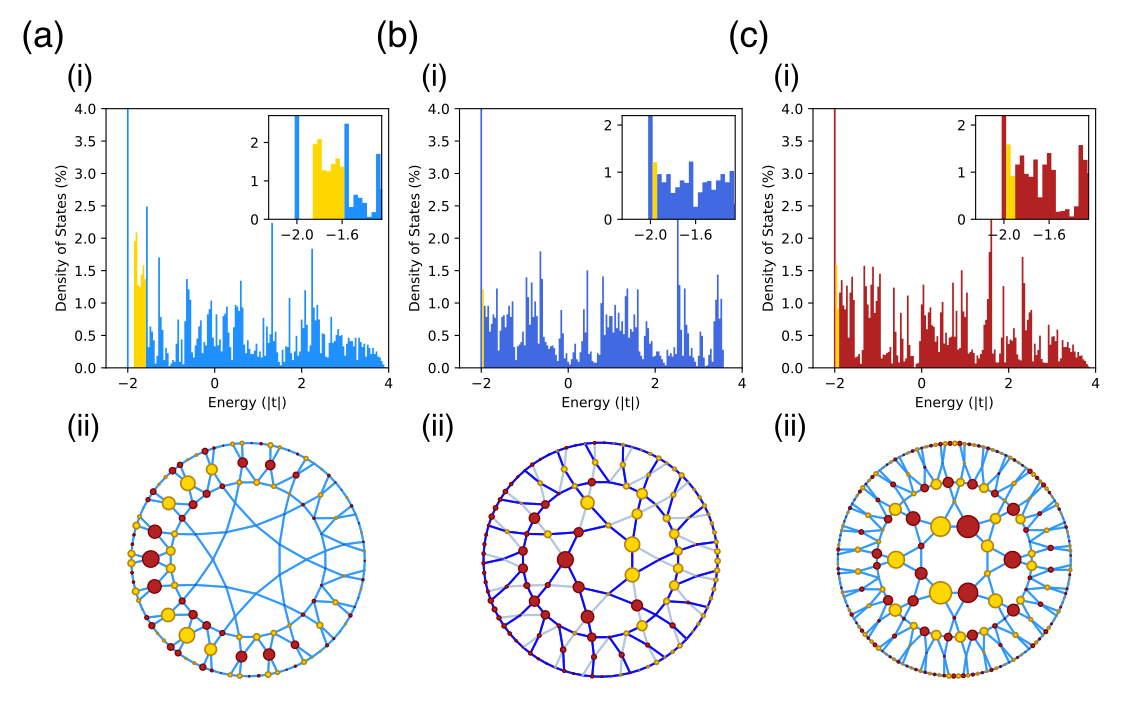}
	\end{center}
	\vspace{-0.6cm}
	\caption{\label{fig:effectiveNumerics} 
    \textbf{Truncated hyperbolic effective lattices.}
     \textbf{a}\textit{i} Numerical DOS for $\bar{H}_s(S_6(T_7))$, where the vertical axis indicates the percentage of the total states found in each energy bin (bin with $= 0.04 |t|$). The DOS of the flat band is off the vertical scale. The maximum eigenvalue is below $4$, and follows the upper gap in $\sigma(A_{S_6(T_7)})$. 
     Because $T_7$ is non-bipartite, there is a finite gap of $\sim 0.17$ at $-2$. However, this is considerably smaller than the $\sim 0.41$ expected for the infinite heptagon-kagome lattice. All the states in this window are therefore edge-induced, and the contribution to the total DOS due to them is indicated in yellow, and shown in more detail in the inset. The surface to volume ratio in hyperbolic space guarantees that these states are macroscopic in number and do not vanish from the DOS of $\bar{H}_s(S_r(T_7))$, no matter the system size.
     The lowest energy state in this window is plotted in \textbf{a}\textit{ii}. It is clearly a whispering-gallery-like edge mode and is typical of the low-lying ``mid-gap'' states observed for non-bipartite layouts. Bulk-like states occur only significantly higher up.
     \textbf{b}\textit{i}-\textit{ii} Corresponding plots for $\bar{H}_a(S_6(T_7))$. The gap near $4$ derives from the lower gap of $A(T_7)$, and no edge-induced states appear here. However, as expected from Sec. \ref{sec:gengraphs}, the gap at $-2$ closes completely, indicated in yellow. In this case, the first mid-gap state, shown in \textbf{b}\textit{ii}, displays bulk-like character.
     \textbf{c}\textit{i}-\textit{ii} DOS and first mid-gap state for $\bar{H}_a(S_5(T_8)) = \bar{H}_s(S_5(T_8))$. This layout is closer to the $3$-regular tree and bipartite, so the expected gap at $-2$ is larger than for $\bar{H}_a(T_7)$, but it too is closed by a macroscopic number of edge-induced states, indicated in yellow. As in the case of $\bar{H}_a(T_7)$, the first mid-gap state, shown in \textbf{c}\textit{ii}, is bulk-like. Such bulk-like mid-gap states are typical of what is observed for $\bar{H}_a$.
    } 
\end{figure}

Finally, we use exact diagonalization to compute the density of states for $\bar{H}_s(S_6(T_7))$, $\bar{H}_a(S_6(T_7))$, and $\bar{H}_s(S_5(T_8))$, shown in Fig. \ref{fig:effectiveNumerics} \textbf{a}\textit{i}-\textbf{c}\textit{i}, respectively. As expected from Sec. \ref{sec:gengraphs}, only $\bar{H}_s(S(T_7))$ displays a gapped flat band. Given Eqn. \ref{eqn:shiraispectrum} and the numerical estimates of the limits of the $\ell^2$ spectra of $A_{T_7}$ and $A_{T_8}$, we can identify portions of $\sigma(\bar{H}_\alpha(S(T_k)))$ which lie outside the $\ell^2$ spectrum of $\bar{H}_\alpha(T_k)$. In all three cases we find that the first states above the flat band belong to such regions, indicated in yellow and highlighted in the insets of Fig. \ref{fig:layoutNumerics} \textbf{a}\textit{i}-\textbf{c}\textit{i}. Examples of the first such states for $r=2$ are shown in Fig. \ref{fig:effectiveNumerics} \textbf{a}\textit{ii}-\textbf{c}\textit{ii}. For $k=7$, and for non-bipartite full-wave models in general, the lowest-lying such states are whispering-gallery-like modes which are supported almost entirely on the outermost shell. For half-wave and bipartite models, this first states has more bulk-like character. The full-wave models also display such misplaced bulk-like modes, but they occur only at the very top of this extra interval in the spectrum. Within numerical resolution, there is no corresponding interval at high energy where the finite-size model has eigenstates but the infinite does not. 

While there are gaps in the middle of the numerical spectra for any finite $r$, they all close progressively with system size. Only the gaps at the ends of the spectrum and near the flat band are stable, suggesting that in the limit $r\rightarrow \infty$ $\sigma(A_{T_k})$ consists of only one interval, and that $\sigma(\bar{H}_\alpha(T_k))$ consists of at most two intervals: the flat band and $\sigma(A_{T_k}+ 1)$. This result is known for $k = \infty$ and $k = 6$, but it remains an open conjecture for $k\geq 7$.

\section{Conclusion}

In conclusion, we have shown that circuit QED lattice devices naturally produce effective lattices whose sites and connectivity are those of the line graph of their hardware layout $X$. These devices have two sets of resonant modes: symmetric full-wave modes, and antisymmetric half-wave modes. We derived the effective Hamiltonian for the full-wave modes which is an s-wave tight-binding model $\bar{H}_s(X): \ell^2(\mathcal{E}(X)) \rightarrow \ell^2(\mathcal{E}(X))$ and showed that it is equivalent to the graph Laplacian on the line graph of $X$ if all the hopping matrix elements and on-site energies are equal. We also derived the effective p-wave tight-binding model for the half-wave modes $\bar{H}_a(X)$ and showed that it too is a closely related operator on $\ell^2(\mathcal{E}(X))$.

We showed that for the case of constant negative hopping prevalent in CPW lattices, $ \sigma(\bar{H}_\alpha (X)) = \{-2\}^\infty \cup\sigma(D_X \pm A_X)$, where $D_X$ is a diagonal matrix of the degrees of the vertices of $X$, $A_X$ is the adjacency operator on $X$, the $+$ sign is for $\bar{H}_s$, and the $-$ sign for $\bar{H}_a$. 
In particular, this demonstrates that the effective tight binding models $\bar{H}_s(X)$ and $\bar{H}_a(X)$ exhibit flat bands at $-2$ for any layout $X$, finite or infinite, regular or irregular, homogeneous or inhomogeneous.
Using this relation, we have examined the spectra of $\bar{H}_\alpha(x)$ for a variety of $X$'s, including both Euclidean and non-Euclidean lattices, where many aspects of traditional band structure calculations fail
We have derived criteria for the existence and maximization of spectral gaps, concentrating in particular on a potential gap at $-2$. Adding non-linearity and effective photon-photon interactions to such isolated flat bands is an ideal starting point for quantum simulation of strongly correlated many body physics with photons and will be the focus of future experimental and theoretical work.\cite{Bergman:2008es,Kollar:2018vc}

For regular layouts, $ \sigma(\bar{H}_\alpha (X))$ can be completely understood by examining $\sigma(A_X)$. Because of the minus sign, a spectral gap at $-2$ for $\bar{H}_a(X)$ arises only due to an expander gap in the spectrum of $A_X$. Therefore, this case can be understood completely by calling upon the existing graph-theory literature, and no finite and planar layout graph can give rise to a macroscopic gap at $-2$ for $\bar{H}_a(X)$. $\bar{H}_s(X)$ is a less conventional operator largely not covered by the existing literature, and we have shown that for finite layouts it has a gap at $-2$ if and only if $X$ is non-bipartite.

For infinite, regular, homogeneous layouts, the existence of such a gap at $-2$ for $\bar{H}_\alpha$ can be understood from the structure and amenability of the group of isomorphisms of $X$. For amenable groups, such as Euclidean crystallographic groups, the $\ell^2$ spectrum of $A_X$ is gapped away from $-d$ if and only if $X$ is non-bipartite and is never gapped away from $d$. Therefore, $\bar{H}_a(X)$ never has a gap at $-2$ and $\bar{H}_s(X)$ has a gap at $-2$ if and only if $X$ is non-bipartite. Additionally, we have shown how this same result can be derived from the real-space topology technique of Ref.\cite{Bergman:2008es}. For non-amenable groups such as the hyperbolic crystallographic groups, the $\ell^2$ spectrum of $A_X$ is always gapped away from $\pm d$. Induced subgraphs of these non-amenable models are finite and of bounded degree and only exhibit gaps at $-2$ if $\alpha = s$ and $X$ is non-bipartite. In all other cases a macroscopic number of finite-size-induced states fill in the $\ell^2$ gap.

We also examined the largest gap intervals that such layouts and effective lattices can have, and showed that for layouts of degree less than or equal to three, the largest possible gap interval above a flat band at $-2$ is $(-2,-1)$. The layouts that achieve this maximum are $3$-regular Hoffman graphs, and we present an infinite Cayley graph which is a universal cover of all such examples. In addition to a flat band at $-2$ which gives rise to this maximal gap in $\bar{H}_\alpha$, these Hoffman layouts also display a flat band at $0$ which can also be gapped out without compromising its flatness. In this case, the maximum gap interval is $(\frac{1 - \sqrt{5}}{2}, 0) \cup (\frac{1 + \sqrt{5}}{2}, 0)$. It is obtained for special Hoffman layout graphs $X$ such that $X = L(\mathbb{S}(Y))$, where $Y$ is itself a $3$-regular Hoffman graph. Additionally, we present Euclidean examples which achieve both of these maximally gapped flat bands.

For the hyperbolic tilings $T_k$, $k \geq 7$, we derived analytic bounds on the limits of the $\ell^2$ spectrum of $A_{T_k}$ as well as $\bar{H}_s(T_k)$. 
Additionally, we conducted numerical exact diagonalization studies of both $A_{T_k}$ and $\bar{H}_\alpha(T_k)$. Using sparse matrix diagonalization techniques we computed $\lambda_{max}$ and $\lambda_{min}$ for truncated hyperbolic layout graphs $S_r(T_k)$ up to systems sizes $\sim 1,000,000$, and extrapolated to estimates of the $\ell^2$ gap for $r = \infty$. In particular, we obtained a numerical estimate of $\sim 0.41$ for the lower gap of $T_7$ and $\bar{H}_s(T_7)$, even though $\bar{H}_s(S_r(T_7))$ only displays a gap of $\sim 0.17$. Using the numerical estimates of the limits of the $\ell^2$ spectra for $r = \infty$, we identify regions near $-2$ in the DOS of $\bar{H}_\alpha(S_r(T_7))$ which are outside the spectrum for $r = \infty$ and are filled with a macroscopic number of finite-size-induced states. For full-wave non-bipartite models this region does not extend all the way to $-2$, leaving a finite gap, and consists primarily of whispering-gallery-like edge modes. Otherwise, this region fills the $\ell^2$ gap that would otherwise be present in the infinite case and consists primarily of bulk-like modes.

\appendix

\section{Construction of $\bar{H}_a(X)$ from $N$ and $N^t$}\label{app:HWshirai}

The incidence matrices $M$, $M^t$, $N$, and $N^t$ can be understood as instructions for how to navigate a graph $X$. Consider first the simpler case of half-wave modes, and begin with a state $\vec{j}$ on the graph $X$ which is one on the $j^{th}$ vertex and zero elsewhere. The action of $M$ on this state takes it to all the edges which are incident on $v_j$. Acting next with $M^t$ then goes from these incident edges back to the vertex set and lands on all vertices that touch these chosen edges, including the original source vertex $v_j$. Therefore, $M^t M$ describes how to get from a vertex $v_j$ to neighboring vertices by taking a walk though the edge set; however, it includes the option of going back to the original site. Such an option is not allowed under the action of $A_X$, but the number of times that this mistaken path can occur is precisely the degree of the vertex $v_j$. It therefore follows that $A_X = M^t M - D_X $.

Instructions for navigating the edge set can derived in an analogous manner. Starting from a state $\vec{e}_j$ on $\mathcal{E}(X)$ which is one on the $j^{th}$ edge and zero elsewhere, the action of $M^t$ takes this state to all the vertices which are incident on the jth edge. Acting next with $M$ then goes from these vertices to all the edges that they touch. Analogously to above, $MM^t$ describes how to navigate from an edge to neighboring edges by taking a walk though the vertices, but once again, it allows for the option of returning to the original location. Since each edge has two ends, this incorrect path will occur exactly twice, and we find that $A_{L(X)} = \bar{H}_s(X) = MM^t - 2 I$. 

In the oriented half-wave case, we must navigate with $N$ and $N^t$ instead. The combinatorics of the allowed paths is identical to the full-wave case with $M$ and $M^t$. The only outstanding detail is the additional minus signs. Consider $N^tN$. It allows two types of paths through the vertex set: correct paths to neighboring vertices and extra paths back to the source vertex. A path from the source vertex back to itself will always consist of hopping to one end of an edge and back off of that same edge. It will therefore carry an amplitude $(\pm1)^2 = 1$. Moving from a vertex to its correct neighbors will always involve entering at one end of an edge and exiting at the other, and will therefore have the amplitude $1 \times -1 = -1$. It therefore follows that $A_X = D_X - N^tN$.

Now consider navigating the edge set using $N N^t$. There are once again incorrect paths that return to the source edge, which will always have positive amplitude. Correct paths to neighboring vertices will have variable sign depending on which two types of edge ends are involved. If the transition goes between the ends which have the same sign, the transitions amplitude will be positive. Otherwise it will be negative. This is precisely the desired behavior for $\bar{H}_a(X)$, and it therefore follows that $\bar{H}_a(X) = NN^t - 2 I$.

\section{Compact Support Eigenstates}\label{app:compactSupport}

\begin{figure}[h]
	\begin{center}
		\includegraphics[width=0.9\textwidth]{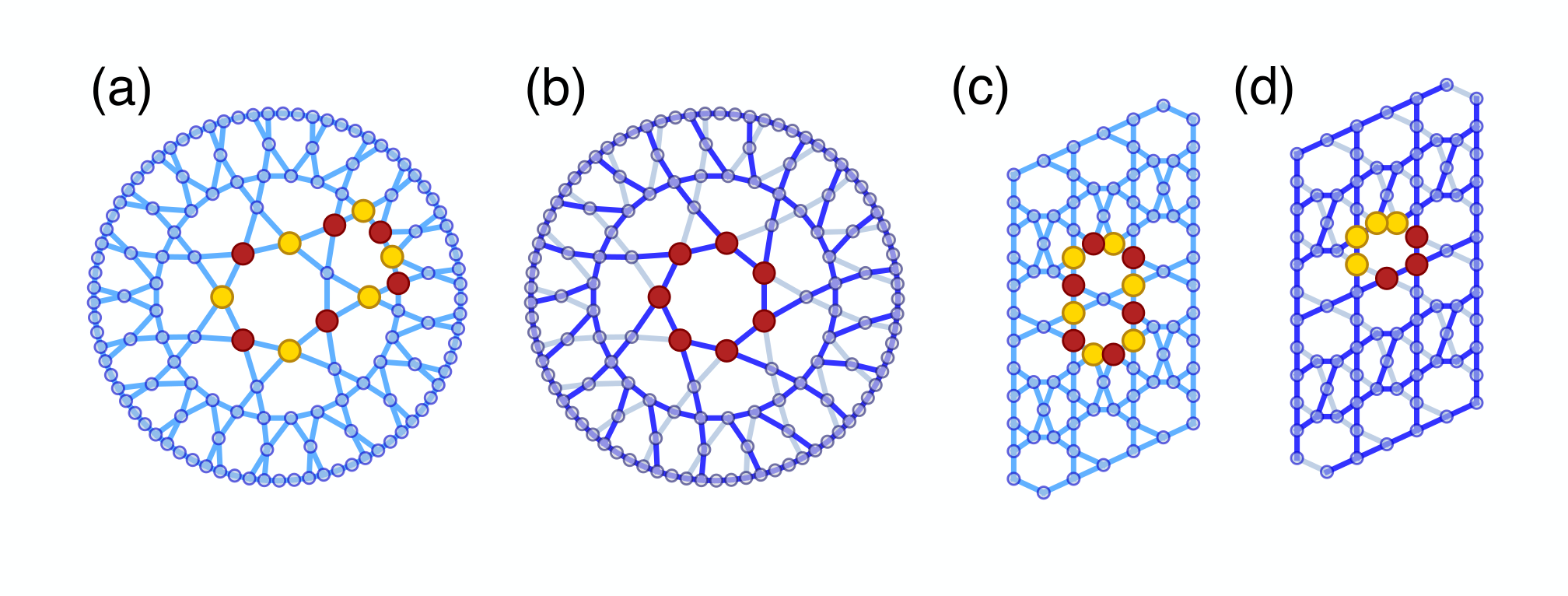}
	\end{center}
	\vspace{-0.6cm}
	\caption{\label{fig:FWHW_FB_States} 
    \textbf{Flat-band states in full-wave and half-wave lattices.} 
    \textbf{a} The smallest flat band state in the heptagon-kagome lattice obtained as the line graph of hyperbolic heptagon-graphene (heptagonal honeycomb) lattice. All hopping matrix elements are negative and shown in light blue. The state is completely alternating and exists on an even cycle enclosing two plaquettes.
    \textbf{b} The smallest flat band state in the heptagon-kagome lattice using half-wave modes. Since the layout is non-bipartite, this Hamiltonian has both positive (dark blue) and negative (light blue) hopping matrix elements. This tight-binding wavefunction encloses a single odd cycle, and the orientation has been chosen to make it particularly simple.
    \textbf{c} A smallest flat band state in a Euclidean tiling obtained as the line graph of $3$-regular tiling of heptagons and pentagons, known as the heptagon-pentgon-kagome lattice. As in its hyperbolic counterpart, this flat band state is alternating and encloses two plaquettes.
    \textbf{d} A smallest flat band state in the heptagon-pentagon-kagome lattice using half-wave modes. This choice of gauge necessitates a sign flip of the tight-binding wavefunction from one side of the plaquette to the other, but this state still encloses only a single plaquette.
    } 
\end{figure}


$\bar{H}_s(X)$ has a compact support eigenstate with eigenvalue $-2$ for every \textit{even} cycle in $X$ and $\bar{H}_a(X)$ has such a state for \textit{every} cycle in $X$.
To see this, we will demonstrate the compact support eigenstates by construction.  Consider first the simpler case of $\bar{H}_s$, and assume that the layout graph $X$ contains an even cycle. This will give rise to a cycle in $L(X)$ with equal length.
Choose a labeling of the vertices of $X$ such that the even cycle is the first $2n$ vertices. 
Letting $\oplus$ and $\ominus$ denote addition and subtraction modulo $2n$, the corresponding cycle in $L(X)$ is then indexed by the unordered pairs of sequential vertices $\{x,y\} = \{y,x\}$ such that $y\equiv x \oplus 1 $. 
Define a state on $L(X)$
\begin{equation}
\psi_c(\{x,y\} ) = \begin{cases} 1, & \mbox{if } x,y \leq 2n \mbox{, } x \mbox{ is even and } y \equiv x \oplus 1, \\
		 -1 & \mbox{if } x,y  \leq 2n \mbox{, } x \mbox{ is odd and } y \equiv x \oplus 1 ,\\
		 0  &  \mbox{otherwise}.
		  \end{cases} 
\end{equation}
The state $\psi_c$ obeys
$$  
\bar{H}_s(X) \psi_c(\{ x,y\} ) = \begin{cases} \psi_c( \{ x\ominus 1,x\} ) + \psi_c(\{ x\oplus 1,x \oplus 2 \} ), & \mbox{if } x,y \leq 2n \mbox{ and } y \equiv x\oplus 1, \\
		\psi_c( \{ x \ominus 1,x \}) + \psi_c(\{ x, x \oplus 1 \} )  &   \mbox{if } x \leq 2n \mbox{ and } y > 2n.\\
		 0 & \mbox{if } x,y > 2n.
		  \end{cases} 
$$
Using the oscillations in $\psi_c(x,y)$, we find $\bar{H}_s \psi_c(x,y) = -2 \psi_c(x,y)$. The state $\psi_c$ is therefore an eigenstate with eigenvalue $-2$ which is localized and of compact support by construction.

Such compact-support eigenstates are highly unusual in generic lattices, but in line-graph lattices they arise due to destructive interference in the plaquettes which surround the vertices of the layout graph because hopping from neighboring sites of opposite sign cancels. In circuit QED lattices this same phenomenon can also be understood at the hardware level as destructive interference between voltages incident on the coupling capacitors at the vertices of the layout. If two of the ports of the coupler have equal and opposite voltage, then the outgoing voltage at the third port is their sum, which is zero. Examples of these compact support eigenstates are shown in Fig. \ref{fig:FWHW_FB_States} \textbf{a},\textbf{c}.
In the absence of even cycles, the states with energy $-2$ are believed to be exponentially localized, as was shown for the Cayley graph of 
$\mathbb{Z}_2 \ast \mathbb{Z}_3$ in Ref.\cite{McLaughlin:1986vb}

The argument for $\bar{H}_a$ is similar, but is simplest in a well chosen orientation of $X$. Assume now that $X$ contains a cycle of any length, even or odd.
As before, this will give rise to a cycle in $L(X)$ of equal length. Consider then states on this cycle.
The analysis in this case is complicated by the need to chose an orientation of $X$ and work in that particular realization of $\bar{H}_a(X)$, but the underlying physics of destructive interference is the same.
To see this, choose a gauge where $\varphi$ goes from negative to positive when moving around the cycle in a clockwise direction. This makes $t_{i,j}$ everywhere positive going around cycle. Resonators that touch the cycle will see one site on the cycle with $\varphi = 1$ and one with $\varphi = -1$, so the $t_{i,j}$ for leaving cycle will come in pairs, one positive and one negative. Consider the state $\psi = 1,1,1,1, \cdots$ around the cycle and zero everywhere else. The tunneling amplitudes onto all neighboring resonators cancel by destructive interference, so $\psi$ is perfectly localized. It therefore follows that 
$$  
\bar{H}_a \psi(i) =2 \psi (i) =  \begin{cases} 2 \psi (i), & \mbox{for } i \mbox{ in the cycle }, \\
		0 &   \mbox{for } i \mbox{ not in the cycle }.
		  \end{cases} 
$$

To prove the existence of a flat band, we need to show that construction of such a localized state can be done simultaneously for many (if not all) cycles in $X$. It is not generally possible to chose a gauge which is as simple as the example above for all cycles simultaneously, but it is not necessary. For any given cycle, the gauge choice described above demonstrates the existence of a localized tight-binding wavefunction with eigenvalue $-2$. Since the tight-binding model is shorthand for the underlying voltage model in Eqn. \ref{HinV2}, the voltage configuration to which this state corresponds exists regardless of the choice of gauge. Therefore, a localized state of compact support on $\mathcal{E}(X)$ which is an eigenstate of $\bar{H}_a(X)$ exists for \textit{every} cycle in $X$, and all 3-regular half-wave lattices have a flat band whether they are Euclidean or hyperbolic; or bipartite, or not. Examples of these states for two non-bipartite examples are shown in Fig. \ref{fig:FWHW_FB_States} \textbf{b} and \textbf{d} alongside their full-wave counterparts.

\section{Reduced $C^*$-algebras}\label{app:Cstar}
In lieu of the band structures coming from Bloch waves in the Abelian setting (see Sec. \ref{sec:euclidean}) the theory of reduced $C^*$-algebras yields some information on the bands and gaps. For $G$ a discrete (finitely-generated) torsion-free group (that is it has no nontrivial elements of finite order), the Kaplansky-Kadison conjecture asserts that the $\ell^2$ reduced group $C^*$-algebra $C_{red}^* (G)$ has no nontrivial idempotents. More precisely, viewing elements $\sum_{g\in G}{c_g g}$ ($c_g \in \mathbb{C}$ and all but finitely many $c_g$'s being zero) in the convolution group algebra $C^* (G)$ as bounded operators on $\ell^2 (G)$ by convolving on the left, $C_{red}^* (G)$ is the closure of these operators in the operator norm of $\ell^2 (G)$. The Kaplansky-Kadison conjecture is known for many such $G$'s and in particular all of the $G$'s encountered in the paper (see Ref.\cite{Puschnigg:2002ju}). An immediate consequence is that the spectrum of any $\mathcal{D}$ in $C^* (G)$ acting on $\ell^2 (G)$ is connected and in particular, if $\mathcal{D}$ is self-adjoint then its spectrum is a single (possibly degenerate) interval. The proof of this idempotent property proceeds by showing that for any idempotent $e$, the (normalized) trace of $e$ is integral and, hence, if it is not $0$, then it must be $1$, and $e = I$. The integrality of the trace extends to $C_{red}^* (G)\otimes \mbox{Mat}_n(\mathbb{C})$, where the trace on the second factor is the usual one on $n\times n$ matrices with complex coefficients. It follows that the spectrum of any self-adjoint $\mathcal{D}$ in $C_{red}^* (G)\otimes \mbox{Mat}_n(\mathbb{C})$ consists of at most $n$ bands.

Following Sunada\cite{Sunada:1992tv} we apply this to our infinite layouts $X$ on which a torsion free $G$ acts as automorphisms with $|G \backslash V(X)| = n$. In this case we can realize $A_X$ with its action on $\ell^2(X)$ as an element of $C_{red}^* (G)\otimes \mbox{Mat}_n(\mathbb{C})$ and conclude that $\sigma(A_X)$ has at most $n$ bands. For example, if $G = F_k$ the free group on generators $g_1,g_2, \ldots g_k$; $\mathcal{D}= g_1 + g_1^{-1} + \cdots g_k + g_k^{-1}$; and $X$ is the $2k$-regular tree realized as the Cayley graph of $G$ with generators $g_1, g_1^{-1}, \ldots g_k, g_k^{-1}$; then $\mathcal{D}$ is identified with $A_X$. According to the above $\sigma(A_X)$ consists of one band. As we have noted before, according to Kesten\cite{Kesten:1959us} $\sigma(A_X) = [-2\sqrt{2k-1}, 2\sqrt{2k-1}]$. Another example is $X = \mathbb{M}$, the McLaughlin graph from Ref.\cite{McLaughlin:1986vb} and Sec. \ref{sec:gengraphs}. The group $G = \left< R\right > * \left< Q \right >$ with $Q^2 = R^3 = 1$ acts simply transitively on $V(\mathbb{M})$. The kernel $\Gamma$ of $\phi: G \rightarrow \mathbb{Z}/2\mathbb{Z} \times \mathbb{Z}/3\mathbb{Z}$ where $Q\rightarrow (1,0)$, $R\rightarrow (0,1)$ has index $6$ in $G$ and is torsion free. Hence $|\Gamma \backslash \mathbb{M} |= 6$, and we can apply the idempotent theorem to conclude that $A_\mathbb{M}$ has at most $6$ bands. In fact, as McLaughlin showed, $A_\mathbb{M}$ has $4$ bands. Without a further understanding of the projections this technique of embedding $\mathcal{D}$ in such $C^*$-algebras gives upper bounds on the number of bands; bounds which we do not expect to be sharp.

\begin{figure}[h]
	\begin{center}
		\includegraphics[width=0.35\textwidth]{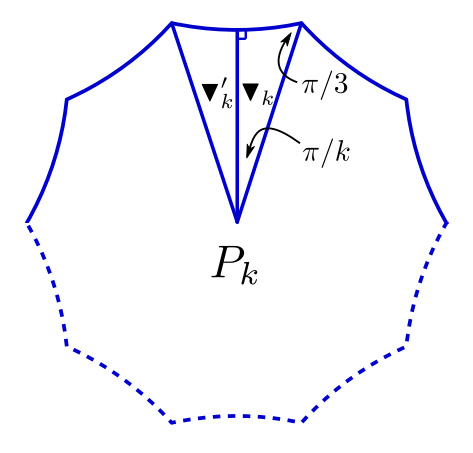}
	\end{center}
	\vspace{-0.6cm}
	\caption{\label{fig:hyperpoly} 
    \textbf{Fundamental domains and symmetry generators.} 
    Schematic drawing of a single plaquette of $T_k$, showing the two hyperbolic triangles $\blacktriangle_k$ and $\blacktriangle_k^\prime$ whose reflections generate the symmetry group $G_k$, and whose union constitutes the fundamental domain for the largest torsion-free subgroup $\Gamma_k$ in $G_k$.
    } 
\end{figure}

We apply this to the tessellation graphs $T_k$, $k \geq 7$ for which this $C^*$-algebra method is the only one that we know of that controls the number of bands. The full symmetry group $G_k$ of $T_k$ is the Coxeter reflection group $[k,3]$
that is the reflection group of a hyperbolic triangle $\blacktriangle_k$ with angles $(\pi/2, \pi/3, \pi/k)$, shown in Fig. \ref{fig:hyperpoly}.
$G_k$ has a presentation with generators $R_1,R_2,R_3$ and relations $R_1^2 = R_2^2 = R_3^2 = (R_1 R_2)^2 = (R_2 R_3)^3 = (R_3 R_1)^k = 1$. The index-$2$ subgroup $\mathcal{D}_k$ of $G_k$ consisting of even words in the $R$'s, and consists of the orientation preserving isometries of the hyperbolic plane $\mathbb{H}$ that preserve $T_k$. $\mathcal{D}_k$ acts transitively on $V(T_k)$ with the stabilizer of any $v \in V(T_k)$ being order $3$. $\mathcal{D}_k$ acts on $\mathbb{H}$ with a fundamental domain $\blacktriangle_k \cup \blacktriangle^\prime_k$ of area $(k-6) \pi /3k$ (shown in Fig. \ref{fig:hyperpoly}) and it has a presentation with generators $A,B,C$ and relations 
\begin{equation}\label{eqn:abc}
A^2 = B^3  = C^k = ABC = 1.
\end{equation}

In order to apply the $C^*$-algebra results above we seek the largest torsion-free subgroup $\Gamma_k$ of $\mathcal{D}_k$. If $\Gamma$ is  such a group of index $m = m_\Gamma$, then the compact orientable hyperbolic surface $\Gamma \backslash \mathbb{H}$ has genus $g = g_\Gamma \geq 2$. The Gauss-Bonnet formula relates the genus to its area.
\begin{equation}\label{eqn:area}
\mbox{Area}(\Gamma \backslash \mathbb{H} )= 4 \pi (g_\Gamma - 1).
\end{equation}
On the other hand, a fundamental domain for the action of $\Gamma$ on $\mathbb{H}$ consists of $m$ copies of $\blacktriangle_k \cup \blacktriangle_k ^\prime$. The area of $\blacktriangle_k$ is $(\pi - \pi/2-\pi/3 - \pi/k)$ which leads to the relation
\begin{equation}\label{eqn:mkg}
m(k-6) = 12(g-1).
\end{equation}
The smallest integer solution to Eqn. \ref{eqn:mkg} depends on the factorization of $k$ and is given by $m_k$ in Table \ref{table:numbands}  assuming the factorization $k = 2^a 3^b k_1$, with $k > 6$ and $k_1 \equiv 1 \mod 6$.
\begin{table}[ht]
\caption[A]{}
\centering
\begin{tabular}{ c| c| c| c}
a & b & $m_k$ &$g_k$ \\ 
  \hline
 0 & 1 & 12k & k-5 \\ 
 0 & $\geq$ 1 & 4k & (k-6)/3+1 \\ 
1 & 0 & 3k & (k-6)/4 +1 \\ 
1 & $\geq$ 1 & k & (k-6)/12 +1\\
$\geq$ 2 & 0 &6k & (k-6)/2 +1 \\
$\geq$ 2 & $\geq$ 1 & 2k & (k-6)/6 + 1\\
\end{tabular}
\label{table:numbands}
\end{table}
Thus, the smallest index of a $\Gamma$ in $\mathcal{D}_k$ which is torsion free is at least $m_k$. In fact, one can find a torsion free $\Gamma_k$ in $\mathcal{D}_k$ with index $m_k$ (see Ref.\cite{Edmonds:1982vv} where this smallest index is resolved for a Fuchsian group).

Now $\Gamma_k$ acts freely on $V(T_k)$ and since the stabilizer of any $v \in V(T_k)$ has order $3$ in $\mathcal{D}_k$, it follows that the number of orbits of $\Gamma_k$ on $V(T_k)$ is 
\begin{equation}\label{eqn:maxnumbands}
|\Gamma_k \backslash V(T_k)| = \frac{m_k}{3}.
\end{equation}
According to the idempotent results applied to $C_{red}^* (\Gamma_k) \otimes \mbox{Mat}_{m_k/3}(\mathbb{C})$, we deduce that $\sigma(T_k)$ has at most $m_k/3$ bands.

\section{The Gap at $-2$ for $\bar{H}_s(X)$ and finite $X$}\label{app:Hsgap}

Our aim is to prove the lower bound of Eqn. \ref{thm:cycledensity} which according to Eqn. \ref{eqn:simgaH1} amounts to giving a lower bound for $\lambda_{\text{min}} \left(D_X + A_X \right) $ for a layout $X$. One could proceed as in the proof of Eqn. \ref{eqn:ksquaredbound}, however there is an illuminating combinatorial characterization of $\lambda_{\text{min}}$ being bounded from below (uniformly as $|X| \rightarrow \infty$) that we use instead. For $S \subset V(X)$, let cut($S$) be the set of edges of $X$ with one end point in $S$ and the other outside $S$. Let $e_{\text{min}}(S)$ be the minimum number of edges of $S$ that need to be removed (here $S$ is the induced subgraph) so that the resulting graph is bipartite. Set 
\begin{equation}\label{equ:d1}
    \digamma (X) = \min_{\phi \neq S \subset V(X)} \frac{e_{\min}(S) + |\text{cut} (S)|}{|S|}
\end{equation}
The following is a bipartite analogue of the combinatorial Cheeger inequality\cite{Alon:1985wi} and is formulated and proved in Ref.\cite{Desai:1994uv}
\begin{equation}\label{equ:d2}
    \frac{\digamma ^2(X)}{4 \, d^{*}(X)} \leq \lambda_{\text{min}} (D_X + A_X) \leq 4 \digamma (X),
\end{equation}
where $d^{*}(X)$ is the maximum degree of any vertex of $X$.

We use this to prove that if $r \geq 2$ is fixed and $X$ is any layout for which the induced subgraphs on $B_r (x) = \left\{ y \in X : d(y, x) \leq r \right\} $ are non-bipartite for all $x \in X$, then 
\begin{equation}\label{equ:d3}
    \lambda_{\text{min}} (D_X + A_X) \geq \left[48 \left(3.2^{2 r -1} - 1 \right)^2 \right]^{-1}.
\end{equation}
To apply Eqn. \ref{equ:d2} we estimate $\digamma(X)$ for $X$'s which satisfy this local non-bipartite condition. Let $S \subset V(X)$ and choose $W \subset S$ a maximally $2r$-separated subset, that is for $v, w\in W$ $d_X (v, w) > 2r$. Then 
\begin{equation}\label{equ:d4}
    |W| \geq \frac{|S|}{M_{2r}}, \ \ \text{with} \ \   M_t = 3.2^{t} - 2.
\end{equation}
If this is not true then
\begin{equation}
    |W| < \frac{|S|}{M_{2r}},
\ \ \text{with} \ \
    |\bigcup\limits_{w\in W} B_{2r}(w)| < |W| M_{2r} < |S|,
\end{equation}
where we have used the fact that for a layout $|B_t (x)| \leq M_t$.
In this case there would exist $s \in S$ such that $s\notin B_{2r} (w)$ for any $w \in W$, but then $s \cup W$ is $2r$-separated and larger than $W$, which contradicts the assumption that $W$ was maximal. Thus, Eqn. \ref{equ:d4} holds.

For $w \in W$ either
\begin{itemize}
    \item[(i)] $B_r (w) \cap S = B_r (w)$, in which case this induced subgraph is non-bipartite by assumption, and hence this local contribution to $e_{\text{min}}(S)$ in Eqn. \ref{equ:d1} is at least 1, or
    \item[(ii)] $B_r(w)\cap S \subsetneqq B_r(w)$ in which case the local contribution to $\text{cut}(S)$ is at least 1, and this edge is in $B_r(w)$.
\end{itemize}

Thus, in either case the local contribution to the numerator in Eqn. \ref{equ:d1} is at least 1. Since the different $B_r(w)$'s with $w\in W$ are disjoint it follows from above that 
\begin{equation}
    \digamma (X) \geq \frac{|S|}{M_{2r}|S|} = (M_{2r})^{-1}
\end{equation}
which establishes Eqn. \ref{equ:d3}.

\section{Classification of Hoffman Layouts}\label{app:hoffman}
$X$ is a layout (i.e. has degree at most 3) and $A_X$ its adjacency operator. Thanks to the results of Refs. \cite{Hoffman:1977uw, Cameron:1975vu,Doob:1979vt}, it is known that if $\lambda_{\text{min}}(A_X)> -2$, then $X$ is either a generalized line graph (see below for the definition) $L(T; 1, 0, ..., 0)$ with $T$ a tree; a line graph $L(\mathcal{K})$ with $\mathcal{K}$ a tree or an odd cycle; or it is one of a finite list of $X$'s (see Theorem 2.1 of Ref. \cite{Doob:1979vt}). In the case that $X = L(\mathcal{K})$ and $\mathcal{K}$ is a tree, it follows from Eqn. \ref{eqn:simgaH1} that $\sigma (A_X) = -2 + \sigma (D_\mathcal{K} + A_\mathcal{K})$ and since $\mathcal{K}$ is bipartite $\sigma(A_X) = -2 + \sigma(D_\mathcal{K} - A_\mathcal{K})$. Moreover, $\mathcal{K}$ is planar and of bounded degree and hence from Ref. \cite{Lipton:1980wt} it follows that
\begin{equation}\label{equ:e1}
    \lim_{|\mathcal{K}|\rightarrow \infty} \lambda_{\text{min}}\left(A_{L(\mathcal{K})} \right) \leq -2.
\end{equation}
 In as much as $T$ is an induced subgraph of $L(T; 1, 0,...,0)$ it follows that Eqn. \ref{equ:e1} continues to hold for $X = L(T; 1, 0,...,0)$ when $|T|\rightarrow \infty$. Eqn. \ref{equ:e1} also holds for $\mathcal{K}$ an odd cycle and $|\mathcal{K}|\rightarrow \infty$. Putting these together we deduce that for $X$ a layout 
 \begin{eqnarray}\label{equ:e2}
 \limsup_{|X| \rightarrow \infty } \lambda_{\text{min}} (A_X) \leq -2.
 \end{eqnarray}
 
 Write $\mathbb{E}_X = 3 I - D_X$, then $\mathbb{E}_X$ is diagonal with nonnegative entries and hence 
 \begin{eqnarray}
     \lambda_{\text{min}}(D_X + A_X) &=& \lambda_{\text{min}} (3I - \mathbb{E}_X + A_X) \\
     &\leq& \lambda_{\text{min}} (3I + A_X) \leq 3 + \lambda_{\text{min}}(A_X).
 \end{eqnarray}
 Therefore from Eqn. \ref{equ:e2}
\begin{equation}
    \limsup_{|X|\rightarrow \infty} \lambda_{\text{min}} (D_X + A_X) \leq 3-2 = 1.
\end{equation}
 From Eqn. \ref{eqn:simgaH1} this gives that 
 \begin{equation} \label{equ:e3}
 \limsup_{|X|\rightarrow \infty} \lambda (\bar{H}_s (X)) \leq -1,
 \end{equation}
proving Eqn. \ref{eqn:Haextremalgap}.

For the rest of this Appendix, $X$ is a 3-regular layout. For these one can strengthen Eqn. \ref{equ:e2} to (see Thm. 2.5 in Ref. \cite{Doob:1979vt}):
\begin{equation}
    \lambda_{\text{min}} (A_X) \leq -2,
\end{equation}
except for the case that $X$ is the complete graph $T_3$. The case that $\lambda_{\text{min}}(A_X) = -2$, that is when $X$ is a 3-regular Hoffman graph, can be characterized. According to Ref. \cite{Cameron:1975vu}, any such graph is one of a finite number of graphs or a generalized line graph.
In more detail the latter are given as follows: there is a connected graph $Y$ with $n$ vertices $v_1,v_2,...,v_n$ and non-negative integers $a_1,a_2,...,a_n$ and the ``cocktail party'' graphs $CP(a_j)$ with $2a_j$ vertices and degree $2a_j -2$. ($CP(0)$ is the empty set, $CP(1) =  \includegraphics[height=0.5cm]{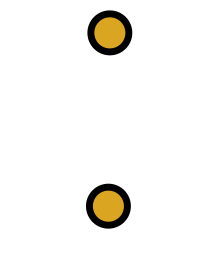}$,
$CP(2) =  \includegraphics[height=0.5cm]{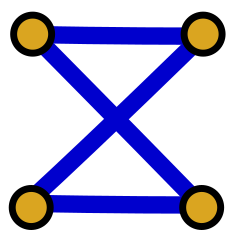}$, ...) The generalized line graph $X = L(Y; a_1,...,a_n)$ is defined by: $X$ has as its vertices those of $L(Y)$ as well as those of $CP(a_1)$,..., $CP(a_n)$, and also all the edges of these graphs together with extra edges joining any edge $e$ of $Y$ (i.e. a vertex of $L(Y)$) to all the vertices of $CP(a_j)$ if $v_j$ is an end point of $e$.

In order that $L(Y; a_1, \ldots , a_n)$ be $3$-regular we must have $a_j = 0$ or $1$ for all $j$. For if $a_j \geq2$ for some $j$ then any edge emanating from $v_j$ will have degree bigger than $3$ in $X$. Actually, $a_j = 1$ is also impossible since $CP(1) =  \includegraphics[height=0.5cm]{cp1.png}$ and if $v_j \in Y$ has degree $3$ or more than any edge $e$ emanating from $v_j$ would have degree at least $4$ in $X$. On the other hand, if $v_j$ in $Y$ has degree $1$ or $2$ then the vertices of $X$ corresponding to $CP(a_j)$ have degree less than three. It follow that all the $a_j$'s are $0$ and hence a large $3$-regular Hoffman graph $X$ is of the form $X = L(Y)$. In order that $X$ be $3$-regular one checks that $Y$ has to be $3,2$-biregular. A $3,2$-biregular graph $Y$ is obtained as a subdivision graph of a $3$-regular graph $Z$. We conclude that a $3$-regular Hoffman graph $X$ is of the form
\begin{equation}\label{eqn:e4}
X = L(\mathbb{S}(Z)) \mbox{ for some 3-regular } Z.
\end{equation}
This competes the proof of the claims in Sec. \ref{subsec:finitelayouts} that all regular Hoffman layouts are achieved by the process leading to Eqn. \ref{eqn:Hsgapbound}.

We conclude this Appendix with proofs of Eqns. \ref{eqn:McLaughlinGaps} and \ref{eqn:CharonGaps}. Firstly, if $X$ is $[-3,-2)$-gapped and is large, then according to the discussion above, $X = L(\mathbb{S}(Z))$ for a cubic $Z$. From the Eqn. \ref{eqn:XLSX} in Appendix \ref{app:subdivisionGraphs} it follows that the rest of the gap set in Eqn. \ref{eqn:McLaughlinGaps} corresponds exactly to $Z$ being $[-3,-2\sqrt{2}) \cup (2 \sqrt{2}, 3]$-gapped. Moreover, non-bipartite Ramanujan $Z$'s achieve these gaps. Equation \ref{eqn:McLaughlinGaps} then follows by combining these observations. Equation \ref{eqn:CharonGaps} is obtained similarly by choosing $Z$ to be a Hoffman graph.

\section{Subdivision Graphs and their Line Graphs} \label{app:subdivisionGraphs}

Let $X$ be a regular graph of degree three with $n = 2\nu$ vertices. Its subdivision graph $\mathbb{S}(X)$ is a $3,2$-biregular graph  with $m = 3\nu$ vertices of degree $2$ and $n = 2\nu$ vertices of degree $3$ with $6\nu$ edges. Next, form the line graph $L(\mathbb{S}(X))$. It is a cubic graph with $6\nu$ vertices. The adjacency matrices of these three graphs are closely related and their characteristic polynomials can be related to one another using block matrix identities. The treatment we give here is a summary of the proofs given Chapter 2 of Ref.\cite{Cvetkovic:1980}, specializing to the case of a starting graph of degree three.

Let $M$ and $N$ be the incidence operators of the starting graph $X$ as defined in Sec.\ref{sec:gengraphs}. The adjacency matrices of $X$, $\mathbb{S}(X)$ and $L(\mathbb{S}(x))$ can be written as
$$A_X = M^t M - D_X, $$
\begin{equation}\label{eqn:splitAdjacency}
A_{\mathbb{S}(X)} = 
\begin{bmatrix}
    0 & M \\
    M^t & 0,
\end{bmatrix}
\end{equation}
\begin{equation}\label{eqn:linegraphofsplitAdjacency}
A_{L(\mathbb{S}(X))} = 
\begin{bmatrix}
    M \left(\frac{M^t + N^t}{2} \right) - I_{3\nu} &  I_{3\nu} \\
    I_{3\nu}                   & M \left(\frac{M^t - N^t}{2}\right) - I_{3\nu},
\end{bmatrix}
\end{equation}
where $I_{l}$ is the $l\times l$ identity matrix. 

Two lemmas from linear algebra are required to relate the characteristic polynomials of these matrices.
First, given an $m \times n$ matrix $B$, the characteristic polynomials of $B^t B$ and $B^t$ are related by
\begin{equation}\label{eqn:BtBBBt}
\lambda^n P_{BB^t}(\lambda) = \lambda^m P_{B^t B}(\lambda).
\end{equation}
Second, given a $2\times 2$ block matrix with square diagonal blocks, its determinant can be computed from the blocks if one of the diagonal blocks is non-singular:
\begin{equation}\label{eqn:BMidentity}
\left|\begin{bmatrix}
    B_1 & B_2 \\
    B_3 & B_4,
\end{bmatrix}
\right| = 
|B_1| \times |B_4 - B_3 B_1^{-1}B_2|.
\end{equation}
Applying these identities to $X$, $\mathbb{S}(X)$, and $L(\mathbb{S}(X))$ yields three relations.
\begin{equation}\label{eqn:XSX}
P_{\mathbb{S}(X)}(\lambda) = (\lambda)^\nu \times P_{X}(\lambda^2 -3),
\end{equation}
and
 \begin{eqnarray}\label{eqn:XLSX}
P_{L(\mathbb{S}(X))}(\lambda) &=&(\lambda+2)^\nu \times (\lambda)^\nu \times P_{X}(\lambda^2 -\lambda - 3)\\
 & = &P_{L(X)}(\lambda^2 -\lambda - 2). \nonumber
 \end{eqnarray}
These in turn can be converted to the eigenvalue relations
\begin{equation}\label{eqn:EXESX}
E_{\mathbb{S}(X)}= \begin{cases} \pm\sqrt{E_X +3}\\
		 0,
		  \end{cases} 
\end{equation}
and 
\begin{eqnarray}\label{eqn:EXSXLSX}
E_{L(\mathbb{S}(X))} &=& \begin{cases} \frac{1 \pm \sqrt{1 + 4(E_X + 3)}}{2}\\
		0,\\
		 -2
		  \end{cases}  \\
          &=& \frac{1 \pm \sqrt{1 + 4(E_{L(X)} + 2)}}{2}. \nonumber
\end{eqnarray}

\begin{acknowledgements}
We thank David Huse, J\'{a}nos Koll\'{a}r, Charles Fefferman, Siddharth Parameswaran, and P\'{e}ter Csikv\'{a}ri for helpful discussions.
This work was supported by the NSF, the Princeton Center for Complex Materials DMR-1420541, and by the MURI W911NF-15-1-0397.
\end{acknowledgements}


%

\end{document}